\documentclass[sn-basic]{sn-jnl}

\usepackage{graphicx}
\usepackage{multirow}
\usepackage{amsmath,amssymb,amsfonts}
\usepackage{amsthm}
\usepackage{mathrsfs}
\usepackage[title]{appendix}
\usepackage{xcolor}
\usepackage{textcomp}
\usepackage{manyfoot}
\usepackage{booktabs}
\usepackage{algorithm}
\usepackage{algorithmicx}
\usepackage{algpseudocode}
\usepackage{listings}
\theoremstyle{thmstyleone}

\theoremstyle{thmstyletwo}

\theoremstyle{thmstylethree}

\begin{document}

\title[Classification  of carbon stars from HES survey]{Classification and characterization using HCT/HFOSC spectra of carbon stars selected from the HES survey}

\author*[1,2]{\fnm{Meenakshi} \sur{Purandardas}}\email{meenakshi.p@christuniversity.in}

\author[1]{\fnm{Aruna} \sur{Goswami}}\email{aruna@iiap.res.in}

\affil*[1]{\orgname{Indian Institute of Astrophysics}, \orgaddress{\street{, Koramangala}, \city{, Bangalore}, \postcode{ 560034}, \state{Karnataka}, \country{India}}}
\affil*[2]{\orgname{Department of Physics and Electronics, CHRIST (Deemed to be University)}, \orgaddress{\city{Bangalore}, \postcode{ 560029}, \state{Karnataka}, \country{India}}} 

\abstract{We present results from the analysis of 88 carbon stars selected from Hamburg/ESO (HES) survey using low-resolution spectra (R$\sim$1330 \& 2190). The spectra were obtained with the Himalayan Faint Object Spectrograph Camera (HFOSC) attached to the 2-m Himalayan Chandra Telescope (HCT). Using a well-defined spectral criteria based on the strength of carbon molecular bands, the stars are classified into different groups. In our sample, we have identified 53  CH stars, four C-R stars, and two C-N type stars. Twenty-nine stars could not be classified due to the absence of prominent C$_{2}$ molecular bands in their spectra. 
We could derive the atmospheric parameters for 36 stars. The surface temperature is determined using photometric calibrations and synthesis of the H-alpha line profile. The surface gravity log g estimates are obtained  using parallax estimates from the Gaia DR3 database whenever possible. Microturbulent velocity ($\zeta$) is derived using calibration equation of log g \& ${\zeta}$. We could  determine metallicity for 48 objects  from near-infrared Ca II triplet features using calibration equations. The derived metallicity  ranges from  $-$0.43$\leq$[Fe/H]$\leq$$-$3.49. Nineteen objects are found to be  metal-poor  ([Fe/H] $\leq$$-$1), 14 very metal-poor ([Fe/H]$\leq$$-$2), and five extremely metal-poor ([Fe/H]$\leq$$-$3.0) stars. Eleven objects are found to have a metallicity in the range $-$0.43 $\leq$[Fe/H]$\leq$$-$0.97. We could derive the carbon abundance for 25 objects using the spectrum synthesis calculation of the C$_{2}$ band around 5165\AA. The most metal-poor objects found will make important targets for follow-up detailed chemical composition studies based on high-resolution spectroscopy, that are likely to provide  insight into the  Galactic chemical evolution.}
\keywords{ stars: Population II,stars: Late-type, stars: carbon, stars: atmospheric parameters, stars: low-mass, stars: metallicity}

\maketitle

\section{Introduction}\label{sec1}
Very metal-poor stars play a crucial role in unraveling the chemical evolution of the  early Galaxy, as they retain the chemical signatures from the nucleosynthesis processes of preceding stellar generations.
Among the metal-poor stars, stars that exhibit enhancement of carbon ([C/Fe] $>$ 1, \cite{Beers_2005} ; [C/Fe] $>$ 0.7, \cite{Aoki_2007}) are known as carbon-enhanced metal-poor (CEMP) stars, and are  excellent probes to study the early galactic nucleosynthesis and  chemical evolution. These stars are believed to have originated from the remnants of Population III stars and act as the fossil records of their nucleosynthetic products. Particularly, the CEMP-no stars are considered bonaﬁde second-generation stars \citep{Placco_2015,hansenetal.2016A&A...588A...3H,Yoon_2016}. This group exhibits the enhancement of carbon without any enhancement of heavy elements. In contrast, other sub-classes of CEMP stars such as CEMP-s, CEMP-r, and CEMP-r/s stars exhibit enhancement of neutron-capture elements. 

\par While CEMP-s stars show enhancement of slow-neutron capture(s-) process elements, CEMP-r stars exhibit enhancement of rapid neutron-capture(r-) process elements. CEMP-r/s sub-group shows enhancements of both s- and r-process elements. Studies on CEMP-s and CEMP-r/s stars can give insight into the formation and evolution of low- and intermediate-mass stars. Many studies have shown that these stars are found in binary systems \citep{McClure&Woodsworth1990ApJ...352..709M, preston&sneden2001AJ....122.1545P, hansenetal.2016A&A...588A...3H, jorrisen.2016A&A...586A.159J,Purandardas_2021b}, and their peculiar surface chemical compositions are attributed to the mass transfer from their binary companions. Understanding of the formation and evolution of low- and intermediate- mass stars are of special significance as they can give important clues regarding the galactic chemical evolution. About half of the heavy elements in our galaxy were produced by the low- and intermediate-mass stars during their evolution and while passing through  their asymptotic giant branch (AGB) phase of evolution \citep{lugaro.2003ApJ...586.1305L,herwig.2005ARA&A..43..435H,karakas.2014PASA...31...30K}.  The origin of CEMP-r stars remains unclear, potentially linked to core-collapse supernovae and neutron star mergers  that are the known sources of the r-process elements\citep{arcones.2013JPhG...40a3201A, rosswog.2014MNRAS.439..744R, abbott.2017ApJ...848L..13A, drout.2017Sci...358.1570D, shappee.2017Sci...358.1574S}. Thus CEMP stars play an important role as the tracers of neutron-capture  nucleosynthesis. CEMP stars, especially CEMP-s stars, outnumber CEMP-r/s and CEMP-r stars. CEMP-r stars are particularly rare\citep{hansen.2011ApJ...743L...1H}. \cite{Beers_2005} provide an extensive review on  metal-poor stars and classification of CEMP stars based on the abundances of barium and europium.  \cite{Partha_2021} present a recent discussion on the classification scheme of CEMP stars, distinguishing between CEMP-s and CEMP-r/s stars.

\par CEMP stars represent a subgroup of carbon stars, easily identified through low-resolution spectra. The  spectra of carbon stars are characterized by the dominant Swan bands of  C$_{2}$ molecules and other carbon compounds such as CH and CN, and these features are  easily detectable  on the spectra of cool carbon stars. \cite{Bidelman_1956} introduced CH stars as a distinct group of carbon stars that show 
a strong CH band in the spectra. CEMP-s stars are recognized  as the  metal-poor counterparts of CH stars \citep{Lucatello_2005, Abate_2016}. Furthermore, \cite{Jorrisen_2016} showed that CH and CEMP-s stars fall in the same region of the period–eccentricity diagram and represent similar mass-function distributions. Based on these findings, they suggest treating CH and CEMP-s stars as a unique stellar family. These groups of stars are found to exhibit enhanced neutron-capture elements, particularly s-process elements. The origin of their peculiar surface chemical composition is attributed to their binary companion. Many studies have shown that most of the CH stars are in binary systems in which the primary companion is now an invisible white dwarf that once transferred the nucleosynthesis materials to the secondary star (CH/CEMP-s stars) during its AGB phase \citep{McClure&Woodsworth1990ApJ...352..709M, preston&sneden2001AJ....122.1545P, hansenetal.2016A&A...588A...3H, jorrisen.2016A&A...586A.159J,Purandardas_2021b}.

\par Besides CH stars, carbon stars are further classified into C-N, C-J, and C-R groups based on spectral properties. C-N stars exhibit strong depression of light in the violet region of their spectra, lower effective temperatures, and strong molecular bands compared to C-R stars  (\cite{Barnbaum_1996}, \cite{Goswami_2005}, \cite{Goswami_2010}). The carbon isotopic ratios exhibited by most of the C-N stars lie in the range 30 to 100, while CH, and C-R stars 
show $^{12}$C/$^{13}$C in the ranges 20-100, and 4-9 respectively \citep{Lambert_1986}. Classification based on isotopic ratios would however require high-resolution spectra.

\par Among the  C-R stars and  CH stars, the presence of the strong Ca I line at 4226 \AA\, in C-R stars distinguishes them from CH stars \citep{Goswami_2005}. High-resolution spectroscopy indicates that C-R stars exhibit heavy element abundances similar to the Sun, unlike CH stars.

\par Carbon stars that are characterized by strong Merrill-Sanford (M-S) bands due to SiC$_{2}$ in 
their spectra are known as C-J group of stars.  These bands are generally absent in the spectra of CH stars, and a few C-N stars show the presence of SiC$_{2}$ band. Similar to C-R stars, C-J stars are also found to exhibit a low carbon isotopic  ratio, typically below 15 \citep{Goswami_2005,Goswami_2007,goswami_2010b}. 
Although many high-resolution spectroscopic analyses are available for CH stars \citep{Perea.2003,Aoki_2007,Purandardas_2019a,Karinkuzhi_2014,Karinkuzhi_2015}, such studies are quite scanty for C-R, and C-J stars. 

\par   \cite{Christlieb_2001} compiled a list of faint high-latitude carbon (FHLC) stars, potentially containing various sub-groups of carbon stars. Distinguishing between these classes is crucial for understanding their astrophysical implications. Hence, we have undertaken to identify stars belonging to  the different  sub-groups of carbon stars and to derive their stellar atmospheric parameters, and carbon abundances by selecting objects from this list. 

In this work, we have sampled 88 carbon stars taken from \cite{Christlieb_2001}, and classified 65 of them for the first time using low-resolution spectra and estimated their atmospheric parameters and carbon abundances whenever possible. The remaining 23 stars were already classified into different sub-groups of carbon stars by \cite{Goswami_2005} and \cite{Goswami_2007,goswami_2010b}. We have derived stellar atmospheric parameters for these stars and also the carbon abundance whenever possible. In this paper,  we present the results of these analyses. The paper is arranged as follows: observations and data reductions are described in Section \ref{sec2}. Spectral characteristics and classifications of these
stars are presented in Section \ref{sec3}. In Section \ref{sec4} we have described methods used
for the determination of the stellar atmospheric parameters. Determination of carbon abundance is discussed in Section \ref{sec5}, and discussions and conclusions are presented in Section \ref{sec6}.

\section{Observations and Data Reduction}\label{sec2}
The programme stars are selected from the list of stars from HES survey and particularly from the list of faint high-latitude carbon stars of \cite{Christlieb_2001}. The  spectra  were obtained using the Himalayan Faint Object Spectrograph Camera (HFOSC) attached to 2-m Himalayan Chandra Telescope (HCT) at the Indian Astronomical Observatory (IAO), Hanle during 2010 to 2021. The stars were selected based on their observability using this facility. HFOSC is an optical imager cum spectrograph used for low- and medium- resolution grism spectroscopy. The grism and the camera combinations used for the observations provide a spectral resolution R (${\lambda}/{\delta\lambda}$) $\sim$ 1330 for grism 7 and $\sim$ 2190 for grism 8. We have obtained spectra using  grism 7 and 8 that cover the wavelength regions from 3800 - 6840 {\rm \AA} and 5800 - 8350 {\rm \AA} respectively. Observations of Fe-Ar and Fe-Ne hollow cathode lamps taken immediately before and after the stellar exposures are used for  wavelength calibrations. The spectra of the stars used for the comparison and identification of the spectral type of the program stars were obtained during earlier observation cycles  using the same observational set-up. Data reduction was performed following the standard procedures using IRAF\footnote{IRAF is distributed by the National Optical Astronomical Observatories, which is operated by the Association for Universities for Research in Astronomy, Inc., under contract to the National Science Foundation} software spectroscopic reduction package. For each object, minimum of two spectra each of 30 minutes exposures were
taken and combined to increase the signal-to-noise ratio. 

{\footnotesize
\begin{table*}
\caption{HE stars with prominent C$_{2}$ molecular bands}\label{tab1}
\resizebox{\textwidth}{!}{\begin{tabular}{lcccccccccccc}
\hline
Star No.     & RA$(2000)$ & Dec.$(2000)$ & $l$  & b  & B & V   &  J     & H      & K & Bands & Obs. date & Class \\
             &            &              &      &    &   &     &        &        &    & noticed & & \\
\hline
HE 0002+0053$^{\ast}$	&	00 05 24.99	&	01 10 03.82	&	99.71	&	-59.61	&	15.02	&	13.3	&	11.02	&	10.38	&	10.12 & C$_{2}$, CH, CN	& 06.11.04	 & C-R	\\
             &            &              &      &    &   &     &        &        &    &  & 20.10.20 & \\
             &            &              &      &    &   &     &        &        &    &  & 21.07.21 & \\             
HE 0017+0055$^{\ast}$	&	00 20 21.59	&	01 12 06.81	&	106.9	&	-60.69	&	12.99	&	11.66	&	9.31	&	8.68	&	8.49 & C$_{2}$, CH, CN	&	15.11.03 & CH/CEMP-r/s	\\
             &            &              &      &    &   &     &        &        &    &  & 21.10.20 & \\
             &            &              &      &    &   &     &        &        &    &  & 14.09.21 & \\
HE 0228-0256$^{\ast \ast}$	&	02 31 15.6	&	-02 43 06.45	&	171.59	&	-55.85	&	16.69	&	14.7	&	12.51	&	11.81	&	11.53 & C$_{2}$, CH, CN	&	18.01.09 & CN	\\
HE 0037-0654$^{\ast}$ & 00 40 01.98 &  -06 38 12.39 & 114.85 & -69.33 & 16.69 & 15.50 & 14.15 & 13.71 & 13.72 & C$_{2}$, CH, CN & 21.10.20 & CH/CEMP-s \\
HE 0039-2635 & 00 41 39.7 & -26 18 53  &52.81&-87.67& 13.10  & -   &10.571 & 10.11 & 9.99 & C$_{2}$, CH, CN & 06.12.16 & CH/CEMP-r/s \\ 
HE 0113+0110 & 01 15 52.2 & +01 26 21  &135.54&-60.83& 16.30	 & 15.00  &  13.02 & 12.36 & 12.23 & C$_{2}$, CH, CN & 06.12.16&CH/CEMP-r/s\\ 
HE 0155-0221 & 01 57 33.7 & -22 07 06  &198.09&-74.17& 17.00  & 15.30&  12.73 & 12.07 & 11.92 & C$_{2}$, CH, CN & 06.12.16&CH/CEMP-r/s\\
HE 0206-1916$^{\ast \ast \ast}$ & 02 09 19.63 & -19 01 55.45 & 192.69 & -70.37 & 15.13 & 13.99 & 12.24 & 11.76 & 11.66 & C$_{2}$, CH, CN & 23.10.05 & CH/CEMP-s  \\  
HE 0237-0835 & 02 40 13.6 & -08 22 18  &181.93&-58.16& 16.80	 & 15.60& 13.71 & 13.14 & 12.99 & C$_{2}$, CH, CN& 19.12.12&CH/CEMP-r/s\\
HE 0251-2118 & 02 53 42.6 & -21 05 59  &207.41&-61.53& 14.30	 & 13.30& 11.48 & 10.96 & 10.82 & C$_{2}$, CH, CN& 06.12.16&CH/CEMP-s\\
HE 0258-0218 & 03 01 04.9 & -02 06 17  &179.62&-50.14& 15.90  & 14.80& 13.48 & 12.88 & 12.79 & C$_{2}$, CH, CN& 21.12.12&CH/CEMP-r/s\\
HE 0319-0215$^{\ast}$ & 03 21 46.26 & -02 04 33.95 & 184.58 & -46.16 & 15.03 & 13.60 & 11.78 & 11.22 & 11.06 &C$_{2}$, CH, CN& 17.11.20 & CH/CEMP-r/s \\
HE 0322-1504$^{\ast}$	&	03 24 40.10	&	-14 54 24.35	&	201.89	&	-52.38	&	15.81	&	14.29	&	12.1	&	11.53	&	11.34	& C$_{2}$, CH, CN &	06.11.04&CH/CEMP-s\\
HE 0323-2702 & 03 26 04.6 & -26 51 38  & 221.55 & -55.69  & 16.60 & 15.10 & 13.61 & 13.07 & 12.97 & C$_{2}$, CH, CN& 07.12.16 &CH/CEMP-r/s\\
HE 0326-2603 & 03 28 19.68 & -25 53 19.01 & 220.02 & -55.02 & 16.12 & 15.10 & 13.67 & 13.24 & 13.17 & C$_{2}$, CH, CN& 20.10.20 & CH/CEMP-r/s\\
             &            &              &      &    &   &     &        &        &    &  & 14.09.21& \\
HE 0333-1819$^{\ast}$ & 03 35 18.87 & -18 09 53.30 & 208.37 & -51.32 & 13.17 & 11.53 & 9.43 & 8.86 & 8.68 & C$_{2}$, CH, CN& 20.10.20 & CH/CEMP-s\\
             &            &              &      &    &   &     &        &        &    &  & 15.09.21 & \\
HE 0422-2518 & 04 24 38.42 & -25 12 10.04 & 223.27 & -42.48 & - & - & 10.81 & 10.29 & 10.10 & C$_{2}$, CH, CN & 16.09.03 &CH /CEMP-s\\
HE 0429+0232$^{\ast}$	&	04 31 53.74	&	02 39 00.49	&	192.72	&	-29.17	&	14.65	&	13.3	&	11.09	&	10.52	&	10.32& C$_{2}$, CH, CN	&	16.09.03&CH/CEMP-s\\
HE 0507-1653$^{\ast}$	&	05 09 16.56	&	-16 50 04.69	&	217.54	&	-29.95	&	13.63	&	12.51	&	10.88	&	10.43	&	10.31& C$_{2}$, CH, CN	&	06.11.04 &CH/CEMP-r/s\\
             &            &              &      &    &   &     &        &        &    &  & 17.11.20 & \\
HE 0507-1430 & 05 10 07.6 & -14 26 32  &215.09&-28.84& 13.20	 & 14.40 &12.32 & 11.72 & 11.57 & C$_{2}$, CH, CN& 20.12.12&CH/CEMP-r/s\\
HE 0516-2515 & 05 18 09.4 & -25 12 25  &227.49&-30.86& 15.20	 & 13.90& 11.25 & 10.59 & 10.35 & C$_{2}$, CH, CN& 06.12.16 & CH/CEMP-r/s\\
HE 0518-2322$^{\ast}$	&	05 20 35.57	&	-23 19 14.26	&	225.62	&	-29.74	&	14.07	&	12.79	&	11.15	&	10.67	&	10.57	& C$_{2}$, CH, CN&	15.11.03&CH/CEMP-s	\\
HE 1008-0946 & 10 11 22.4 & -10 01 13  &251.16&36.29   & 16.80	 & 15.80& 13.22 & 12.63 & 12.49 & C$_{2}$, CH, CN& 06.12.16 & CH/CEMP-r/s\\
HE 1023-1504$^{\ast \ast \ast}$ & 10 25 55.55 & -15 19 17.08 & 258.78 & 34.79 & 16.26 & 14.40 & 12.32 & 11.61 & 11.42 & C$_{2}$, CH, CN & 30.03.05& CH/CEMP-s\\
HE 1045-1434$^{\ast \ast}$ & 10 47 44.18 & -14 50 22.52 & 263.59 & 38.40 & 15.83 & 14.60 & 12.93 & 12.44 & 12.24 & C$_{2}$, CH, CN & 24.02.20 & CH/CEMP-r/s\\
             &            &              &      &    &   &     &        &        &    &  & 21.4.21 & \\
HE 1104-0957$^{\ast}$ & 11 07 19.40 & -10 13 15.89 & 265.35 & 44.92 & 12.12 & 10.76 & 8.26 & 7.56 & 7.32 & C$_{2}$, CH, CN &24.02.20 & C-R\\
HE 1112-2557 & 11 15 14.10 & -26 13 27.76 & 277.44 & 31.84 & 14.89 & 13.60 & 11.54 & 11.00 & 10.85 & C$_{2}$, CH, CN & 17.05.16 & CH/CEMP-s\\
HE 1150-2546 & 11 53 15.48 & -26 03 41.43 & 286.94 & 34.99 & 13.17 & 11.93 & 9.89 & 9.35 & 9.23 & C$_{2}$, CH, CN & 17.05.16 & CH/CEMP-s\\
HE 1152-2432 & 11 54 34.92 & -24 48 44.11 & 286.88 & 36.28 & 12.45 & 10.91 & 9.41 & 8.83 & 8.68 & C$_{2}$, CH, CN & 30.03.17 & CH/CEMP-s\\
HE 1152-0355$^{\ast \ast \ast}$ & 11 55 06.05 & -04 12 24.59 & 277.32 & 55.84 &13.89 & 11.43 & 10.24 & 8.66 & 8.43 & C$_{2}$, CH, CN & 24.02.20 & CH/CEMP-s \\
HE 1157-0518 & 12 00 18.1 & -05 34 43  &280.33&55.03   & 15.00  & 15.30& 13.42 & 12.92 & 12.85 & C$_{2}$, CH, CN& 17.05.16&CH/CEMP-r/s\\
HE 1158-0708 & 12 00 49.2 & -07 25 33  &281.59&53.34   & 16.20	 & 15.00&  13.38 & 12.62 & 12.48& C$_{2}$, CH, CN & 30.03.17&CH/CEMP-r/s\\
HE 1205-0417 & 12 07 51.8 & -04 34 39  &282.92&56.59   & 16.70	 & 15.70&  12.63 & 11.88 & 11.68  & C$_{2}$, CH, CN& 01.03.13&CH/CEMP-r/s\\
HE 1205-0849 & 12 08 29.0 & -09 05 50  &285.42&52.30   & 13.90  & 12.60&  10.80  & 10.18  & 10.03 & C$_{2}$, CH, CN& 01.03.13&CH/CEMP-s\\ 
HE 1205-0521 & 12 07 53.08 & -05 37 50.90 & 283.50 & 55.58 & 15.49 & 14.40 & 12.42 & 11.98 & 11.80 & C$_{2}$, CH, CN& 24.02.20 & C-R \\
HE 1212-0323 & 12 15 29.1 & -03 40 23  &285.82&57.99   & 12.70	 & 15.0&  13.04 & 12.43 & 12.29 & C$_{2}$, CH, CN & 30.03.17&CH/CEMP-r/s\\
HE 1221-0651 & 12 23 50.0 & -07 07 54  &290.91&55.09   & 15.80	 & 14.8&  13.04 & 12.42 & 12.31 & C$_{2}$, CH, CN& 05.03.10&CH/CEMP-r/s\\
HE 1236-0337 & 12 39 04.6 & -03 54 25  &296.97&58.82   & 14.10	 & 15.40&  13.23 & 12.61 & 12.43& C$_{2}$, CH, CN & 17.04.17 & CH/CEMP-r/s \\
HE 1241-0337 & 12 44 27.20 & -03 54 01.19 & 299.55 & 58.92 & 15.80 & 14.30 & 11.87 & 11.23 & 11.01 & C$_{2}$, CH, CN & 14.04.20 & CH/CEMP-r/s \\
HE 1251-2313 & 12 54 31.0 & -23 29 35  &303.84&39.37   & 14.40	 & 13.50&  11.85 & 11.38 & 11.25 & C$_{2}$, CH, CN& 17.04.17&CH /CEMP-s\\
HE 1255-2324 & 12 58 01.2 & -23 40 24  &304.87&39.17   & 11.70  &  -  & 8.66  & 8.17  & 08.00  & C$_{2}$, CH, CN& 24.01.10&CH/CEMP-s\\
HE 1305+0007$^{\ast\ast\ast}$ & 13 08 03.85 & -00 08 47.50 & 311.94 & 62.43 & 13.98 & 12.22 & 10.24 & 9.75 & 9.60 & C$_{2}$, CH, CN & 29.01.05& CH/CEMP-s  \\
HE 1308-1012 & 13 11 10.9 & -10 28 35  &310.85&52.09   & 14.60	 & 13.70&  12.42 & 11.97 & 11.87 & C$_{2}$, CH, CN & 17.04.17&CH/CEMP-r/s\\
HE 1318-1657 & 13 21 19.4 & -17 13 40  &313.06&45.05   & 13.30	 & 14.40&  12.51 & 11.99 & 11.81 & C$_{2}$, CH, CN& 20.12.12&CH/CEMP-r/s\\
HE 1319-1935 & 13 22 38.70 & -19 51 11.61 & 312.89 & 42.41 & 15.69 & 14.19 & 12.33 & 11.76 & 11.68 & C$_{2}$, CH, CN & 13.05.20 & CH/CEMP-r/s\\
HE 1336+0200 & 13 38 41.2 & +01 45 23  &328.97&62.21   & 17.30	 & 14.90& 11.76 & 10.92 & 10.67 & C$_{2}$, CH, CN& 18.04.16&C-R\\
HE 1406-2016 & 14 09 44.17 & -20 30 56.31 & 326.64 & 38.72 & 15.48 & 14.20 & 12.09 & 11.55 & 11.42 & C$_{2}$, CH, CN& 17.05.16 &CH/CEMP-s\\
HE 1429-0551$^{\ast}$ & 14 32 31.29 & -06 05 00.20 & 343.01 & 48.76 & 14.01 & 12.61 & 10.73 & 10.27 & 10.07 & C$_{2}$, CH, CN & 13.05.20 & CH /CEMP-r/s\\  
&            &              &      &    &   &     &        &        &    &  & 07.04.21 & \\
HE 1430+0227 & 14 32 46.5 & +02 14 44  &351.53&55.25   & 17.10	 & 15.90&  14.02 & 13.48 & 13.31& C$_{2}$, CH, CN& 17.05.16&CH/CEMP-r/s\\
HE 1431-0245 & 14 33 54.2 & -02 58 33  &346.32&51.05   & 16.20	 & 15.30&  13.57 & 13.01 & 12.99 & C$_{2}$, CH, CN& 19.02.17&CH/CEMP-r/s\\
HE 1442-0346 & 14 45 02.1	& -03 58 46	 &348.10&48.53   & 16.30  & 15.40  & 13.75 & 13.25 & 13.11 & C$_{2}$, CH, CN& 17.05.16 & CH/CEMP-r/s\\
HE 1523-1155$^{\ast}$ & 15 26 41.04 & -12 05 42.66 & 351.87 & 35.63 & 14.57 & 13.22 & 11.37 & 10.85 & 10.75 & C$_{2}$, CH, CN& 13.04.20 & CH/CEMP-r/s \\
HE 1528-0409$^{\ast}$ & 15 30 54.30 & -04 19 40.36 & 359.86 & 40.29 & 16.00 & 14.76 & 12.94 & 12.45 & 12.36 & C$_{2}$, CH, CN& 13.05.20 & CH/CEMP-r/s \\
HE 2145-1715$^{\ast}$	&	21 48 44.46	&	-17 01 02.46	&	36.63	&	-46.73	&	14.59	&	13.20	&	11.03	&	10.36	&	10.25& C$_{2}$, CH, CN	&	17.09.03&CH/CEMP-s	\\
HE 2150-1800 & 21 53 17.55 & -17 46 38.92 & 36.16 & -48.01 & 15.76 & 14.78 & 13.24 & 12.79 & 12.70 & C$_{2}$, CH, CN & 20.10.20 & CH/CEMP-s\\
HE 2218+0127$^{\ast}$	&	22 21 26.05	&	01 42 19.81	&	65.46	&	-43.8	&	14.80	&	14.00	&	11.83	&	11.51	&	11.43	& C$_{2}$, CH, CN&	16.09.03& CH/CEMP-s	\\
HE 2319-0852 & 23 22 17.32 & -08 36 16.94 & 70.07 & -61.93 & - & 15.20 & 13.46 & 12.99 & 12.82 & C$_{2}$, CH, CN & 20.10.2020 & CH/CEMP-s \\
HE 2331-1329$^{\ast}$ & 23 33 44.51 & -13 12 33.74 & 66.55 & -67.12 & 16.79 & 14.50 & 11.84 & 10.99 & 10.65 & C$_{2}$, CH, CN& 17.11.20 & CN \\
             &            &              &      &    &   &     &        &        &    &  & 3.12.20 & \\
HE 2339-0837$^{\ast}$ & 23 41 59.93 & -08 21 18.60 & 78.51 & -65.05 & 15.32 & 14.00 & 12.63 & 12.11 & 12.03 & C$_{2}$, CH, CN& 21.10.20 & CH/CEMP-r/s\\
\hline
\end{tabular}}

$^{\ast}$ \cite{Goswami_2005}, $^{\ast \ast}$ \cite{goswami_2010b}, $^{\ast \ast \ast}$ \cite{Goswami_2007} 
\end{table*}
}

{\footnotesize
\begin{table*}
\caption{HE stars without prominent C$_{2}$ molecular bands}\label{tab2}
\resizebox{\textwidth}{!}{\begin{tabular}{lccccccccccc}
\hline
Star No.     & RA$(2000)$ & Dec.$(2000)$ & $l$  & b  & B & V   &  J      & H      & K      & Bands & Obs. date\\
             &            &              &      &    &         &     &         &        &        & noticed &     \\
\hline
HE 0341-0314 & 03 44 13.5 & -03 04 39	 &190.36&-42.25  & 17.0  & 16.4& 14.245 & 13.667 & 13.450 & CH,CN & 23.10.15 \\
HE 0359-0141$^{\ast}$	&	04 02 21.30	&	-01 33 03.96	&	192.03	&	-37.64	&	14.66	&	13.3	&	11.07	&	10.44	&	10.25	& CH,CN &	15.11.03	\\
HE 0417-0513$^{\ast}$	&	04 19 46.83	&	-05 06 17.21	&	198.66	&	-35.82	&	15.01	&	13.7	&	11.17	&	10.52	&	10.38	& CH,CN &	15.11.03	\\
HE 0432-0923 & 04 34 25.67 & -09 16 50.51 & 205.26 & -34.60 & - & 15.16 & 13.52 & 13.04 & 12.97 &CH,CN & 18.03.21 \\
HE 0443-1847$^{\ast}$	&	04 46 10.89	&	-18 41 39.63	&	217.23	&	-35.75	&	14.17	&	12.9	&	10.7	&	10.17	&	10.01	& CH,CN &	16.09.03	\\
HE 0503-2009$^{\ast \ast \ast}$	&	05 06 02.95	&	-20 05 57.35	&	220.77	&	-31.84	&	14.3	&	13.1	&	11.15	&	10.62	&	10.49	&CH, CN&	23.10.05\\
HE 0508-1604$^{\ast}$	&	05 10 47.01	&	-16 00 39.39	&	216.82	&	-29.31	&	13.25	&	11.65	&	9.9	&	9.33	&	9.16	 & CH,CN  &	20.12.04	\\
HE 0518-1751$^{\ast}$	&	05 20 28.46	&	-17 48 42.67	&	219.71	&	-27.84	&	13.85	&	12.8	&	10.65	&	10.11	&	9.97	& CH,CN &	07.11.04	\\
HE 0519-2053$^{\ast}$	&	05 21 54.42	&	-20 50 35.31	&	223.06	&	-28.62	&	14.88	&	13.7	&	11.96	&	11.39	&	11.29	& CH,CN &	15.11.03	\\
HE 0919+0200$^{\ast}$	&	09 22 13.07	&	01 47 55.78	&	230.52	&	33.93	&	13.91	&	12.6	&	10.67	&	10.1	&	9.95	& CH,CN  &	03.03.04	\\
HE 0926-0417 & 09 29 10.3 & -04 30 44  &237.91&31.84   & 14.1	 & 13.3& 11.631 & 11.107 & 10.999 & CH,CN & 06.12.16\\
HE 0930-0018$^{\ast}$	&	09 33 24.63	&	-00 31 44.62	&	234.74	&	35.01	&	16.13	&	14.7	&	12.19	&	11.54	&	11.33& CH,CN	&	02.03.04	\\
HE 0930-0945 & 09 32 40.6 & -09 58 48  &243.56&29.21   & 14.5  & 13.4& 11.66  & 11.11  & 11.00 & CH,CN  & 19.02.17\\
HE 1032-1655 & 10 34 36.5 & -17 10 59  &262.17&34.64   & 13.7  & -   & 10.854 & 10.256 & 10.134 & CH,CN & 19.02.17\\
HE 1058-1300 & 11 00 37.2 & -13 16 55  &265.79&41.46   & 14.1	 & 13.2& 11.231 & 10.659 & 10.497 & CH,CN & 19.02.17\\
HE 1130-1956 & 11 32 52.7	& -20 13 25	 &278.99&38.97   & 15.0  & 13.6& 11.604 & 11.079 & 10.946 & CH,CN & 17.04.17\\
HE 1150-2049 & 11 53 27.7 & -21 05 50  &285.24&39.77   & 15.7	 & 15.0& 13.103 & 12.542 & 12.422 & CH,CN & 06.03.16\\
HE 1208-1247 & 12 10 55.1 & -13 04 09  &287.91&48.62   & 14.4  & 13.5& 11.588 & 10.949 & 10.811 & CH,CN & 30.03.17\\
HE 1212-1414 & 12 14 51.8 & -14 31 15  &289.79&47.42   & 15.9  &  -  & 9.708  & 9.088  & 8.941  & CH,CN & 19.02.17\\
HE 1238-1714 & 12 40 46.3	& -17 31 17  &299.32&45.27   & 15.0  &  -  & 8.924  & 8.391  & 8.253  & CH,CN & 17.04.17\\
HE 1244-3036 & 12 47 39.23 & -30 53 16.06 & 301.97 & 31.97 & 14.53 & 13.60 & 12.03 & 11.54 & 11.46 & CH,CN & 18.03.21 \\
HE 1247-2554 & 12 50 20.7 & -26 10 38  &302.63&36.69   & 13.2  &  -  & 10.153 & 9.554  & 9.377  & CH,CN & 17.04.17\\
HE 1318-2451 & 13 20 57.7 & -25 07 30  &311.33&37.26   & 15.4  &  -  & 10.236 & 9.681  & 9.538  & CH,CN & 19.02.17\\
HE 1319-2340 & 13 21 43.9 & -23 56 27  &311.77&38.41   & 15.3	 & 13.6& 11.461 & 10.937 & 10.795 & CH,CN & 05.10.12\\
HE 1345-2616 & 13 48 02.1 & -26 31 12  &318.35&34.65   & 12.9  &  -  & 9.855  & 9.276  & 9.099  & CH,CN & 30.03.17  \\
HE 1354-2552 & 13 56 55.1 & -26 07 35  &320.80&34.46   & 13.2  &  -  & 10.114 & 9.528  & 9.360  & CH,CN & 17.04.17 \\
HE 2138-1616$^{\ast}$	&	21 41 16.61	&	-16 02 39.85	&	36.95	&	-44.7	&	14.91	&	13.9	&	11.9	&	11.45	&	11.32	& CH,CN &	16.09.03	\\
HE 2224-0330$^{\ast}$	&	22 26 47.82	&	-03 14 57.48	&	61.22	&	-48.01	&	14.58	&	13.5	&	11.82	&	11.36	&	11.26& CH,CN	&	16.09.03 \\
HE 2352-1906$^{\ast}$	&	23 54 49.02	&	-18 49 31.13	&	62.49	&	-74.57	&	13.21	&	12.07	&	10.16	&	9.65	&	9.51	& CH,CN &	16.09.03	\\
            &              &      &    &   &     &        &        &    &  & & 03.12.20  \\
\hline
\end{tabular}}

$^{\ast}$ \cite{Goswami_2005}, $^{\ast \ast \ast}$ \cite{Goswami_2007}
\end{table*}
}

\section{Spectral classification of the programme stars}\label{sec3}

We obtained low-resolution spectra for 65 stars and categorized them into distinct groups using well defined spectral criteria outlined in \cite{Goswami_2005}. Key features considered include the strength of the CH band around 4300 {\rm \AA}, prominence of secondary P-branch head near 4342 {\rm \AA}, strength of the Ca I feature at 4226 {\rm \AA} and isotopic band depths of C$_{2}$ and CN molecular bands. Objects are classified into various sub-groups by a thorough comparison of the program stars' spectra with those of well-known CH, C-R, C-N, and C-J stars obtained under the same observational setup and resolution.
Additionally, low-resolution spectra of carbon stars from the spectral atlas of \cite{Barnbaum_1996} were employed.

\par For comparative analysis and classification, we utilized the low-resolution spectra of established CH stars such as HD 26, HD 5223, and HD 209621, C-R stars RV Sct, HD 156074, and HD 76846, and CN stars Z Psc, and V460 Cyg. From the comparison, we have found that our sample consists of 11 stars showing spectral features similar to those of HD 26, 13 objects showing similarities to HD 5223, and 10 stars resembling HD 209621. Two objects show spectral features similar to the C-R star RV Sct. 
We selected these comparison stars because they have been extensively studied by various authors and classified based on high-resolution spectroscopy. For example, from a detailed high-resolution spectroscopic analysis, \cite{Goswami_2016} have shown that HD~26 displays the characteristic properties of CH stars with a larger 
enhancement of the  second peak s-process elements such as Ba, La, Ce, Nd, Sm  compared
to the first-peak s-process elements Sr, Y and Zr. Compared to the
Sun ([Sr/Ba] = 0.74), this object  shows a much smaller ratio of
[Sr/Ba] $\sim$ $-$0.04. With estimated [Ba/Eu] $\sim$ 1.32  and metallicity
[Fe/H] $\sim$ $-$ 1.13 \citep{Goswami_2016} this object can be placed in the  CEMP-s sub-group following 
the classification criteria of \cite{Beers_2005}. It is to be noted that stars with [C/Fe] 
 $>$ 1.0  and [Fe/H] $<$ $-$1.0 are termed as CEMP stars by \cite{Beers_2005}. High
 resolution spectroscopic analysis 
by \cite{Goswami_2006} have shown that  the object HD 5223 also  exhibits the characteristic properties of CH stars.  However, with an estimated  metallicity [Fe/H] $\sim$ $-2.05$ and following the classification criteria of \cite{Beers_2005}  the object 
can be placed among the CEMP-s sub-group.  
A detailed high  resolution spectroscopic study  by \cite{Goswami_2010} have shown 
that,  although,  HD 209621 shows characteristics of  CH  stars in terms of enhancement
of heavy s-process elements relative to the lighter s-process
elements,  the estimated [Ba/Eu]=+0.35 places the star in the group of CEMP-r/s stars.
High-resolution spectroscopic analysis of RV Sct by \cite{Zamora_2009} showed that this 
object is a late-R type star with solar metallicity.

\par Out of the 65 objects we have analysed, the spectra of 29 objects are found to exhibit no prominent C$_{2}$ molecular bands, rendering them unclassifiable based on the adopted spectral criteria.
Among the rest which were previously classified into different sub-groups of carbon stars by \cite{Goswami_2005} and \cite{Goswami_2007,goswami_2010b}, the objects HE 0017+0055, HE 0206$-$1916, HE 0228$-$0256, HE 0037-0654, HE 0319-0215, HE 0322$-$1504, HE 0333-1819, HE 0429+0232, HE 0507$-$1653, HE 1023$-$1504, HE 1045-1434, HE 1104$-$0957, HE 1152$-$0355, HE 1305+0007, HE 1429$-$0551, HE 1523$-$1155, HE 1528-0409, HE 2145$-$1715, HE 0518$-$2322, HE 2218+0127, HE 2331$-$1329, and HE 2339$-$0837 were classified as potential CH star candidates. While two objects  HE 0002+0053 and HE 1104-0957 were classified as C-R stars, HE 0228$-$0256 and HE 2331-1329 were classified as CN stars. These authors could not classify the objects into CEMP sub-groups due to the unavailability of metallicity estimates for their sample as well as the lack of high-resolution abundance results
for comparison. In this study, we have reevaluated their classification, and assigned the objects to distinct CEMP sub-groups. We have used the criteria [Fe/H] $\leq$ $-1$  of \cite{Beers_2005} for classifying the potential CH star candidates into different CEMP sub-groups using our metallicity estimates as discussed in detail in Section \ref{sec4}. Furthermore, we confirmed the CEMP sub-groups of the program stars by comparing their spectra with those of the comparison stars, as discussed in the subsequent sub-sections. Thus, this classification relies solely on visual comparisons between the spectra of the program stars and the spectra of the comparison stars, as well as the metallicity estimates. Additional high-resolution measurements of various elemental abundance ratios would be useful  to confirm their classification.  

\par The basic data of the programme stars with prominent C$_{2}$ molecular bands in their spectra and their class are listed in Table \ref{tab1} and  those without are presented in Table \ref{tab2}. Classification and the description of the spectra of the objects listed in Table \ref{tab1}
are presented in the following sub-sections.

\subsection{Candidate CH, CEMP-s and CEMP-r/s stars : Description of the spectra}\label{subsec1}

{\bf HE 1406-2016, HE 1150$-$2546, HE 1112$-$2557, HE 0422$-$2518, HE 1205$-$0849, HE 1152$-$2432, HE 1251$-$2313, HE 1255$-$2324, and HE 0251$-$2118, HE 2150$-$1800, HE 2319$-$0852} \\
The spectra of these stars closely resemble the spectrum of the star HD~26. The strength of Ca I line at 4226 \AA\ is found to be similar in all these stars. The CH bands around 4300 {\rm \AA} are found to be of similar strength in all these objects and closely match with that of HD 26. The other carbon molecular bands such as CN band around 4215 \AA, and C$_{2}$ bands around 4730, 5165, and 5635 {\rm \AA} in these stars are also of similar strength to that of HD 26 except for a few objects. While the objects HE 0251$-$2118, HE 1251$-$2313, HE 1255$-$2324, and HE 2218$-$0127 exhibit relatively stronger C$_{2}$ band around 4730 \AA, the object HE 1406$-$2016 shows stronger C$_{2}$ band around 5165 \AA. The recent high-resolution studies by \cite{roriz.2023} also confirms the star HE 1255-2324 as a CH star.
\par
\begin{figure}
\centering
\includegraphics[width=10cm,height=9cm]{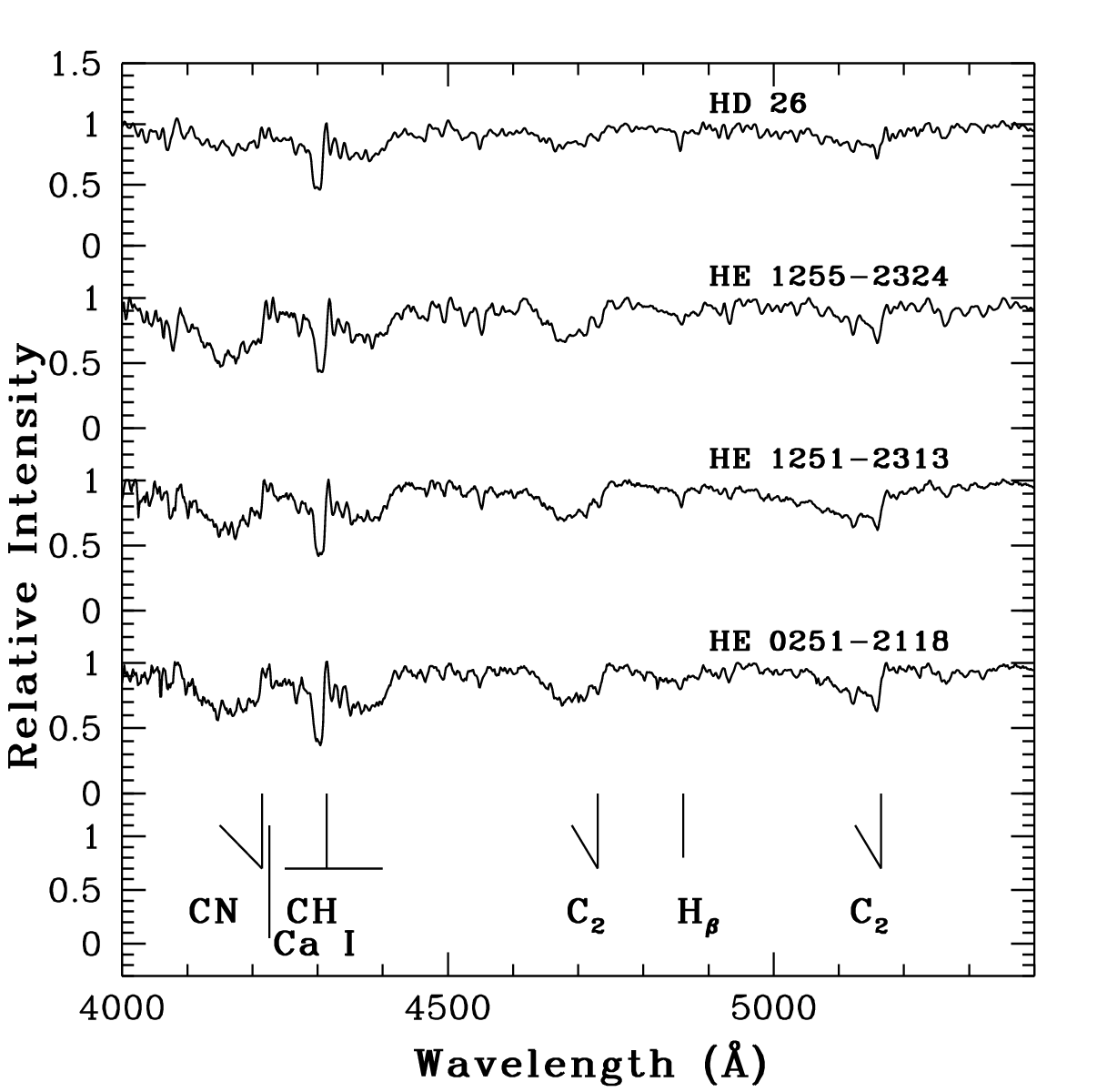}
\caption{Examples of the spectra of potential CH star candidates, in the wavelength region 4000 to 5400 {\rm \AA} that resemble closely the spectrum of the CH star HD 26}\label{fig1}
\end{figure}

 HE 1406$-$2016, HE 1150$-$2546, HE 1112$-$2557, HE 1205$-$0849, HE 1255$-$2324, and HE 2319$-$0852 show relatively stronger Na D line at 5890 {\rm \AA} in their spectra than that in HD 26. In HE 1251$-$2313, and HE 0251$-$2118, the Na D line is found to be weaker than that in HD 26. The strengths of H-$\alpha$ and H-$\beta$ lines are of similar strengths in these stars, except for HE 0251$-$2118, HE 1406$-$2016, HE 1150$-$2546, HE 1205$-$0849, and HE 1255$-$2324 where these lines are found to be weaker. Ba II at 6496 {\rm \AA} is found to be very strong in HE 1112$-$2557, HE 1205$-$0849, and HE 1255$-$2324. Examples of the spectra of the program stars that resemble closely the spectrum of the CH star HD 26 are shown in Figure \ref{fig1}.\\

{\bf \noindent HE 0155$-$0221, HE 0258$-$0218, HE 0237$-$0835, HE 0323$-$2702, HE 1442$-$0346, HE 1318$-$1657, HE 1431$-$0245, HE 1221$-$0651, HE 1308$-$1012, HE 1212$-$0323, HE 1430+0227, HE 1157$-$0518, and HE 0507$-$1430.}\\

The spectra of these stars show close match with  the spectrum of HD 5223. All these stars show carbon molecular bands of similar strengths in their spectra except for HE 0237$-$0835, and HE 1212$-$0323 whose spectra show weak C$_{2}$ band around 5635 \AA. The spectra of the objects HE 0237$-$0835, HE 0258$-$0218, HE 1212$-$0323, and HE 1318$-$1657 show relatively strong Ca I line at 4226 \AA\ than that in HD 5223. 
\par
\begin{figure}
\centering
\includegraphics[width=10cm,height=9cm]{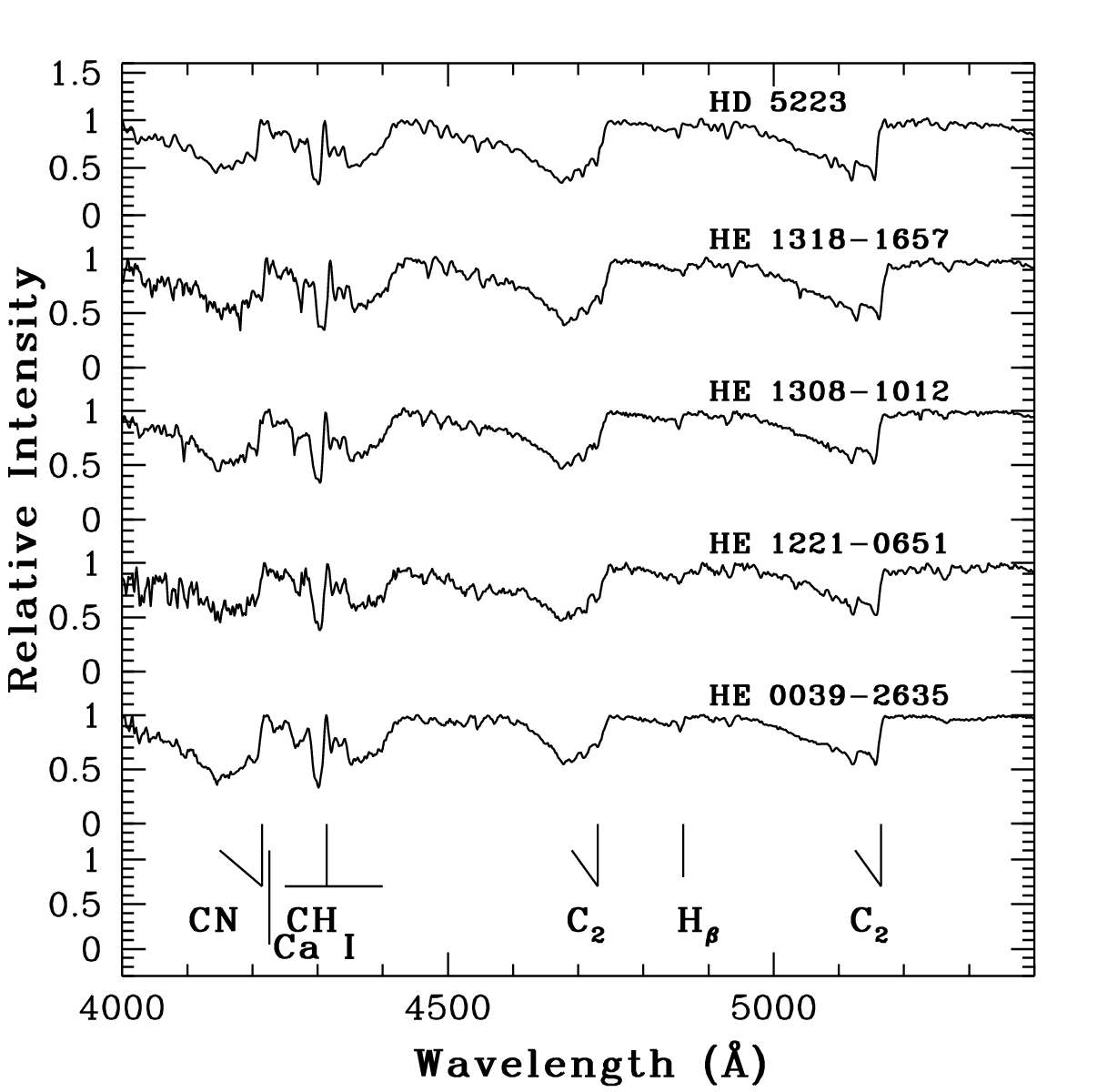}
\caption{ Examples of the spectra of potential CH star candidates, in the wavelength region 4000 to 5400 
{\rm \AA} that resemble closely the spectrum of  HD 5223. }\label{fig2}
\end{figure}

 While the stars HE 0155$-$0221, HE 0237$-$0835, and HE 1430+0227 show very strong Ba II lines at 4554 \AA, and 6496 {\rm \AA} in their spectra, HE 1308$-$1012 and HE 1442$-$0346 show relatively weaker Ba II lines. The spectra of the star HE 1442$-$0346 shows very weak H-$\alpha$ and H-$\beta$ lines. Examples of the spectra of a few of these  stars that resemble closely the spectrum of the star HD 5223 are shown in Figure \ref{fig2}.\\

{\bf \noindent HE 0326-2603, HE 1158$-$0708, HE 1008$-$0946, HE 0516$-$2515, HE 0113+0110, HE 0039$-$2635, HE~1205$-$0417, HE 1236$-$0337, HE 1241-0337, and HE 1319-1935} \\

The spectra of these stars show close resemblances with the spectrum of the star HD~209621. HE 0113+0110 shows a very strong Ca I line at 4226 \AA\ than that in HD 209621. The carbon molecular bands in these stars show similar strengths except HE 0039$-$2635, HE 0113+0110, HE 0516$-$2515, HE 1158$-$0708, and HE~1205$-$0417 where the C$_{2}$ band around 5635 {\rm \AA} is found to be relatively weaker. 
\par
\begin{figure}
\centering
\includegraphics[width=10cm,height=9cm]{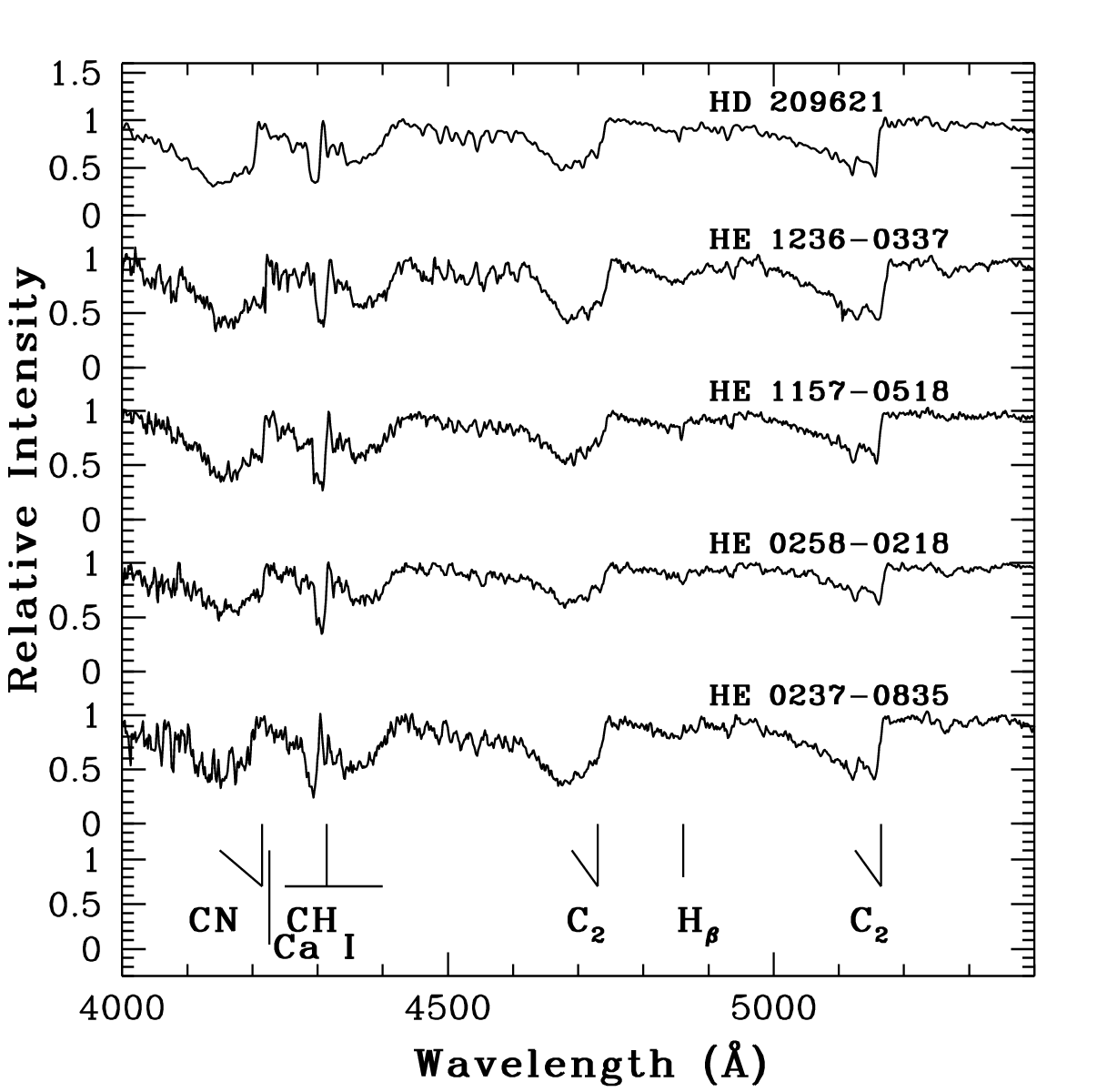}
\caption{ Examples of the spectra of potential CH star candidates, in the wavelength region 4000 to 5400   {\rm \AA} that resemble closely the spectrum of  HD 209621. }\label{fig3}
\end{figure}

 Ba II at 4554 {\rm \AA} is found to be marginally weaker in HE 0039$-$2635, HE 0516$-$2515, HE 1008$-$0946, and HE 1158$-$0708 than in HD 209621. But this line along with Ba II at 6496 {\rm \AA} are very strong in HE 0113+0110. The spectra of HE 1158$-$0708, HE~1205$-$0417, and HE 1236$-$0337 show relatively weak H-$\alpha$ and H-$\beta$ lines. While Na D lines in the spectra of HE~1205$-$0417 are found to be strong, the spectra of HE 0039$-$2635 shows a very weak Na D line. Examples of the spectra of the program stars that resemble closely the spectrum of the  star HD 209621 are shown in Figure \ref{fig3}.

\begin{figure}
\centering
\includegraphics[width=10cm,height=9cm]{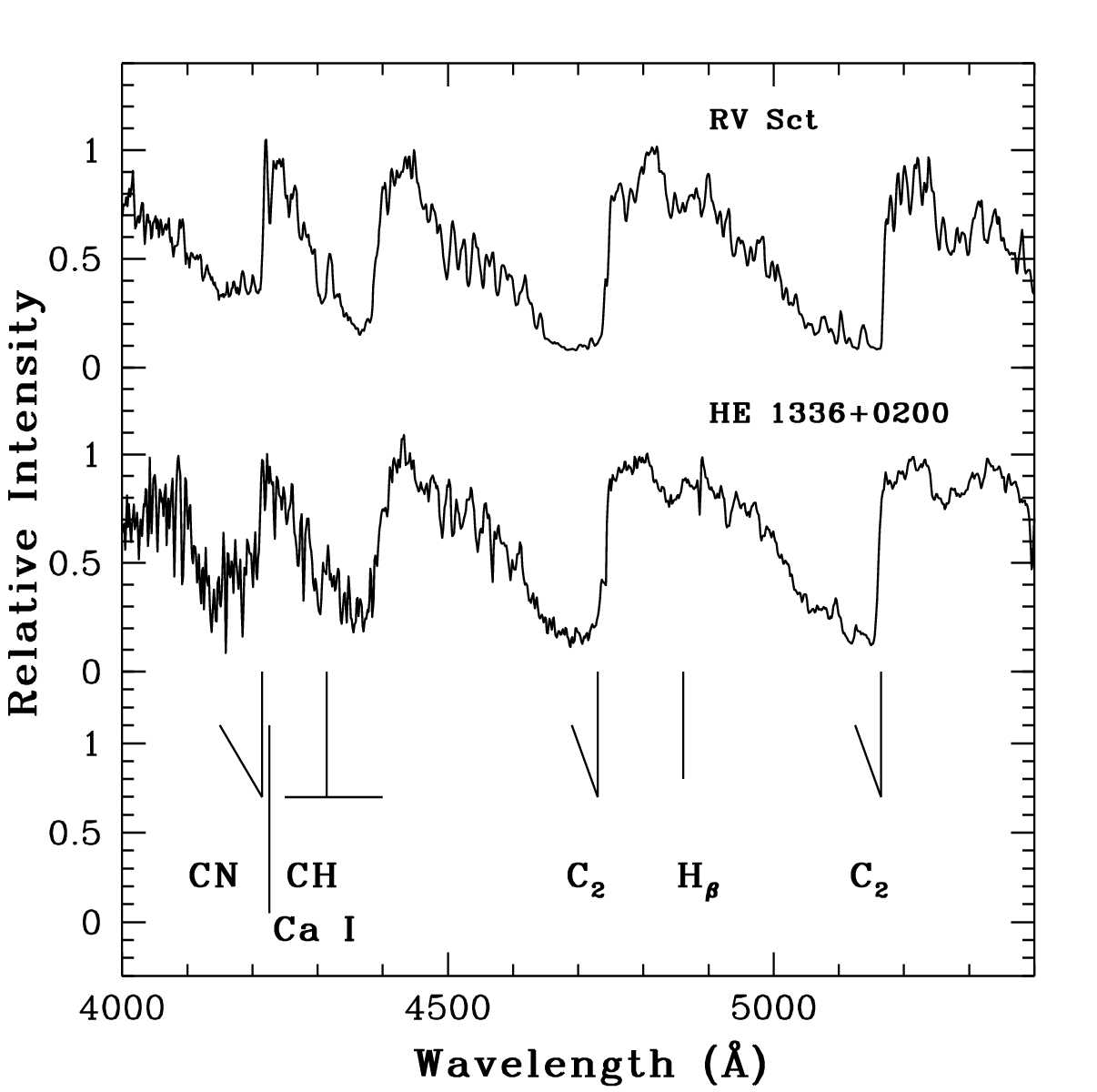}
\caption{The spectra of HE 1336+0200, in the wavelength region 4000 to
5400 Å that resemble closely the spectrum of the C-R star RV Sct.}\label{fig4}
\end{figure}

\subsection{Candidate C-R stars : Description of the spectra}\label{subsec2}
{\bf HE 1205-0521, and HE~1336+0200}\\
The spectra  of these objects resemble the  spectrum of the C-R star RV Sct (Figure \ref{fig4}). Both of them show carbon molecular bands of similar strengths. HE 1205-0521 shows relatively weak Ca I line at 4226 \AA\,. While the spectrum of HE~1336+0200 shows relatively stronger Ba II line at 4554 {\rm \AA} and 6496 {\rm \AA} than that in RV Sct star, these lines are marginally detectable in the spectrum of HE 1205-0521. Na D line is also found to be very weak in the spectrum of HE 1205-0521. HE~1336+0200 shows relatively strong H-$\beta$ line in the spectra.

\section{Determination of Atmospheric parameters}\label{sec4}
Estimates of stellar atmospheric parameters the effective temperature (T$_{eff}$), surface gravity (log\,{g}), microturbulance (${\xi}$) and metallicity ([Fe/H]) are are crucial for the comprehensive characterization and classification of stars. These parameters play a pivotal role in unraveling the nucleosynthesis processes and understanding the various evolutionary stages. We could derive stellar atmospheric parameters for 36 stars in our sample following the methods as described in the following sub-sections. However, for the remaining stars, such determinations were challenging due to various factors.
For instance, some stars fell outside the available evolutionary tracks, precluding the derivation of their mass-a key factor for determining both surface gravity and micro-turbulent velocity. Additionally, in certain cases, the H-$\alpha$ line was too strong and saturated, and it posed challenges in obtaining a reliable temperature value. Other instances involved the unavailability of grism 8 spectra, preventing the derivation of metallicity from the Ca II triplet (CaT) lines.

\subsection{Effective temperature}\label{subsec1}
Among the methods, the ones that are commonly used to derive effective temperatures are, using photometric colours, flux calibrated low-resolution spectra, and in high-resolution spectra using fitting shape of the balmer lines, and through excitation balance, i.e., forcing no trend of Fe I abundances with the excitation potential of the lines. 

\par We have determined preliminary estimates of the effective temperatures of the  programme stars  using the calibration equations of \cite{Alonso_1999}. These estimates are further verified through spectral synthesis of the H-$\alpha$ line at 6562.8 {\rm \AA}, whenever it was possible, using the model atmospheres selected from the Kurucz grid of model atmospheres with no convective overshooting (\url{http://kurucz.harvard.edu/grids.html}). Solar abundances are taken from \cite{Asplund_2009}. Some examples of spectrum synthesis fits of H$_{\alpha}$ line are shown in Figures \ref{fig5}. The photometric estimates for the programme stars and their comparison with the temperature derived from the H-$\alpha$ line are presented in Table \ref{tabA1}.

\begin{figure}
\centering
\includegraphics[width=0.8\columnwidth]{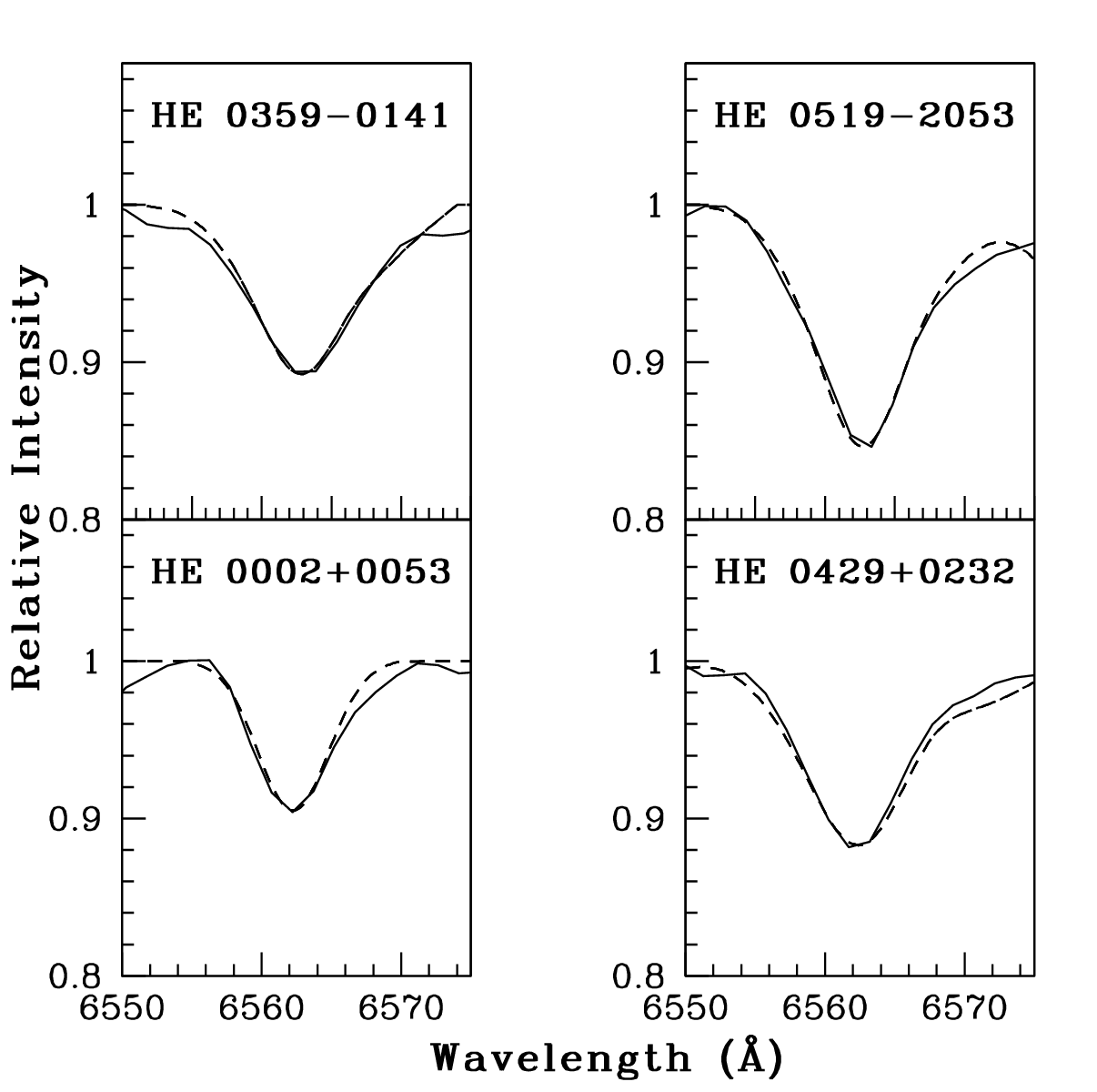}\label{fig5}
\caption{ Examples of spectrum synthesis fits  of H$_{\alpha}$ line profile  at 6562.8 {\rm \AA}. Short dashed lines represent synthesized spectra and the solid lines indicate the observed spectra.}
\label{fig5}
\end{figure}

\subsection{Surface gravity}\label{subsec2}
The surface gravity log\,g is calculated from the mass of the star using the relation\\\\
log (g/g$_{\odot}$) = log (M/M$_{\odot}$) + 4log (T$_{eff}$/T$_{eff\odot}$) + 0.4(M$_{bol}$ $-$ M$_{bol\odot}$).\\\\
The adopted values for the Sun are log g$_{\odot}$ = 4.44, T$_{eff\odot}$ = 5770 K, and M$_{bol\odot}$ = 4.75 mag \citep{Yang_2016}. Masses of the stars are determined from their positions in the H-R diagram. For this estimation, we made use of the \cite{Girardi_2000} database  (\url{http://pleiadi.pd.astro.it/}) of evolutionary tracks corresponding to the metallicity of the star. The metallicity of the star is taken to be those estimated from the CaT calibration. Parallaxes are taken from Gaia DR3 (\cite{Gaia_2016, Gaia_2018} \url{https://gea.esac.esa.int/archive/}). The estimated masses of our program stars range from 0.83 M$_{\odot}$ to 2.2 M$_{\odot}$.

{\footnotesize
\begin{table*}
\caption{Metallicity estimates from CaT lines for the programme stars}\label{tab3}
\resizebox{0.8\textheight}{!}{\begin{tabular}{lccccccccccccc}
\hline
Star name	&	V	&	Parallax  &	Mv	&	W8498	&	W8542	&	W8662	&	[Fe/H] 	&	[Fe/H] 	&	[Fe/H] 	&	[Fe/H] 	& Lit.value & Reference	\\
          &   &  (mas)    &     & (m\AA)      &  (m\AA) & (m\AA)  & (W8498+W8542+W8662) & (W8498+W8542) & (W8542+W8662) & (W8498+W8662)  & & \\ 
          &   &           &     &             &         &         & (CT1)               & (CT2)         & (CT3)         & (CT4)          &  & \\
\hline
HE 0002+0053	&	13.3	&	0.07$\pm$0.04	&	-2.55$\pm$1.41	&	1.45	&	2.65	&	1.77	&	-1.77$\pm$0.38	&	-2.44$\pm$0.33	&	-2.32$\pm$0.34	&	-2.78$\pm$0.31	& -2.18	&	1	\\
HE 0017+0055	&	11.7	&	0.25$\pm$0.05	&	-1.31$\pm$0.39	&	1.5	&	4.63	&	1.27	&	-0.83$\pm$0.11	&	-1.33$\pm$0.11	&	-1.42$\pm$0.10	&	-2.7$\pm$0.08	& -2.4	& 2		\\
HE 0037-0654    &   15.5    &   0.15$\pm$0.03   &   1.47$\pm$0.50       &  1.73  &   2.02   &  1.22  & -1.08$\pm$0.13   &  -1.65$\pm$0.12   &  -1.88$\pm$0.11   &  -2.03$\pm$0.11  & -&- \\
 HE 0039-2635 & 12.2 & 0.19$\pm$0.05 & -1.28$\pm$2.78 & 0.66 & 1.25 & 0.92 & -2.67$\pm$1.49 & -3.11$\pm$1.18 & -2.98$\pm$1.54 & -3.31$\pm$1.34 & -3.60 & 3 \\
HE 0155-2221	&	15.3	&	0.04$\pm$0.03	&	-1.84$\pm$1.55	&	1.49	&	3.15	&	2.48	&	-1.1$\pm$0.74	&	-2.06$\pm$1.19	&	-1.67$\pm$1.00	&	-2.32$\pm$1.31	& -2.28	&	1	\\
HE 0206-1916    & 13.9 & 0.14$\pm$0.02 & -0.28$\pm$0.28 & 0.52 & 1.36 & 0.94 & -2.46$\pm$0.06 &-2.94$\pm$0.05 & -2.71$\pm$0.05 & -3.20$\pm$0.05 & -2.52 & 4 \\
HE 0228-0256	&	14.7	&	0.008$\pm$0.04	&	-5.68$\pm$4.77	&	1.68	&	3.3	&	2.47	&	-2.13$\pm$1.64	&	-2.9$\pm$1.94	&	-2.65$\pm$1.84	&	-3.16$\pm$2.04 &-	&	-	\\
HE 0251-2118 & 13.3 & 0.20$\pm$0.02 & -0.19$\pm$0.25 & 1.09 & 2.71 & 1.97 & -1.17$\pm$0.07 & -2.0$\pm$0.06 & -1.63$\pm$0.06 & -2.33$\pm$0.05 & -1.50 & 1\\  
HE 0258-0218	&	14.8	&	0.08$\pm$0.06	&	-0.56$\pm$0.67	&	0.82	&	2.75	&	1.48	&	-1.57$\pm$0.43	&	-2.19$\pm$0.38	&	-1.91$\pm$0.40	&	-2.77$\pm$0.34	& -	&	-	\\
HE 0319-0215    &   13.6    &  0.08$\pm$0.01    &  -1.77$\pm$0.64      &  0.55  &  1.41  &  0.74  & -2.83$\pm$0.09 &  -3.19$\pm$0.08  &  -3.09$\pm$0.08  & -3.61$\pm$0.07 & -2.42 & 1 \\
HE 0322-1504	&	14.3	&	1.26$\pm$0.28	&	4.8$\pm$0.47	&	1.48	&	3.04	&	2.07	&	-0.60$\pm$0.14	&	-0.47$\pm$0.12	&	-0.15$\pm$0.12	&	-0.97$\pm$0.12	&	-2.00	&	4	\\
HE 0323-2702	&	15.1	&	0.04$\pm$0.03	&	-1.89$\pm$0.75	&	0.94	&	3.54	&	1.82	&	-1.44$\pm$0.94	&	-2.14$\pm$1.26	&	-1.8$\pm$1.10	&	-2.84$\pm$1.58	&	 -	&	-	\\
HE 0326-2603    &   15.1    & 0.20$\pm$0.02     &  1.66$\pm$0.38    &  0.43  &  0.76 &  0.26 & -2.85$\pm$0.04 & -3.06$\pm$0.04 & -3.25$\pm$0.04 & -3.82$\pm$0.04 & - & - \\
HE 0333-1819    &   11.5    &   0.88$\pm$0.01   &    1.26$\pm$0.03      &  1.46     &   3.43    &   2.78    &   0.06$\pm$0.01       &  -1.17$\pm$0.01    &   -0.58$\pm$0.01   &  -1.46$\pm$0.01   & - & - \\
HE 0359-0141	& 13.4	& 0.43$\pm$0.02 &	1.56$\pm$0.12 &	1.78  &	3.24  &	2.52 &	0.09$\pm$0.03	& -1.04$\pm$0.03 &	-0.70$\pm$0.03 & 	-1.32$\pm$0.03 & - & - \\
HE 0417-0513	&	13.7	&	0.43$\pm$0.04	&	1.87$\pm$0.20	&	1.78	& 3.56	&	2.27	&	-0.21$\pm$0.06	&	-0.81$\pm$0.05	&	-0.59$\pm$0.06	&	-1.41$\pm$0.05	&	-1.88	&	1	\\
HE 0429+0232 & 13.3 & 0.97$\pm$0.05 & 3.23$\pm$0.3 & 1.44 & 3.09 & 1.82 & 0.02$\pm$0.67 & -0.85$\pm$0.57 & -0.67$\pm$0.64 & -1.48$\pm$0.57 & -2.05 & 1 \\
HE 0443-1847	&	12.9	&	0.45$\pm$0.03	&	1.19$\pm$0.06	&	2.25	&	3.73	&	2.90	&	-0.55$\pm$0.02	&	-0.71$\pm$0.02	&	-0.43$\pm$0.02	&	-1.08$\pm$0.01	&-	& -		\\
HE 0503-2009	&	13.1	&	0.28$\pm$0.03	&	0.33$\pm$0.18	&	1.73	&	3.95	&	2.42	&	-0.05$\pm$0.07	&	-1.09$\pm$0.05	&	-0.78$\pm$0.06	&	-1.73$\pm$0.05	& -	&	-	\\
HE 0507-1653	&	12.5	&	0.52$\pm$0.03	&	1.08$\pm$0.12	&	0.05	&	2.32	&	1.61	&	-1.63$\pm$0.03	&	-2.4$\pm$0.03	&	-1.65$\pm$0.03	&	-2.81$\pm$0.02	& -1.43	&	5	\\
HE 0507-1430	&	14.4	&	0.05$\pm$0.02	&	-2.06$\pm$0.92	&	1.08	&	3.55	&	1.22	&	-1.66$\pm$0.25	&	-2.13$\pm$0.23	&	-2.01$\pm$0.27	&	-3.09$\pm$0.22	&-2.4 &	4	\\
HE 0508-1604	&	11.6	&	0.62$\pm$0.03	&	0.62$\pm$0.12	&	1.09	&	3.25	&	2.55	&	-0.47$\pm$0.03	&	-1.57$\pm$0.03	&	-0.94$\pm$0.03	&	-1.89$\pm$0.02 & -	& -		\\
HE 0518-1751	&	12.8	&	1.16$\pm$0.03	&	3.12$\pm$0.06	&	1.20	&	2.79	&	1.87	&	-0.24$\pm$0.01	&	-1.14$\pm$0.01	&	-0.81$\pm$0.01	&	-1.61$\pm$0.01 & -1.9	&	1	\\
HE 0518-2322	&	12.8	&	0.25$\pm$0.02	&	-0.21$\pm$0.22	&	1.28	&	2.92	&	1.95	&	-1.02$\pm$0.06	&	-1.84$\pm$0.05	&	-1.55$\pm$0.05	&	-2.26$\pm$0.04	&	-	&	-	\\
HE 0519-2053	&	13.7	&	0.19$\pm$0.02	&	0.15$\pm$0.16	&	0.72	&	1.06	&	3.35	&	-1.35$\pm$0.04	&	-2.91$\pm$0.03	&	-1.66$\pm$0.04	&	-1.81$\pm$0.04& -1.45	&	1	\\
HE 0919+0200	&	12.6	&	0.51$\pm$0.04	&	1.13$\pm$0.17	&	2.18	&	3.53	&	2.67	&	-0.33$\pm$0.05	&	-0.83$\pm$0.05	&	-0.64$\pm$0.05	&	-1.22$\pm$0.04	& -	& -		\\
HE 0926-0417	&	13.3	&	0.58$\pm$0.03	&	2.13$\pm$0.09	&	1.74	&	3.32	&	2.01	&	0.04$\pm$0.03	&	-0.88$\pm$0.02	&	-0.75$\pm$0.03	&	-1.49$\pm$0.02	& -	& -		\\
HE 0930-0018	&	14.7	&	5.68$\pm$0.03	&	8.47$\pm$0.01	&	0.91	&	2.08	&	1.83	&	-0.61$\pm$0.005	&	-0.49$\pm$0.005	&	0.009$\pm$0.005	&	-0.68$\pm$0.01	& -	&	-	\\
HE 1045-1434    &   14.6    &   0.07$\pm$0.02   &   -1.02$\pm$0.60      &   0.91    &   1.19    &   0.82    &   -2.58$\pm$0.12      &   -2.97$\pm$0.12      &   -3.01$\pm$0.11  &  -3.17$\pm$ 0.11 & -2.50 & 4 \\
HE 1104-0957    &   10.8    &   0.26$\pm$0.02   &  -2.17$\pm$0.17       &   1.14    &   2.11    &   1.09    &    -2.25$\pm$0.04   &   -2.69$\pm$0.04  &  -2.71$\pm$0.03  &  -3.13$\pm$0.03  & - & - \\
HE 1152-0355    &   11.4    &   0.13$\pm$0.02   &   -2.99$\pm$0.38     &   0.89     &   2.37   &   1.38    & -2.34$\pm$0.09  & -2.87$\pm$0.08 & -2.67$\pm$0.09  &  -3.28$\pm$0.08  & -1.27 & 6 \\
HE 1157-0518	&	15.1	&	0.08$\pm$0.04	&	-0.42$\pm$0.22	&	1.62	&	1.80	&	1.55	&	-1.56$\pm$0.30	&	-2.23$\pm$0.27	&	-2.26$\pm$0.27	&	-2.34$\pm$0.27	& -2.39	&	5	\\
HE 1205-0521    &   14.4    &  0.03$\pm$0.02    & -3.14$\pm$0.30        & 0.90    &  2.07  & 1.27 & -2.52$\pm$0.52 & -3.01$\pm$0.47 & -2.87$\pm$0.48  & -3.36$\pm$0.43 & - & - \\
HE 1205-0849    & 12.6      & 0.35$\pm$0.01     & 0.34$\pm$0.09         &   1.44    &   3.24    &   2.27    &   -0.52$\pm$0.03      &   -1.49$\pm$0.02      &   -1.14$\pm$0.02  &   -1.91$\pm$0.02  & - & - \\
HE 1319-1935    &  14.2     & 0.03$\pm$0.02     &  -3.68$\pm$0.90       & 1.03  &  2.48 & 1.73 & -2.29$\pm$0.63  &  -2.91$\pm$0.54  & -2.67$\pm$0.58  & -3.22$\pm$0.51 & -2.22 & 7 \\
HE 1429-0551    &  12.6  &  0.21$\pm$0.01    &  -0.75$\pm$0.15   &  0.52 & 1.49 & 0.64 & -2.64$\pm$0.03  & -2.96$\pm$0.03  & -2.89$\pm$0.03  & -3.53$\pm$0.03 & -2.45 & 8 \\
HE 1442-0346	&	15.4	&	0.07$\pm$0.06	&	-0.35$\pm$2.78	&	0.64	&	1.88	&	1.78	&	-1.83$\pm$0.67	&	-2.62$\pm$0.57	&	-2.1$\pm$0.64	&	-2.67$\pm$0.57	& - & -\\
HE 1523-1155    &  13.2     &  0.22$\pm$0.01    & -0.05$\pm$0.13       &    1.63    &   1.58    &   1.04    &   -1.78$\pm$0.03      &   -2.24$\pm$0.02      &   -2.51$\pm$0.03  &   -2.48$\pm$0.02   & -2.42 & 8 \\
HE 1528-0409    &   14.8    &  0.18$\pm$0.03   &  0.99$\pm$0.32  &   0.19  &  1.41  &  0.66 &  -2.48$\pm$0.06 &  -2.86$\pm$0.06 &  -2.58$\pm$0.06  & -3.61$\pm$0.06 & -2.35 & 8\\
HE 2138-1616	&	13.9	&	0.69$\pm$0.04	&	3.11$\pm$0.14	&	1.69	&	2.51	&	2.07	&	-0.05$\pm$0.03	&	-1.05$\pm$0.03	&	-0.86$\pm$0.03	&	-1.26$\pm$0.03	&	-0.5	&	9	\\
HE 2145-1715	&	13.2	&	0.03$\pm$0.05	&	-4.28$\pm$3.14	&	0.59	&	2.46	&	1.86	&	-2.56$\pm$1.53	&	-3.23$\pm$1.83	&	-2.77$\pm$1.62	&	-3.46$\pm$1.93 & -1.5	&	1	\\
HE 2150-1800 & 14.8 & 0.17$\pm$0.02 & 0.89$\pm$0.05 & 0.34 & 1.49 & 0.90 & -2.25$\pm$0.06 & -2.73$\pm$0.05 & -2.42$\pm$0.05 & -3.15$\pm$0.05 & - & - \\
HE 2218+0127	& 14.0  &	1.49$\pm$0.03 &	4.86$\pm$0.06  &	1.32  &	2.44  &	1.66  &	0.01$\pm$0.03 & -0.85$\pm$0.02 &	-0.67$\pm$0.02 &	-1.28$\pm$0.02 & - & - \\
HE 2224-0330	&	13.5	&	0.50$\pm$0.03	&	2.0$\pm$0.13	&	1.24	&	3.09	&	2.25	&	-0.21$\pm$0.03	&	-1.24$\pm$0.03	&	-0.77$\pm$0.03	&	-1.64$\pm$0.49 &	-1.2	&	1	\\
HE 2319-0852 & 15.2 & 0.06$\pm$0.03 & -0.63$\pm$0.22 & 0.32 & 0.60 & 0.57 & -3.24$\pm$0.22 & -3.77$\pm$0.21 & -3.49$\pm$0.22 & -3.81$\pm$0.21 & - & - \\
HE 2331-1329    &   14.5    & 0.22$\pm$0.02    &  -3.78$\pm$0.51           & 1.52 & 3.58  &  1.51  &  -1.85$\pm$0.43  &  -2.37$\pm$0.38  & -2.38$\pm$0.39  &  -3.13$\pm$0.33 & - & - \\
HE 2339-0837    &   14.0    & 0.11$\pm$0.02     &   -0.82$\pm$0.30    &  0.59  & 1.17 & 0.76  &  -2.71$\pm$0.08  &  -3.10$\pm$0.07  &  -3.01$\pm$0.07  & -3.38$\pm$0.07  & -2.71 & 1\\
HE 2352-1906	&	12.1	&	1.28$\pm$0.03	&	2.6$\pm$0.05	&	2.15	&	3.58	&	1.99	&	-0.49$\pm$0.01	&	-0.43$\pm$0.02	&	-0.51$\pm$0.01	&	-1.19$\pm$0.01 & -	&	-	\\	
\hline

\end{tabular}
}

1. \cite{Kennedy_2011}, 2. \cite{Jorrisen_2016}, 3. \cite{Barbuy_1997}, 4. \cite{Abate_2015}, 5. \cite{Aoki_2007}, 6. \cite{Goswami_2006}, 7. \cite{Yong_2013}, 8. \cite{Purandardas_2021a}, 9. \cite{Hansen_2016c} 
\end{table*}
}

\subsection{Microturbulence}\label{subsec3}
The HERES collaboration \citep{Barklem_2005} could measure the microturbulent velocity $\xi$, for 254 metal-poor stars based on the analysis of high-resolution spectra. \cite{Johnson_2007} fitted the log g and $\xi$ values obtained by \cite{Barklem_2005} with a second order polynomial, \\\\
$\xi$ = 2.822-0.669log g+0.080(log g)$^{2}$ km s$^{-1}$.\\\\
We made use of this relation for the estimation of the microturbulent velocity. This equation is valid for the surface gravity in the range 1 $<$ log g $<$ 4.2. The HERES data show an rms scatter of 0.17 kms$^{-1}$ around this relation. \cite{Johnson_2007} found that the uncertainties in $\xi$ do not contribute significantly to the final abundance errors.

\begin{figure}
\centering
\includegraphics[width=10cm,height=9cm]{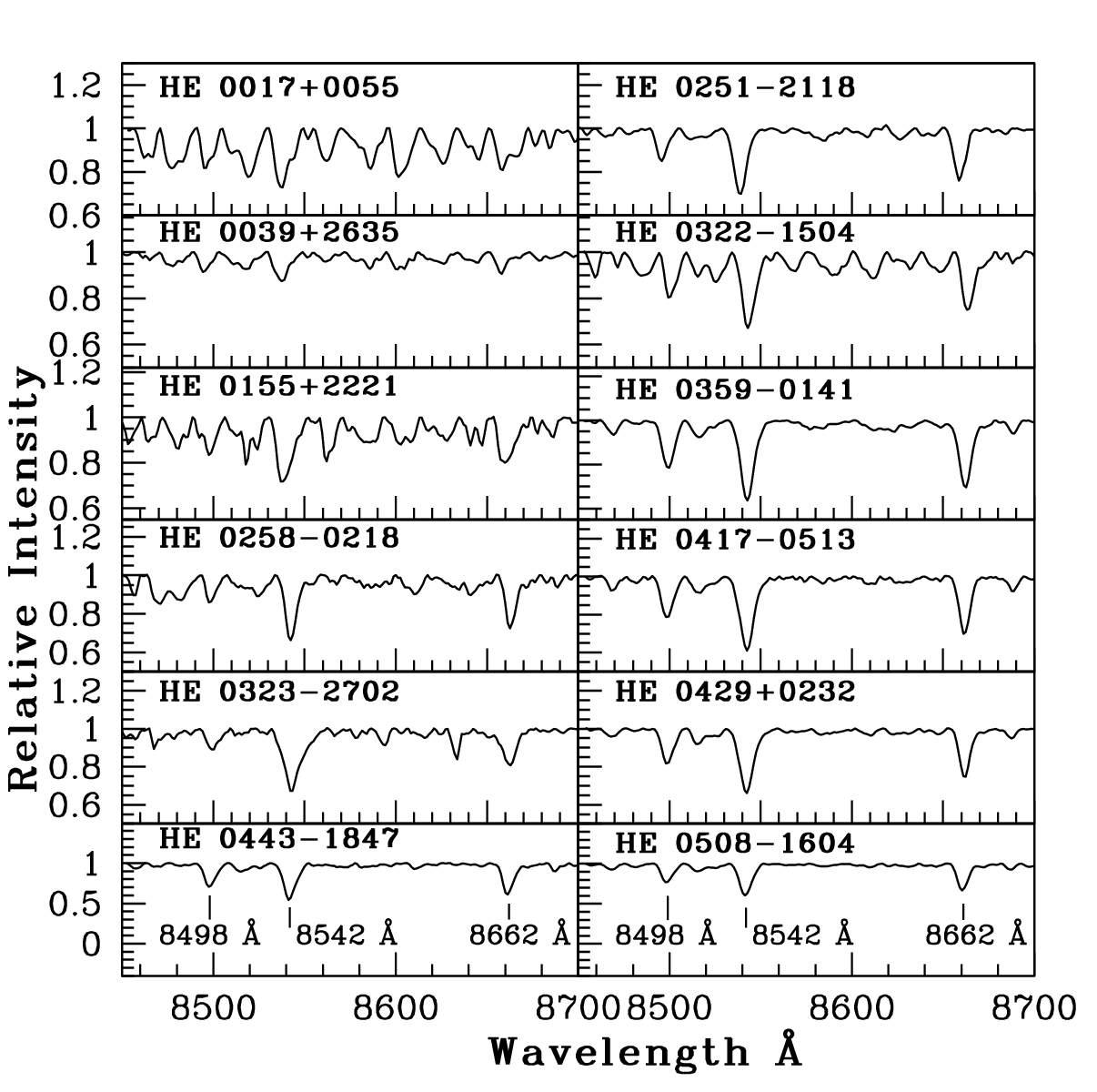}
\includegraphics[width=10cm,height=9cm]{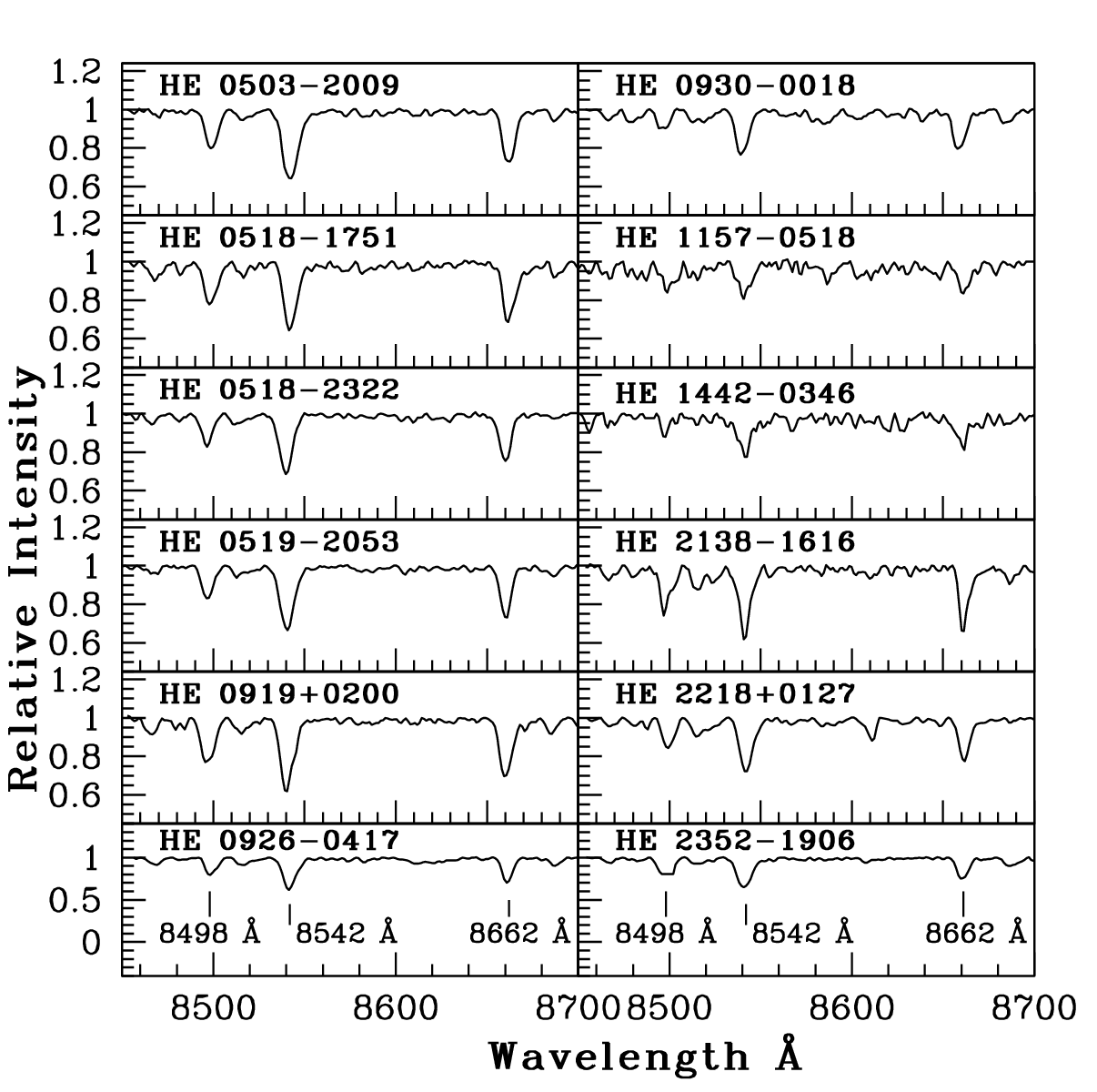}
\caption{Examples of the sample spectra of programme stars showing the CaT lines in the wavelength region 8450 to 8700 Å.}\label{fig6}
\end{figure}

\begin{figure}
\centering
\includegraphics[width=9cm,height=9cm]{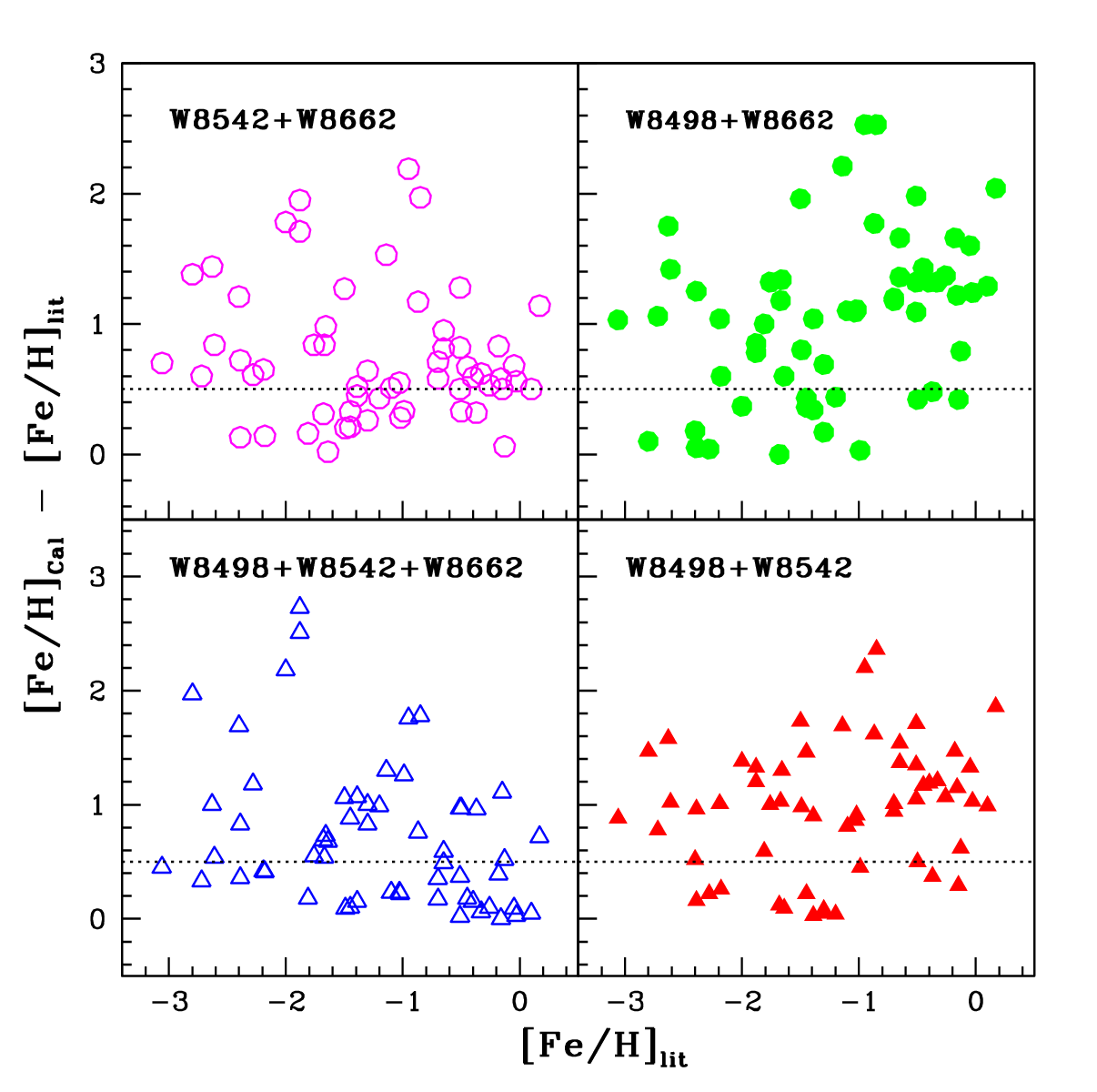}
\caption{Plot of metallicity, [Fe/H] of stars determined from the high-resolution spectroscopic analysis (compiled from various literature, Table \ref{tableB2}) vs. difference between their metallicities calculated from the CaT lines and that from high-resolution spectroscopic analysis. The combination of CaT lines used for the calculation of the metallicity is displayed in the corresponding panels. The short dashed line corresponds to the aforementioned difference equal to 0.5dex}
\label{fig7}
\end{figure}

\subsection{Metallicity }\label{subsec4}
The metallicities of the programme stars are determined from the CaT lines using the calibration equation provided by \cite{Carrera_2013}. They have presented a calibration of the CaT lines as a metallicity indicator in the metallicity range  $-$4.0 $\leq$ [Fe/H] $\leq$ +0.5, based on a tight correlation observed between [Fe/H] values derived from the CaT and [Fe/H] measured from high-resolution spectra as follows:\\\\
$[Fe/H] = a+b\times Mag+c\times \sum Ca+d\times \sum Ca^{-1.5}+e\times \sum Ca \times Mag$,\\\\
where $Mag$ represents the luminosity indicator. The values of the terms, a,b,c,d, and e for each luminosity indicators are available in \cite{Carrera_2013}. $\sum$ Ca is the CaT index which represents the sum of the equivalent widths of the CaT features.

\par We have used this relation to derive the metallicity of our program stars whenever the grism 8 spectra are available for them.  These values are presented in Table \ref{tab3}. We have used absolute magnitude in V (M$_{v}$) as the luminosity indicator. Examples of the sample spectra of the programme stars showing the CaT lines in the wavelength region 8450 to 8700 {\rm \AA} are shown in Figure \ref{fig6}. We have also determined the CaT metallicity for different combinations of equivalent widths of CaT features for 43 stars using high-resolution spectra (R $\sim$ 30000, 42000 and 60000) for which high-resolution metallicity estimates are available (Table \ref{tableB2}). We have plotted the metallicity of 43 stars, determined through high-resolution spectroscopy (values compiled from various sources in literature, as indicated below Table \ref{tableB2}), in relation to the variance between their metallicities calculated from the CaT lines and the corresponding high-resolution estimates (Figure \ref{fig7}). From this figure, we found that the metallicities corresponding to W8498+W8542+W8662 and W8542+W8662 show good agreement with the literature values.

{\footnotesize												
\begin{table*}																							
\caption{Derived atmospheric parameters of the programme stars. }\label{tab4}												
\resizebox{0.4\textheight}{!}{\begin{tabular}{lccccc}												
\hline												
Star  	&	T$_{eff}$          	&	T$_{eff}$          	&	 log g           	&	 $\zeta$        	&	 [Fe/H] \\
      	&	(J-K)	&	 (H$_{\alpha}$) 	&	 (Parallax) 	&	 (log g-Parallax) 	&	 (CaT) 	         \\
      	&	 (K)  	&	 (K)  	&	  (cgs) 	&	 (km s$^{-1}$) 	&	 		 \\
\hline	
HE 0002+0053 	& 4070.0	&	4250	&	1.28	&	1.28	&	  $-1.77$$\pm$0.38 (CT1)                  		 \\
HE 0017+0055 	& 4261.4&	4500	&	0.94	&	2.26	&	 $-1.42$$\pm$0.10 (CT3) 	  \\
HE 0037-0654	& 5458.0  &	-       &	3.07	&	1.52	&	$-1.88$$\pm$0.11(CT3)		\\
HE 0039-2635 	& 4911.7&	5100	&	1.68	&	1.92	&	  $-2.67$$\pm$1.49 (CT1)            	  \\
HE 0155-2221 	& 4266.0	&	4500	&	2.00	&	1.82	&	 $-1.67$$\pm$1.0 (CT3)  		\\ 
HE 0206-1916	    & 4895.0	&	4650	&	1.84	&	1.86	&	$-2.71$$\pm$0.05 (CT3)		\\
HE 0228-0256    & 3926.2& 4200  & 1.50 & 1.99  & $-$2.65$\pm$1.84(CT3)\\
HE 0251-2118 	& 4655.0	&	4400	&	1.71	&	1.91	&	 $-1.63$$\pm$0.06 (CT3) 		 \\
HE 0258-0218 	& 4586.9&	4800	&	1.39	&	2.04	&	  $-1.91$$\pm$0.40 (CT3)      		  \\
HE 0319-0215	&	-	&	4700	&	1.30	&	2.08	&	$-3.09$$\pm$0.08(CT3)		\\
HE 0322-1504 & 4370.3& 4600 & 1.20 & 2.13 & $-$0.15$\pm$0.12 (CT3) \\
HE 0323-2702 	& 4701.8&	4500	&	1.32	&	2.08	&	  $-1.80$$\pm$1.10 (CT3)                    	 \\
HE 0359-0141 	&4243.2	&	4450	&	2.09	&	1.77	&	 $-1.04$$\pm$0.03 (CT2) 		   \\
HE 0443-1847 	&4559.4	&	4300	&	2.34	&	1.69	&	 $-0.43$$\pm$0.02 (CT3) 		 \\
HE 0503-2009 	&4678.3	&	4450	&	2.31	&	1.7	&	 $-1.09$$\pm$0.05 (CT2) 		 \\
HE 0507-1430 	&4407.5	&	4600	&	0.78	&	2.35	&	 $-2.01$$\pm$0.27 (CT3)                   		 \\
HE 0507-1653 	&4935.0	&	5200	&	2.27	&	1.72	&	 $-1.63$$\pm$0.03 (CT1)                    		 \\
HE 0508-1604 	&4432.7	&	4650	&	2.05	&	1.77	&	 $-0.94$$\pm$0.03 (CT3) 		 \\
HE 0518-1751 	&4589.8	&	4350	&	2.95	&	1.54	&	 $-1.14$$\pm$0.01 (CT2)         		 \\
HE 0518-2322 	&4885.4	&	5100	&	1.99	&	1.81	&	 $-1.02$$\pm$0.06 (CT1)       		 \\
HE 0519-2053 	&4623.5	&	4400	&	2.00	&	1.82	&	 $-1.35$$\pm$0.04 (CT1) 		 \\
HE 0926-0417 	&4731.7	&	4500	&	2.78	&	1.58	&	 $-0.88$$\pm$0.02 (CT2)                   		  \\
HE 1045-1434    &4564.8	&	4800	&	1.61	&	1.95	&	$-3.01$$\pm$0.11	(CT3)	\\
HE 1157-0518 	& 4921.7&	4700	&	1.88	&	1.85	&	 $-1.56$$\pm$0.30 (CT1) 		 \\
HE 1205-0521	&	4768.0	&	5000	&	1.50	&	1.99	&	$-$2.87$\pm$0.48(CT3)		\\
HE 1319-1935	&	4678.0	&	-	&	1.01	&	2.23	&	$-2.67$$\pm$0.58	(CT3)	\\
HE 1429-0551 	&	4609.4	&	4800	&	1.61	&	1.96	&	$-2.89$$\pm$0.03	(CT3)	\\
HE 1442-0346 	&	4719.7	&	4950	&	1.62	&	1.95	&	  $-1.83$$\pm$0.67 (CT1)            \\         	
HE 1523-1155	&	4768.2	&	5000	&	1.81	&	1.87	&	$-$2.51$\pm$0.03(CT3)\\	
HE 2138-1616 	&	4861.9	&	4600	&	3.48	&	1.46	&	 $-0.86$$\pm$0.03 (CT3) 		 \\
HE 2145-1715 	&	4341.2	&	4100	&	0.43	&	2.55	&	 $-2.56$$\pm$1.53 (CT1)        		  \\
HE 2150-1800    &   5017.1  &   4900    &   1.20    &   2.13    &    $-2.25$$\pm$0.06 (CT1) \\
HE 2218+0127 	&	5623.3	&	5400	&	4.35	&	1.43	&	  $-0.67$$\pm$0.02 (CT3)                    		  \\
HE 2224-0330 	&	4968.8	&	4750	&	2.70	&	1.57	&	 $-0.77$$\pm$0.03 (CT3)                   		 \\
HE 2339-0837	&	4830.7	&	4600	&	1.66	&	1.93	&	$-3.01$$\pm$0.07(CT3)		\\
HE 2352-1906 	&	4672.4	&	4450	&	2.91	&	1.55	&	 $-0.51$$\pm$0.01 (CT3)                   	  \\
												
\hline												
\end{tabular}												
}												
									
\end{table*}												
} 	

\section{Determination of carbon abundance}\label{sec5}
\begin{figure}
\centering
\includegraphics[width=0.8\columnwidth]{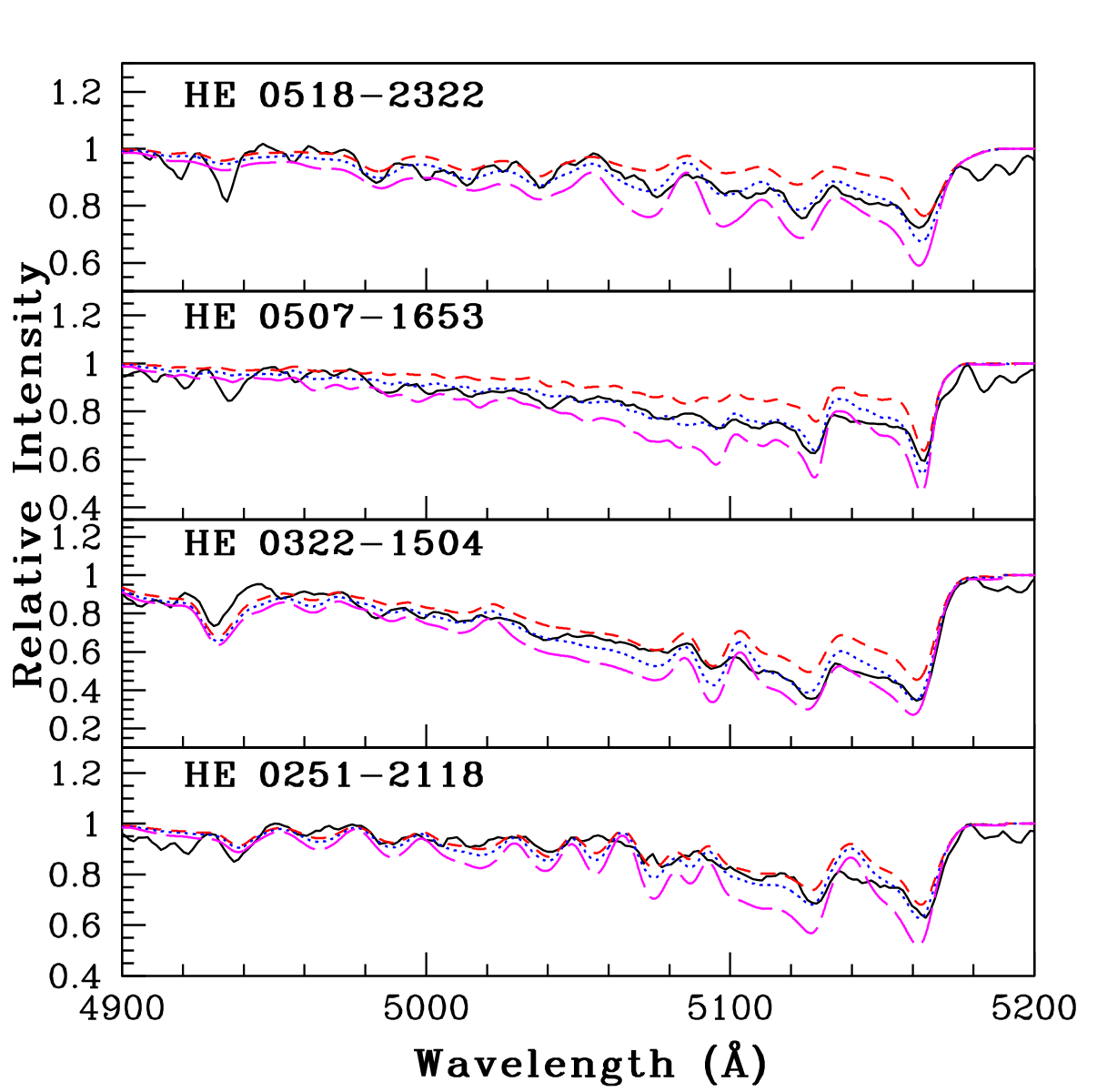}
\caption{Synthesis of C$_{2}$ band around 5165 {\rm \AA}. Dotted line represents synthesized spectra and the solid line indicates the observed spectra. Short dashed line represents the synthetic spectra corresponding to $\Delta$ [C/Fe] = -0.3 and long dashed line corresponds to $\Delta$[C/Fe] = +0.3}\label{fig8}
\end{figure}

We determined the carbon abundance in potential CH star candidates within our sample through spectrum synthesis calculations of the carbon band at 5165 {\rm \AA} (Figure \ref{fig8}). We have made use of MOOG (\cite{Sneden_1973}, updated version 2019) for the analysis under the assumption of local thermodynamic equilibrium. In most cases, CH band around 4300{\rm \AA}, and C$_{2}$ band around 5635 {\rm \AA} are found to be saturated, and hence we used the carbon band at 5165 {\rm \AA} whenever possible. We estimated the abundances of elements such as Mg, Ca, Ti and Ni, which have the maximum contributions in the spectral region of our interest, and are taken from Figure 7 of \cite{Goswami_2000}. The metallicity used for the estimation of carbon abundance is listed in Table \ref{tab4}. We have used the metallicities corresponding to either W8498+W8542+W8662 or W8542+W8662 based on the availability of clean and symmetric CaT lines. We have also used the metallicity derived from the combination W8498+W8542 in cases where the line at 8662 {\rm \AA} is either blended or asymmetric. The derived carbon abundances are listed in Table \ref{tab5}. In many cases, it was not possible to derive the carbon abundance due to various reasons, such as the carbon bands being too strong and saturated, and/or unavailability of reliable  stellar atmospheric parameters for objects, as discussed in Section \ref{sec4}. 

\par In light of the diagram presented in \cite{Yoon_2016}(their figure 1), our program stars for which carbon abundances were derived fall in the region labeled as "Group 1", just where CEMP-s and CEMP-r/s stars are found. Also, since such stars show [C/Fe]$>$1.0 dex, they lie in the region of CEMP stars in the plot of [C/Fe] vs. log(L/L$_{sun}$) of \cite{Aoki_2007}(their figure 4).

{\footnotesize
\begin{table*}
\caption{Derived carbon abundances of the programme stars. }\label{tab5}
\resizebox{0.35\textheight}{!}{\begin{tabular}{lcccc}
\hline
Star  	& log$\epsilon$(C)	& [C/Fe]	&  [C/Fe] &	 Reference\\
      	&                   &           & (Lit.value) &	         \\
\hline								
HE 0002+0053 	&	8.80	&	2.41	&	2.15	&	 1 \\
HE 0017+0055 	&	8.95	&	1.64	&	 2.17 	&	 2 \\
HE 0037-0654    &   8.55    & 2.00  & - & - \\
HE 0039-2635 	&	8.55	&	2.94	&	  2.32 	&	3  \\
HE 0155-2221 	&	8.82	&	1.77	&	1.78	&	 1\\ 
HE 0206-1916    &   8.55    &  2.83     & - & - \\
HE 0228-0256 	&	8.76	&	2.71	&	 - 	&	 - \\ 
HE 0251-2118 	&	8.67	&	1.16	&	1.13	&	 1\\
HE 0258-0218 	&	8.40	&	1.77	&	 - 	&	 - \\
HE 0319-0215    &  8.55  & 3.21 & - & - \\
HE 0322-1504 	&	8.65	&	2.10	&	 2.40 	&	 4\\
HE 0323-2702 	&	8.58	&	1.77	&	- 	&	- \\
HE 0507-1430 	&	8.80	&	2.2	&	2.7	&	 4\\
HE 0507-1653 	&	8.50	&	1.71	&	 1.30 	&	 5\\
HE 0518-2322 	&	8.30	&	1.15	&	 -    	&	 -\\
HE 1045-1434    &   8.70  & 3.28  & - & - \\
HE 1157-0518 	&	8.65	&	2.13	&	 2.15 	&	 5\\
HE 1319-1935    &   8.55   & 2.79 & - & - \\
HE 1429-0551    &   8.60  & 3.06 & - & - \\
HE 1442-0346 	&	8.43	&	1.96	&	- 	&	- \\
HE 1523-1155    &   8.55    & 2.63 & - & - \\
HE 2145-1715 	&	8.41	&	2.64	&	1.14	&	 1 \\
HE 2150-1800    &   7.68    &  1.50   & - & - \\
HE 2218+0127 	&	8.63	&	1.87	&	 - 	&	 - \\
HE 2339-0837    &   8.60    & 3.18 & - & - \\
\hline
\end{tabular}
}

1. \cite{Kennedy_2011}, 2. \cite{jorrisen.2016A&A...586A.159J}, 3. \cite{Barbuy_1997}, 4. \cite{Abate_2015}, 5. \cite{Aoki_2007}
\end{table*}
}

\section{Discussions and Conclusions}\label{sec6}

Low-resolution HCT/HFOSC spectra of a sample of 88 carbon stars are analyzed. The sample is found to contain 53 CH stars, four C-R stars, and two C-N type stars based on a set of well defined carbon star classification criteria. Twenty-nine objects in our sample do not show  prominent C$_{2}$ molecular bands on their spectra. So we could not classify these stars. We could derive stellar atmospheric parameters for 36 stars in our sample. We could not derive these parameters for the remaining stars due to several reasons, as discussed in Section \ref{sec4}. Effective temperatures were determined using both the photometric calibration equations of \cite{Alonso_1999} and the spectral synthesis of the H$_{\alpha}$ line at 6562 \AA. Discrepancies up to 250 K for J-K temperatures and 480 K for J-H and V-K temperatures were observed between the two methods.

\par Several studies in the past have investigated the dependence of the strength of near-infrared CaT lines at $\lambda$ $\lambda$ 8498, 8542, and 8662 {\rm \AA} on stellar  parameters like luminosity  and metallicity (\cite{Carrera_2007, Carrera_2013} and references therein). These lines formed due to transitions between the upper 4p$^{2}$P$_{1/2,3/2}$ levels and the lower meta-stable  3d$^{2}$D$_{3/2,5/2}$ levels are the most conspicuous features in the near-infrared region of the spectrum and  hence easy to observe even in faint stars. These lines are relatively free from blends unlike Na I lines at $\lambda$ $\lambda$ 8183.25, 8194.82 {\rm \AA} that are contaminated due to the presence of telluric lines. The metallicity of the 49 program stars is determined from the CaT lines using the calibration equation provided by \cite{Carrera_2007}. The derived metallicities range from $-0.43$ $\leq$ [Fe/H] $\leq$ $-$3.49. Our sample consists of 19 metal-poor ([Fe/H] $\leq$ $-1$), 14 very metal-poor ([Fe/H] $\leq$ $-2$), and five extremely metal-poor ([Fe/H] $\leq$ $-3$) stars. Eleven objects are found to have metallicity in the range: $-0.43$ $\leq$ [Fe/H] $\leq$ $-0.97$. 

\par CaT metallicities were determined for 43 stars for which high-resolution spectroscopic analysis results  are available. Comparisons with high-resolution estimates revealed good agreement for W8498+W8542+W8662 and W8542+W8662, with average differences of 0.72 dex and 0.75 dex, respectively. This difference for the combination of W8498+W8542 and W8498+W8662 is determined to be 0.97 dex and 1.05 dex respectively.

\par Carbon abundances were derived for 25 potential CH star candidates using spectrum synthesis calculations of the C$_{2}$ band around 5165 \AA\,. All 25 program stars exhibit enhanced carbon abundance  with [C/Fe] in the range 1.15 $\leq$ [C/Fe] $\leq$ 3.28. We have compared our estimates of carbon abundance with that of high-resolution estimates whenever it is available. Comparisons with high-resolution estimates showed an average difference in carbon abundance ([C/Fe]) of 0.5 dex.

\par In this study, we identified potential CH star candidates and estimated their carbon abundances whenever possible. These stars exhibit enhanced carbon abundance and represent a confirmed sample of CEMP stars. While further detailed high-resolution spectroscopic analysis is needed to confirm sub-groups (CEMP-s, CEMP-r/s, or CEMP-r) to which these stars belong, this sample serves as crucial targets for future detailed chemical composition studies based on high-resolution spectroscopy. Such studies are expected to provide valuable insights into the early Galactic chemical evolution.

\backmatter

\bmhead{Acknowledgments}
This work made use of the SIMBAD astronomical database, operated at CDS, Strasbourg, France, the NASA ADS, USA and data from the European Space Agency (ESA) mission Gaia (\url{https://www.cosmos.esa.int/gaia}), processed by the Gaia Data Processing and Analysis Consortium (DPAC, \url{https://www.cosmos.esa.int/web/gaia/dpac/consortium}). We thank the referee for many
constructive suggestions and useful comments on the manuscript that improved this paper. We thank the staff at IAO, Hanle, and  Centre For Research \& Education in Science \& Technology (CREST), Hoskote, for assistance during the observations. We extend our gratitude to Dr Drisya Karinkuzhi, Dr Shejeelammal J and Dr Partha Pratim Goswami for diligently observing several objects used for this study during their dedicated observing nights at CREST.



\bigskip

\bibliography{Bibliography}

\begin{thebibliography}{81}
\providecommand{\natexlab}[1]{#1}
\providecommand{\url}[1]{{#1}}
\providecommand{\urlprefix}{URL }
\providecommand{\doi}[1]{\url{https://doi.org/#1}}
\providecommand{\eprint}[2][]{\url{#2}}
 \bibcommenthead

\bibitem[{{Abate} et~al(2015){Abate}, {Pols}, {Izzard}, and {Karakas}}]{Abate_2015}
{Abate} C, {Pols} OR, {Izzard} RG, et~al (2015) {Carbon-enhanced metal-poor stars: a window on AGB nucleosynthesis and binary evolution. II. Statistical analysis of a sample of 67 CEMP-s stars}. \aap 581:A22. \doi{10.1051/0004-6361/201525876}, {\href{https://arxiv.org/abs/1507.04662}{{arXiv:1507.04662}}} {[astro-ph.SR]}

\bibitem[{{Abate} et~al(2016){Abate}, {Stancliffe}, and {Liu}}]{Abate_2016}
{Abate} C, {Stancliffe} RJ, {Liu} ZW (2016) {How plausible are the proposed formation scenarios of CEMP-r/s stars?} \aap 587:A50. \doi{10.1051/0004-6361/201527864}, {\href{https://arxiv.org/abs/1601.00976}{{arXiv:1601.00976}}} {[astro-ph.SR]}

\bibitem[{{Abbott} et~al(2017){Abbott}, {Abbott}, {Abbott}, {Acernese}, {Ackley}, {Adams}, {Adams}, {Addesso}, {Adhikari}, {Adya}, {Affeldt}, {Afrough}, {Agarwal}, {Agathos}, {Agatsuma}, {Aggarwal}, {Aguiar}, {Aiello}, {Ain}, {Ajith}, {Allen}, {Allen}, {Allocca}, {Aloy}, {Altin}, {Amato}, {Ananyeva}, {Anderson}, {Anderson}, {Angelova}, {Antier}, {Appert}, {Arai}, {Araya}, {Areeda}, {Arnaud}, {Arun}, {Ascenzi}, {Ashton}, {Ast}, {Aston}, {Astone}, {Atallah}, {Aufmuth}, {Aulbert}, {AultONeal}, {Austin}, {Avila-Alvarez}, {Babak}, {Bacon}, {Bader}, {Bae}, {Baker}, {Baldaccini}, {Ballardin}, {Ballmer}, {Banagiri}, {Barayoga}, {Barclay}, {Barish}, {Barker}, {Barkett}, {Barone}, {Barr}, {Barsotti}, {Barsuglia}, {Barta}, {Bartlett}, {Bartos}, {Bassiri}, {Basti}, {Batch}, {Bawaj}, {Bayley}, {Bazzan}, {B{\'e}csy}, {Beer}, {Bejger}, {Belahcene}, {Bell}, {Berger}, {Bergmann}, {Bero}, {Berry}, {Bersanetti}, {Bertolini}, {Betzwieser}, {Bhagwat}, {Bhandare}, {Bilenko}, {Billingsley}, {Billman}, {Birch}, {Birney},
  {Birnholtz}, {Biscans}, {Biscoveanu}, {Bisht}, {Bitossi}, {Biwer}, {Bizouard}, {Blackburn}, {Blackman}, {Blair}, {Blair}, {Blair}, {Bloemen}, {Bock}, {Bode}, {Boer}, {Bogaert}, {Bohe}, {Bondu}, {Bonilla}, {Bonnand}, {Boom}, {Bork}, {Boschi}, {Bose}, {Bossie}, {Bouffanais}, {Bozzi}, {Bradaschia}, {Brady}, {Branchesi}, {Brau}, {Briant}, {Brillet}, {Brinkmann}, {Brisson}, {Brockill}, {Broida}, {Brooks}, {Brown}, {Brown}, {Brunett}, {Buchanan}, {Buikema}, {Bulik}, {Bulten}, {Buonanno}, {Buskulic}, {Buy}, {Byer}, {Cabero}, {Cadonati}, {Cagnoli}, {Cahillane}, {Calder{\'o}n Bustillo}, {Callister}, {Calloni}, {Camp}, {Canepa}, {Canizares}, {Cannon}, {Cao}, {Cao}, {Capano}, {Capocasa}, {Carbognani}, {Caride}, {Carney}, {Casanueva Diaz}, {Casentini}, {Caudill}, {Cavagli{\`a}}, {Cavalier}, {Cavalieri}, {Cella}, {Cepeda}, {Cerd{\'a}-Dur{\'a}n}, {Cerretani}, {Cesarini}, {Chamberlin}, {Chan}, {Chao}, {Charlton}, {Chase}, {Chassande-Mottin}, {Chatterjee}, {Chatziioannou}, {Cheeseboro}, {Chen}, {Chen}, {Chen}, {Cheng},
  {Chia}, {Chincarini}, {Chiummo}, {Chmiel}, {Cho}, {Cho}, {Chow}, {Christensen}, {Chu}, {Chua}, {Chua}, {Chung}, {Chung}, {Ciani}, {Ciolfi}, {Cirelli}, {Cirone}, {Clara}, {Clark}, {Clearwater}, {Cleva}, {Cocchieri}, {Coccia}, {Cohadon}, {Cohen}, {Colla}, {Collette}, {Cominsky}, {Constancio}, {Conti}, {Cooper}, {Corban}, {Corbitt}, {Cordero-Carri{\'o}n}, {Corley}, {Cornish}, {Corsi}, {Cortese}, {Costa}, {Coughlin}, {Coughlin}, {Coulon}, {Countryman}, {Couvares}, {Covas}, {Cowan}, {Coward}, {Cowart}, {Coyne}, {Coyne}, {Creighton}, {Creighton}, {Cripe}, {Crowder}, {Cullen}, {Cumming}, {Cunningham}, {Cuoco}, {Dal Canton}, {D{\'a}lya}, {Danilishin}, {D'Antonio}, {Danzmann}, {Dasgupta}, {Da Silva Costa}, {Dattilo}, {Dave}, {Davier}, {Davis}, {Daw}, {Day}, {De}, {DeBra}, {Degallaix}, {De Laurentis}, {Del{\'e}glise}, {Del Pozzo}, {Demos}, {Denker}, {Dent}, {De Pietri}, {Dergachev}, {De Rosa}, {DeRosa}, {De Rossi}, {DeSalvo}, {de Varona}, {Devenson}, {Dhurandhar}, {D{\'\i}az}, {Di Fiore}, {Di Giovanni}, {Di
  Girolamo}, {Di Lieto}, {Di Pace}, {Di Palma}, {Di Renzo}, {Doctor}, {Dolique}, {Donovan}, {Dooley}, {Doravari}, {Dorrington}, {Douglas}, {Dovale {\'A}lvarez}, {Downes}, {Drago}, {Dreissigacker}, {Driggers}, {Du}, {Ducrot}, {Dupej}, {Dwyer}, {Edo}, {Edwards}, {Effler}, {Eggenstein}, {Ehrens}, {Eichholz}, {Eikenberry}, {Eisenstein}, {Essick}, {Estevez}, {Etienne}, {Etzel}, {Evans}, {Evans}, {Factourovich}, {Fafone}, {Fair}, {Fairhurst}, {Fan}, {Farinon}, {Farr}, {Farr}, {Fauchon-Jones}, {Favata}, {Fays}, {Fee}, {Fehrmann}, {Feicht}, {Fejer}, {Fernandez-Galiana}, {Ferrante}, {Ferreira}, {Ferrini}, {Fidecaro}, {Finstad}, {Fiori}, {Fiorucci}, {Fishbach}, {Fisher}, {Fitz-Axen}, {Flaminio}, {Fletcher}, {Fong}, {Font}, {Forsyth}, {Forsyth}, {Fournier}, {Frasca}, {Frasconi}, {Frei}, {Freise}, {Frey}, {Frey}, {Fries}, {Fritschel}, {Frolov}, {Fulda}, {Fyffe}, {Gabbard}, {Gadre}, {Gaebel}, {Gair}, {Gammaitoni}, {Ganija}, {Gaonkar}, {Garcia-Quiros}, {Garufi}, {Gateley}, {Gaudio}, {Gaur}, {Gayathri}, {Gehrels}, {Gemme},
  {Genin}, {Gennai}, {George}, {George}, {Gergely}, {Germain}, {Ghonge}, {Ghosh}, {Ghosh}, {Ghosh}, {Giaime}, {Giardina}, {Giazotto}, {Gill}, {Glover}, {Goetz}, {Goetz}, {Gomes}, {Goncharov}, {Gonz{\'a}lez}, {Gonzalez Castro}, {Gopakumar}, {Gorodetsky}, {Gossan}, {Gosselin}, {Gouaty}, {Grado}, {Graef}, {Granata}, {Grant}, {Gras}, {Gray}, {Greco}, {Green}, {Gretarsson}, {Groot}, {Grote}, {Grunewald}, {Gruning}, {Guidi}, {Guo}, {Gupta}, {Gupta}, {Gushwa}, {Gustafson}, {Gustafson}, {Halim}, {Hall}, {Hall}, {Hamilton}, {Hammond}, {Haney}, {Hanke}, {Hanks}, {Hanna}, {Hannam}, {Hannuksela}, {Hanson}, {Hardwick}, {Harms}, {Harry}, {Harry}, {Hart}, {Haster}, {Haughian}, {Healy}, {Heidmann}, {Heintze}, {Heitmann}, {Hello}, {Hemming}, {Hendry}, {Heng}, {Hennig}, {Heptonstall}, {Heurs}, {Hild}, {Hinderer}, {Hoak}, {Hofman}, {Holt}, {Holz}, {Hopkins}, {Horst}, {Hough}, {Houston}, {Howell}, {Hreibi}, {Hu}, {Huerta}, {Huet}, {Hughey}, {Husa}, {Huttner}, {Huynh-Dinh}, {Indik}, {Inta}, {Intini}, {Isa}, {Isac}, {Isi}, {Iyer},
  {Izumi}, {Jacqmin}, {Jani}, {Jaranowski}, {Jawahar}, {Jim{\'e}nez-Forteza}, {Johnson}, {Johnson-McDaniel}, {Jones}, {Jones}, {Jonker}, {Ju}, {Junker}, {Kalaghatgi}, {Kalogera}, {Kamai}, {Kandhasamy}, {Kang}, {Kanner}, {Kapadia}, {Karki}, {Karvinen}, {Kasprzack}, {Kastaun}, {Katolik}, {Katsavounidis}, {Katzman}, {Kaufer}, {Kawabe}, {K{\'e}f{\'e}lian}, {Keitel}, {Kemball}, {Kennedy}, {Kent}, {Key}, {Khalili}, {Khan}, {Khan}, {Khan}, {Khazanov}, {Kijbunchoo}, {Kim}, {Kim}, {Kim}, {Kim}, {Kim}, {Kim}, {Kimbrell}, {King}, {King}, {Kinley-Hanlon}, {Kirchhoff}, {Kissel}, {Kleybolte}, {Klimenko}, {Knowles}, {Koch}, {Koehlenbeck}, {Koley}, {Kondrashov}, {Kontos}, {Korobko}, {Korth}, {Kowalska}, {Kozak}, {Kr{\"a}mer}, {Kringel}, {Krishnan}, {Kr{\'o}lak}, {Kuehn}, {Kumar}, {Kumar}, {Kumar}, {Kuo}, {Kutynia}, {Kwang}, {Lackey}, {Lai}, {Landry}, {Lang}, {Lange}, {Lantz}, {Lanza}, {Lartaux-Vollard}, {Lasky}, {Laxen}, {Lazzarini}, {Lazzaro}, {Leaci}, {Leavey}, {Lee}, {Lee}, {Lee}, {Lee}, {Lee}, {Lehmann}, {Lenon},
  {Leonardi}, {Leroy}, {Letendre}, {Levin}, {Li}, {Linker}, {Littenberg}, {Liu}, {Lo}, {Lockerbie}, {London}, {Lord}, {Lorenzini}, {Loriette}, {Lormand}, {Losurdo}, {Lough}, {Lousto}, {Lovelace}, {L{\"u}ck}, {Lumaca}, {Lundgren}, {Lynch}, {Ma}, {Macas}, {Macfoy}, {Machenschalk}, {MacInnis}, {Macleod}, {Maga{\~n}a Hernandez}, {Maga{\~n}a-Sandoval}, {Maga{\~n}a Zertuche}, {Magee}, {Majorana}, {Maksimovic}, {Man}, {Mandic}, {Mangano}, {Mansell}, {Manske}, {Mantovani}, {Marchesoni}, {Marion}, {M{\'a}rka}, {M{\'a}rka}, {Markakis}, {Markosyan}, {Markowitz}, {Maros}, {Marquina}, {Martelli}, {Martellini}, {Martin}, {Martin}, {Martynov}, {Mason}, {Massera}, {Masserot}, {Massinger}, {Masso-Reid}, {Mastrogiovanni}, {Matas}, {Matichard}, {Matone}, {Mavalvala}, {Mazumder}, {McCarthy}, {McClelland}, {McCormick}, {McCuller}, {McGuire}, {McIntyre}, {McIver}, {McManus}, {McNeill}, {McRae}, {McWilliams}, {Meacher}, {Meadors}, {Mehmet}, {Meidam}, {Mejuto-Villa}, {Melatos}, {Mendell}, {Mercer}, {Merilh}, {Merzougui}, {Meshkov},
  {Messenger}, {Messick}, {Metzdorff}, {Meyers}, {Miao}, {Michel}, {Middleton}, {Mikhailov}, {Milano}, {Miller}, {Miller}, {Miller}, {Millhouse}, {Milovich-Goff}, {Minazzoli}, {Minenkov}, {Ming}, {Mishra}, {Mitra}, {Mitrofanov}, {Mitselmakher}, {Mittleman}, {Moffa}, {Moggi}, {Mogushi}, {Mohan}, {Mohapatra}, {Montani}, {Moore}, {Moraru}, {Moreno}, {Morriss}, {Mours}, {Mow-Lowry}, {Mueller}, {Muir}, {Mukherjee}, {Mukherjee}, {Mukherjee}, {Mukund}, {Mullavey}, {Munch}, {Mu{\~n}iz}, {Muratore}, {Murray}, {Napier}, {Nardecchia}, {Naticchioni}, {Nayak}, {Neilson}, {Nelemans}, {Nelson}, {Nery}, {Neunzert}, {Nevin}, {Newport}, {Newton}, {Ng}, {Nguyen}, {Nichols}, {Nielsen}, {Nissanke}, {Nitz}, {Noack}, {Nocera}, {Nolting}, {North}, {Nuttall}, {Oberling}, {O'Dea}, {Ogin}, {Oh}, {Oh}, {Ohme}, {Okada}, {Oliver}, {Oppermann}, {Oram}, {O'Reilly}, {Ormiston}, {Ortega}, {O'Shaughnessy}, {Ossokine}, {Ottaway}, {Overmier}, {Owen}, {Pace}, {Page}, {Page}, {Pai}, {Pai}, {Palamos}, {Palashov}, {Palomba}, {Pal-Singh}, {Pan},
  {Pan}, {Pang}, {Pang}, {Pankow}, {Pannarale}, {Pant}, {Paoletti}, {Paoli}, {Papa}, {Parida}, {Parker}, {Pascucci}, {Pasqualetti}, {Passaquieti}, {Passuello}, {Patil}, {Patricelli}, {Pearlstone}, {Pedraza}, {Pedurand}, {Pekowsky}, {Pele}, {Penn}, {Perez}, {Perreca}, {Perri}, {Pfeiffer}, {Phelps}, {Piccinni}, {Pichot}, {Piergiovanni}, {Pierro}, {Pillant}, {Pinard}, {Pinto}, {Pirello}, {Pitkin}, {Poe}, {Poggiani}, {Popolizio}, {Porter}, {Post}, {Powell}, {Prasad}, {Pratt}, {Pratten}, {Predoi}, {Prestegard}, {Prijatelj}, {Principe}, {Privitera}, {Prodi}, {Prokhorov}, {Puncken}, {Punturo}, {Puppo}, {P{\"u}rrer}, {Qi}, {Quetschke}, {Quintero}, {Quitzow-James}, {Raab}, {Rabeling}, {Radkins}, {Raffai}, {Raja}, {Rajan}, {Rajbhandari}, {Rakhmanov}, {Ramirez}, {Ramos-Buades}, {Rapagnani}, {Raymond}, {Razzano}, {Read}, {Regimbau}, {Rei}, {Reid}, {Reitze}, {Ren}, {Reyes}, {Ricci}, {Ricker}, {Rieger}, {Riles}, {Rizzo}, {Robertson}, {Robie}, {Robinet}, {Rocchi}, {Rolland}, {Rollins}, {Roma}, {Romano}, {Romel}, {Romie},
  {Rosi{\'n}ska}, {Ross}, {Rowan}, {R{\"u}diger}, {Ruggi}, {Rutins}, {Ryan}, {Sachdev}, {Sadecki}, {Sadeghian}, {Sakellariadou}, {Salconi}, {Saleem}, {Salemi}, {Samajdar}, {Sammut}, {Sampson}, {Sanchez}, {Sanchez}, {Sanchis-Gual}, {Sandberg}, {Sanders}, {Sassolas}, {Sathyaprakash}, {Saulson}, {Sauter}, {Savage}, {Sawadsky}, {Schale}, {Scheel}, {Scheuer}, {Schmidt}, {Schmidt}, {Schnabel}, {Schofield}, {Sch{\"o}nbeck}, {Schreiber}, {Schuette}, {Schulte}, {Schutz}, {Schwalbe}, {Scott}, {Scott}, {Seidel}, {Sellers}, {Sengupta}, {Sentenac}, {Sequino}, {Sergeev}, {Shaddock}, {Shaffer}, {Shah}, {Shahriar}, {Shaner}, {Shao}, {Shapiro}, {Shawhan}, {Sheperd}, {Shoemaker}, {Shoemaker}, {Siellez}, {Siemens}, {Sieniawska}, {Sigg}, {Silva}, {Singer}, {Singh}, {Singhal}, {Sintes}, {Slagmolen}, {Smith}, {Smith}, {Smith}, {Somala}, {Son}, {Sonnenberg}, {Sorazu}, {Sorrentino}, {Souradeep}, {Spencer}, {Srivastava}, {Staats}, {Staley}, {Steinke}, {Steinlechner}, {Steinlechner}, {Steinmeyer}, {Stevenson}, {Stone}, {Stops},
  {Strain}, {Stratta}, {Strigin}, {Strunk}, {Sturani}, {Stuver}, {Summerscales}, {Sun}, {Sunil}, {Suresh}, {Sutton}, {Swinkels}, {Szczepa{\'n}czyk}, {Tacca}, {Tait}, {Talbot}, {Talukder}, {Tanner}, {T{\'a}pai}, {Taracchini}, {Tasson}, {Taylor}, {Taylor}, {Tewari}, {Theeg}, {Thies}, {Thomas}, {Thomas}, {Thomas}, {Thorne}, {Thorne}, {Thrane}, {Tiwari}, {Tiwari}, {Tokmakov}, {Toland}, {Tonelli}, {Tornasi}, {Torres-Forn{\'e}}, {Torrie}, {T{\"o}yr{\"a}}, {Travasso}, {Traylor}, {Trinastic}, {Tringali}, {Trozzo}, {Tsang}, {Tse}, {Tso}, {Tsukada}, {Tsuna}, {Tuyenbayev}, {Ueno}, {Ugolini}, {Unnikrishnan}, {Urban}, {Usman}, {Vahlbruch}, {Vajente}, {Valdes}, {van Bakel}, {van Beuzekom}, {van den Brand}, {Van Den Broeck}, {Vander-Hyde}, {van der Schaaf}, {van Heijningen}, {van Veggel}, {Vardaro}, {Varma}, {Vass}, {Vas{\'u}th}, {Vecchio}, {Vedovato}, {Veitch}, {Veitch}, {Venkateswara}, {Venugopalan}, {Verkindt}, {Vetrano}, {Vicer{\'e}}, {Viets}, {Vinciguerra}, {Vine}, {Vinet}, {Vitale}, {Vo}, {Vocca}, {Vorvick},
  {Vyatchanin}, {Wade}, {Wade}, {Wade}, {Walet}, {Walker}, {Wallace}, {Walsh}, {Wang}, {Wang}, {Wang}, {Wang}, {Wang}, {Ward}, {Warner}, {Was}, {Watchi}, {Weaver}, {Wei}, {Weinert}, {Weinstein}, {Weiss}, {Wen}, {Wessel}, {We{\ss}els}, {Westerweck}, {Westphal}, {Wette}, {Whelan}, {Whitcomb}, {Whiting}, {Whittle}, {Wilken}, {Williams}, {Williams}, {Williamson}, {Willis}, {Willke}, {Wimmer}, {Winkler}, {Wipf}, {Wittel}, {Woan}, {Woehler}, {Wofford}, {Wong}, {Worden}, {Wright}, {Wu}, {Wysocki}, {Xiao}, {Yamamoto}, {Yancey}, {Yang}, {Yap}, {Yazback}, {Yu}, {Yu}, {Yvert}, {Zadro{\.z}ny}, {Zanolin}, {Zelenova}, {Zendri}, {Zevin}, {Zhang}, {Zhang}, {Zhang}, {Zhang}, {Zhao}, {Zhou}, {Zhou}, {Zhu}, {Zhu}, {Zimmerman}, {Zucker}, {Zweizig}, {(LIGO Scientific Collaboration}, {Virgo Collaboration}, {Burns}, {Veres}, {Kocevski}, {Racusin}, {Goldstein}, {Connaughton}, {Briggs}, {Blackburn}, {Hamburg}, {Hui}, {von Kienlin}, {McEnery}, {Preece}, {Wilson-Hodge}, {Bissaldi}, {Cleveland}, {Gibby}, {Giles}, {Kippen}, {McBreen},
  {Meegan}, {Paciesas}, {Poolakkil}, {Roberts}, {Stanbro}, {Gamma-ray Burst Monitor}, {Savchenko}, {Ferrigno}, {Kuulkers}, {Bazzano}, {Bozzo}, {Brandt}, {Chenevez}, {Courvoisier}, {Diehl}, {Domingo}, {Hanlon}, {Jourdain}, {Laurent}, {Lebrun}, {Lutovinov}, {Mereghetti}, {Natalucci}, {Rodi}, {Roques}, {Sunyaev}, {Ubertini}, and {(INTEGRAL}}]{abbott.2017ApJ...848L..13A}
{Abbott} BP, {Abbott} R, {Abbott} TD, et~al (2017) {Gravitational Waves and Gamma-Rays from a Binary Neutron Star Merger: GW170817 and GRB 170817A}. \apjl 848(2):L13. \doi{10.3847/2041-8213/aa920c}, {\href{https://arxiv.org/abs/1710.05834}{{arXiv:1710.05834}}} {[astro-ph.HE]}

\bibitem[{{Adamczak} and {Lambert}(2014)}]{Adamczak_2014}
{Adamczak} J, {Lambert} DL (2014) {Carbon and Oxygen Abundances across the Hertzsprung Gap}. \apj 791(1):58. \doi{10.1088/0004-637X/791/1/58}, {\href{https://arxiv.org/abs/1407.2157}{{arXiv:1407.2157}}} {[astro-ph.SR]}

\bibitem[{{Alonso} et~al(1999){Alonso}, {Arribas}, and {Mart{\'\i}nez-Roger}}]{Alonso_1999}
{Alonso} A, {Arribas} S, {Mart{\'\i}nez-Roger} C (1999) {The effective temperature scale of giant stars (F0-K5). II. Empirical calibration of T$_{eff}$ versus colours and [Fe/H]}. \aaps 140:261--277. \doi{10.1051/aas:1999521}

\bibitem[{{Aoki} et~al(2007){Aoki}, {Beers}, {Christlieb}, {Norris}, {Ryan}, and {Tsangarides}}]{Aoki_2007}
{Aoki} W, {Beers} TC, {Christlieb} N, et~al (2007) {Carbon-enhanced Metal-poor Stars. I. Chemical Compositions of 26 Stars}. \apj 655(1):492--521. \doi{10.1086/509817}, {\href{https://arxiv.org/abs/astro-ph/0609702}{{arXiv:astro-ph/0609702}}} {[astro-ph]}

\bibitem[{{Arcones} and {Thielemann}(2013)}]{arcones.2013JPhG...40a3201A}
{Arcones} A, {Thielemann} FK (2013) {Neutrino-driven wind simulations and nucleosynthesis of heavy elements}. Journal of Physics G Nuclear Physics 40(1):013201. \doi{10.1088/0954-3899/40/1/013201}, {\href{https://arxiv.org/abs/1207.2527}{{arXiv:1207.2527}}} {[astro-ph.SR]}

\bibitem[{{Asplund} et~al(2009){Asplund}, {Grevesse}, {Sauval}, and {Scott}}]{Asplund_2009}
{Asplund} M, {Grevesse} N, {Sauval} AJ, et~al (2009) {The Chemical Composition of the Sun}. \araa 47(1):481--522. \doi{10.1146/annurev.astro.46.060407.145222}, {\href{https://arxiv.org/abs/0909.0948}{{arXiv:0909.0948}}} {[astro-ph.SR]}

\bibitem[{{Barbuy} et~al(1997){Barbuy}, {Cayrel}, {Spite}, {Beers}, {Spite}, {Nordstroem}, and {Nissen}}]{Barbuy_1997}
{Barbuy} B, {Cayrel} R, {Spite} M, et~al (1997) {Analysis of two CH/CN-strong very metal-poor stars.} \aap 317:L63--L66

\bibitem[{{Barklem} et~al(2005){Barklem}, {Christlieb}, {Beers}, {Hill}, {Bessell}, {Holmberg}, {Marsteller}, {Rossi}, {Zickgraf}, and {Reimers}}]{Barklem_2005}
{Barklem} PS, {Christlieb} N, {Beers} TC, et~al (2005) {The Hamburg/ESO R-process enhanced star survey (HERES). II. Spectroscopic analysis of the survey sample}. \aap 439(1):129--151. \doi{10.1051/0004-6361:20052967}, {\href{https://arxiv.org/abs/astro-ph/0505050}{{arXiv:astro-ph/0505050}}} {[astro-ph]}

\bibitem[{{Barnbaum} et~al(1996){Barnbaum}, {Stone}, and {Keenan}}]{Barnbaum_1996}
{Barnbaum} C, {Stone} RPS, {Keenan} PC (1996) {A Moderate-Resolution Spectral Atlas of Carbon Stars: R, J, N, CH, and Barium Stars}. \apjs 105:419. \doi{10.1086/192323}

\bibitem[{{Battistini} and {Bensby}(2016)}]{2016A&A...586A..49B}
{Battistini} C, {Bensby} T (2016) {The origin and evolution of r- and s-process elements in the Milky Way stellar disk}. \aap 586:A49. \doi{10.1051/0004-6361/201527385}, {\href{https://arxiv.org/abs/1511.00966}{{arXiv:1511.00966}}} {[astro-ph.SR]}

\bibitem[{{Beers} and {Christlieb}(2005)}]{Beers_2005}
{Beers} TC, {Christlieb} N (2005) {The Discovery and Analysis of Very Metal-Poor Stars in the Galaxy}. \araa 43(1):531--580. \doi{10.1146/annurev.astro.42.053102.134057}

\bibitem[{{Beers} et~al(2014){Beers}, {Norris}, {Placco}, {Lee}, {Rossi}, {Carollo}, and {Masseron}}]{Beers_2014}
{Beers} TC, {Norris} JE, {Placco} VM, et~al (2014) {Population Studies. XIII. A New Analysis of the Bidelman-MacConnell ``Weak-metal'' Stars{\textemdash}Confirmation of Metal-poor Stars in the Thick Disk of the Galaxy}. \apj 794(1):58. \doi{10.1088/0004-637X/794/1/58}, {\href{https://arxiv.org/abs/1408.3165}{{arXiv:1408.3165}}} {[astro-ph.GA]}

\bibitem[{{Bidelman}(1956)}]{Bidelman_1956}
{Bidelman} WP (1956) {The carbon stars{\textemdash}An astrophysical enigma}. Vistas in Astronomy 2(1):1428--1437. \doi{10.1016/0083-6656(56)90071-X}

\bibitem[{{Boeche} and {Grebel}(2016)}]{Boeche_2016}
{Boeche} C, {Grebel} EK (2016) {SP\_Ace: a new code to derive stellar parameters and elemental abundances}. \aap 587:A2. \doi{10.1051/0004-6361/201526758}, {\href{https://arxiv.org/abs/1512.01546}{{arXiv:1512.01546}}} {[astro-ph.IM]}

\bibitem[{{Brewer} et~al(2016){Brewer}, {Fischer}, {Valenti}, and {Piskunov}}]{Brewer_2016}
{Brewer} JM, {Fischer} DA, {Valenti} JA, et~al (2016) {Spectral Properties of Cool Stars: Extended Abundance Analysis of 1,617 Planet-search Stars}. \apjs 225(2):32. \doi{10.3847/0067-0049/225/2/32}, {\href{https://arxiv.org/abs/1606.07929}{{arXiv:1606.07929}}} {[astro-ph.SR]}

\bibitem[{{Carrera} et~al(2007){Carrera}, {Gallart}, {Pancino}, and {Zinn}}]{Carrera_2007}
{Carrera} R, {Gallart} C, {Pancino} E, et~al (2007) {The Infrared Ca II Triplet as Metallicity Indicator}. \aj 134(3):1298. \doi{10.1086/520803}, {\href{https://arxiv.org/abs/0705.3335}{{arXiv:0705.3335}}} {[astro-ph]}

\bibitem[{{Carrera} et~al(2013){Carrera}, {Pancino}, {Gallart}, and {del Pino}}]{Carrera_2013}
{Carrera} R, {Pancino} E, {Gallart} C, et~al (2013) {The near-infrared Ca II triplet as a metallicity indicator - II. Extension to extremely metal-poor metallicity regimes}. \mnras 434(2):1681--1691. \doi{10.1093/mnras/stt1126}, {\href{https://arxiv.org/abs/1306.3883}{{arXiv:1306.3883}}} {[astro-ph.GA]}

\bibitem[{{Cenarro} et~al(2007){Cenarro}, {Peletier}, {S{\'a}nchez-Bl{\'a}zquez}, {Selam}, {Toloba}, {Cardiel}, {Falc{\'o}n-Barroso}, {Gorgas}, {Jim{\'e}nez-Vicente}, and {Vazdekis}}]{Cenarro_2007}
{Cenarro} AJ, {Peletier} RF, {S{\'a}nchez-Bl{\'a}zquez} P, et~al (2007) {Medium-resolution Isaac Newton Telescope library of empirical spectra - II. The stellar atmospheric parameters}. \mnras 374(2):664--690. \doi{10.1111/j.1365-2966.2006.11196.x}, {\href{https://arxiv.org/abs/astro-ph/0611618}{{arXiv:astro-ph/0611618}}} {[astro-ph]}

\bibitem[{{Christlieb} et~al(2001){Christlieb}, {Green}, {Wisotzki}, and {Reimers}}]{Christlieb_2001}
{Christlieb} N, {Green} PJ, {Wisotzki} L, et~al (2001) {The stellar content of the Hamburg/ESO survey II. A large, homogeneously-selected sample of high latitude carbon stars}. \aap 375:366--374. \doi{10.1051/0004-6361:20010814}, {\href{https://arxiv.org/abs/astro-ph/0106240}{{arXiv:astro-ph/0106240}}} {[astro-ph]}

\bibitem[{{da Silva} et~al(2015){da Silva}, {Milone}, and {Rocha-Pinto}}]{Silva_2015}
{da Silva} R, {Milone} AdC, {Rocha-Pinto} HJ (2015) {Homogeneous abundance analysis of FGK dwarf, subgiant, and giant stars with and without giant planets}. \aap 580:A24. \doi{10.1051/0004-6361/201525770}

\bibitem[{{Delgado Mena} et~al(2017){Delgado Mena}, {Tsantaki}, {Adibekyan}, {Sousa}, {Santos}, {Gonz{\'a}lez Hern{\'a}ndez}, and {Israelian}}]{Delgado_2017}
{Delgado Mena} E, {Tsantaki} M, {Adibekyan} VZ, et~al (2017) {Chemical abundances of 1111 FGK stars from the HARPS GTO planet search program. II. Cu, Zn, Sr, Y, Zr, Ba, Ce, Nd, and Eu}. \aap 606:A94. \doi{10.1051/0004-6361/201730535}, {\href{https://arxiv.org/abs/1705.04349}{{arXiv:1705.04349}}} {[astro-ph.SR]}

\bibitem[{{Drout} et~al(2017){Drout}, {Piro}, {Shappee}, {Kilpatrick}, {Simon}, {Contreras}, {Coulter}, {Foley}, {Siebert}, {Morrell}, {Boutsia}, {Di Mille}, {Holoien}, {Kasen}, {Kollmeier}, {Madore}, {Monson}, {Murguia-Berthier}, {Pan}, {Prochaska}, {Ramirez-Ruiz}, {Rest}, {Adams}, {Alatalo}, {Ba{\~n}ados}, {Baughman}, {Beers}, {Bernstein}, {Bitsakis}, {Campillay}, {Hansen}, {Higgs}, {Ji}, {Maravelias}, {Marshall}, {Moni Bidin}, {Prieto}, {Rasmussen}, {Rojas-Bravo}, {Strom}, {Ulloa}, {Vargas-Gonz{\'a}lez}, {Wan}, and {Whitten}}]{drout.2017Sci...358.1570D}
{Drout} MR, {Piro} AL, {Shappee} BJ, et~al (2017) {Light curves of the neutron star merger GW170817/SSS17a: Implications for r-process nucleosynthesis}. Science 358(6370):1570--1574. \doi{10.1126/science.aaq0049}, {\href{https://arxiv.org/abs/1710.05443}{{arXiv:1710.05443}}} {[astro-ph.HE]}

\bibitem[{{Frebel} et~al(2006){Frebel}, {Christlieb}, {Norris}, {Beers}, {Bessell}, {Rhee}, {Fechner}, {Marsteller}, {Rossi}, {Thom}, {Wisotzki}, and {Reimers}}]{Frebel_2006}
{Frebel} A, {Christlieb} N, {Norris} JE, et~al (2006) {Bright Metal-poor Stars from the Hamburg/ESO Survey. I. Selection and Follow-up Observations from 329 Fields}. \apj 652(2):1585--1603. \doi{10.1086/508506}, {\href{https://arxiv.org/abs/astro-ph/0608332}{{arXiv:astro-ph/0608332}}} {[astro-ph]}

\bibitem[{{Fulbright}(2000)}]{Fulbright_2000}
{Fulbright} JP (2000) {Abundances and Kinematics of Field Halo and Disk Stars. I. Observational Data and Abundance Analysis}. \aj 120(4):1841--1852. \doi{10.1086/301548}, {\href{https://arxiv.org/abs/astro-ph/0006260}{{arXiv:astro-ph/0006260}}} {[astro-ph]}

\bibitem[{{Gaia Collaboration} et~al(2016){Gaia Collaboration}, {Prusti}, {de Bruijne}, {Brown}, {Vallenari}, {Babusiaux}, {Bailer-Jones}, {Bastian}, {Biermann}, {Evans}, {Eyer}, {Jansen}, {Jordi}, {Klioner}, {Lammers}, {Lindegren}, {Luri}, {Mignard}, {Milligan}, {Panem}, {Poinsignon}, {Pourbaix}, {Randich}, {Sarri}, {Sartoretti}, {Siddiqui}, {Soubiran}, {Valette}, {van Leeuwen}, {Walton}, {Aerts}, {Arenou}, {Cropper}, {Drimmel}, {H{\o}g}, {Katz}, {Lattanzi}, {O'Mullane}, {Grebel}, {Holland}, {Huc}, {Passot}, {Bramante}, {Cacciari}, {Casta{\~n}eda}, {Chaoul}, {Cheek}, {De Angeli}, {Fabricius}, {Guerra}, {Hern{\'a}ndez}, {Jean-Antoine-Piccolo}, {Masana}, {Messineo}, {Mowlavi}, {Nienartowicz}, {Ord{\'o}{\~n}ez-Blanco}, {Panuzzo}, {Portell}, {Richards}, {Riello}, {Seabroke}, {Tanga}, {Th{\'e}venin}, {Torra}, {Els}, {Gracia-Abril}, {Comoretto}, {Garcia-Reinaldos}, {Lock}, {Mercier}, {Altmann}, {Andrae}, {Astraatmadja}, {Bellas-Velidis}, {Benson}, {Berthier}, {Blomme}, {Busso}, {Carry}, {Cellino}, {Clementini},
  {Cowell}, {Creevey}, {Cuypers}, {Davidson}, {De Ridder}, {de Torres}, {Delchambre}, {Dell'Oro}, {Ducourant}, {Fr{\'e}mat}, {Garc{\'\i}a-Torres}, {Gosset}, {Halbwachs}, {Hambly}, {Harrison}, {Hauser}, {Hestroffer}, {Hodgkin}, {Huckle}, {Hutton}, {Jasniewicz}, {Jordan}, {Kontizas}, {Korn}, {Lanzafame}, {Manteiga}, {Moitinho}, {Muinonen}, {Osinde}, {Pancino}, {Pauwels}, {Petit}, {Recio-Blanco}, {Robin}, {Sarro}, {Siopis}, {Smith}, {Smith}, {Sozzetti}, {Thuillot}, {van Reeven}, {Viala}, {Abbas}, {Abreu Aramburu}, {Accart}, {Aguado}, {Allan}, {Allasia}, {Altavilla}, {{\'A}lvarez}, {Alves}, {Anderson}, {Andrei}, {Anglada Varela}, {Antiche}, {Antoja}, {Ant{\'o}n}, {Arcay}, {Atzei}, {Ayache}, {Bach}, {Baker}, {Balaguer-N{\'u}{\~n}ez}, {Barache}, {Barata}, {Barbier}, {Barblan}, {Baroni}, {Barrado y Navascu{\'e}s}, {Barros}, {Barstow}, {Becciani}, {Bellazzini}, {Bellei}, {Bello Garc{\'\i}a}, {Belokurov}, {Bendjoya}, {Berihuete}, {Bianchi}, {Bienaym{\'e}}, {Billebaud}, {Blagorodnova}, {Blanco-Cuaresma}, {Boch},
  {Bombrun}, {Borrachero}, {Bouquillon}, {Bourda}, {Bouy}, {Bragaglia}, {Breddels}, {Brouillet}, {Br{\"u}semeister}, {Bucciarelli}, {Budnik}, {Burgess}, {Burgon}, {Burlacu}, {Busonero}, {Buzzi}, {Caffau}, {Cambras}, {Campbell}, {Cancelliere}, {Cantat-Gaudin}, {Carlucci}, {Carrasco}, {Castellani}, {Charlot}, {Charnas}, {Charvet}, {Chassat}, {Chiavassa}, {Clotet}, {Cocozza}, {Collins}, {Collins}, {Costigan}, {Crifo}, {Cross}, {Crosta}, {Crowley}, {Dafonte}, {Damerdji}, {Dapergolas}, {David}, {David}, {De Cat}, {de Felice}, {de Laverny}, {De Luise}, {De March}, {de Martino}, {de Souza}, {Debosscher}, {del Pozo}, {Delbo}, {Delgado}, {Delgado}, {di Marco}, {Di Matteo}, {Diakite}, {Distefano}, {Dolding}, {Dos Anjos}, {Drazinos}, {Dur{\'a}n}, {Dzigan}, {Ecale}, {Edvardsson}, {Enke}, {Erdmann}, {Escolar}, {Espina}, {Evans}, {Eynard Bontemps}, {Fabre}, {Fabrizio}, {Faigler}, {Falc{\~a}o}, {Farr{\`a}s Casas}, {Faye}, {Federici}, {Fedorets}, {Fern{\'a}ndez-Hern{\'a}ndez}, {Fernique}, {Fienga}, {Figueras}, {Filippi},
  {Findeisen}, {Fonti}, {Fouesneau}, {Fraile}, {Fraser}, {Fuchs}, {Furnell}, {Gai}, {Galleti}, {Galluccio}, {Garabato}, {Garc{\'\i}a-Sedano}, {Gar{\'e}}, {Garofalo}, {Garralda}, {Gavras}, {Gerssen}, {Geyer}, {Gilmore}, {Girona}, {Giuffrida}, {Gomes}, {Gonz{\'a}lez-Marcos}, {Gonz{\'a}lez-N{\'u}{\~n}ez}, {Gonz{\'a}lez-Vidal}, {Granvik}, {Guerrier}, {Guillout}, {Guiraud}, {G{\'u}rpide}, {Guti{\'e}rrez-S{\'a}nchez}, {Guy}, {Haigron}, {Hatzidimitriou}, {Haywood}, {Heiter}, {Helmi}, {Hobbs}, {Hofmann}, {Holl}, {Holland}, {Hunt}, {Hypki}, {Icardi}, {Irwin}, {Jevardat de Fombelle}, {Jofr{\'e}}, {Jonker}, {Jorissen}, {Julbe}, {Karampelas}, {Kochoska}, {Kohley}, {Kolenberg}, {Kontizas}, {Koposov}, {Kordopatis}, {Koubsky}, {Kowalczyk}, {Krone-Martins}, {Kudryashova}, {Kull}, {Bachchan}, {Lacoste-Seris}, {Lanza}, {Lavigne}, {Le Poncin-Lafitte}, {Lebreton}, {Lebzelter}, {Leccia}, {Leclerc}, {Lecoeur-Taibi}, {Lemaitre}, {Lenhardt}, {Leroux}, {Liao}, {Licata}, {Lindstr{\o}m}, {Lister}, {Livanou}, {Lobel}, {L{\"o}ffler},
  {L{\'o}pez}, {Lopez-Lozano}, {Lorenz}, {Loureiro}, {MacDonald}, {Magalh{\~a}es Fernandes}, {Managau}, {Mann}, {Mantelet}, {Marchal}, {Marchant}, {Marconi}, {Marie}, {Marinoni}, {Marrese}, {Marschalk{\'o}}, {Marshall}, {Mart{\'\i}n-Fleitas}, {Martino}, {Mary}, {Matijevi{\v{c}}}, {Mazeh}, {McMillan}, {Messina}, {Mestre}, {Michalik}, {Millar}, {Miranda}, {Molina}, {Molinaro}, {Molinaro}, {Moln{\'a}r}, {Moniez}, {Montegriffo}, {Monteiro}, {Mor}, {Mora}, {Morbidelli}, {Morel}, {Morgenthaler}, {Morley}, {Morris}, {Mulone}, {Muraveva}, {Musella}, {Narbonne}, {Nelemans}, {Nicastro}, {Noval}, {Ord{\'e}novic}, {Ordieres-Mer{\'e}}, {Osborne}, {Pagani}, {Pagano}, {Pailler}, {Palacin}, {Palaversa}, {Parsons}, {Paulsen}, {Pecoraro}, {Pedrosa}, {Pentik{\"a}inen}, {Pereira}, {Pichon}, {Piersimoni}, {Pineau}, {Plachy}, {Plum}, {Poujoulet}, {Pr{\v{s}}a}, {Pulone}, {Ragaini}, {Rago}, {Rambaux}, {Ramos-Lerate}, {Ranalli}, {Rauw}, {Read}, {Regibo}, {Renk}, {Reyl{\'e}}, {Ribeiro}, {Rimoldini}, {Ripepi}, {Riva}, {Rixon},
  {Roelens}, {Romero-G{\'o}mez}, {Rowell}, {Royer}, {Rudolph}, {Ruiz-Dern}, {Sadowski}, {Sagrist{\`a} Sell{\'e}s}, {Sahlmann}, {Salgado}, {Salguero}, {Sarasso}, {Savietto}, {Schnorhk}, {Schultheis}, {Sciacca}, {Segol}, {Segovia}, {Segransan}, {Serpell}, {Shih}, {Smareglia}, {Smart}, {Smith}, {Solano}, {Solitro}, {Sordo}, {Soria Nieto}, {Souchay}, {Spagna}, {Spoto}, {Stampa}, {Steele}, {Steidelm{\"u}ller}, {Stephenson}, {Stoev}, {Suess}, {S{\"u}veges}, {Surdej}, {Szabados}, {Szegedi-Elek}, {Tapiador}, {Taris}, {Tauran}, {Taylor}, {Teixeira}, {Terrett}, {Tingley}, {Trager}, {Turon}, {Ulla}, {Utrilla}, {Valentini}, {van Elteren}, {Van Hemelryck}, {van Leeuwen}, {Varadi}, {Vecchiato}, {Veljanoski}, {Via}, {Vicente}, {Vogt}, {Voss}, {Votruba}, {Voutsinas}, {Walmsley}, {Weiler}, {Weingrill}, {Werner}, {Wevers}, {Whitehead}, {Wyrzykowski}, {Yoldas}, {{\v{Z}}erjal}, {Zucker}, {Zurbach}, {Zwitter}, {Alecu}, {Allen}, {Allende Prieto}, {Amorim}, {Anglada-Escud{\'e}}, {Arsenijevic}, {Azaz}, {Balm}, {Beck}, {Bernstein},
  {Bigot}, {Bijaoui}, {Blasco}, {Bonfigli}, {Bono}, {Boudreault}, {Bressan}, {Brown}, {Brunet}, {Bunclark}, {Buonanno}, {Butkevich}, {Carret}, {Carrion}, {Chemin}, {Ch{\'e}reau}, {Corcione}, {Darmigny}, {de Boer}, {de Teodoro}, {de Zeeuw}, {Delle Luche}, {Domingues}, {Dubath}, {Fodor}, {Fr{\'e}zouls}, {Fries}, {Fustes}, {Fyfe}, {Gallardo}, {Gallegos}, {Gardiol}, {Gebran}, {Gomboc}, {G{\'o}mez}, {Grux}, {Gueguen}, {Heyrovsky}, {Hoar}, {Iannicola}, {Isasi Parache}, {Janotto}, {Joliet}, {Jonckheere}, {Keil}, {Kim}, {Klagyivik}, {Klar}, {Knude}, {Kochukhov}, {Kolka}, {Kos}, {Kutka}, {Lainey}, {LeBouquin}, {Liu}, {Loreggia}, {Makarov}, {Marseille}, {Martayan}, {Martinez-Rubi}, {Massart}, {Meynadier}, {Mignot}, {Munari}, {Nguyen}, {Nordlander}, {Ocvirk}, {O'Flaherty}, {Olias Sanz}, {Ortiz}, {Osorio}, {Oszkiewicz}, {Ouzounis}, {Palmer}, {Park}, {Pasquato}, {Peltzer}, {Peralta}, {P{\'e}turaud}, {Pieniluoma}, {Pigozzi}, {Poels}, {Prat}, {Prod'homme}, {Raison}, {Rebordao}, {Risquez}, {Rocca-Volmerange}, {Rosen},
  {Ruiz-Fuertes}, {Russo}, {Sembay}, {Serraller Vizcaino}, {Short}, {Siebert}, {Silva}, {Sinachopoulos}, {Slezak}, {Soffel}, {Sosnowska}, {Strai{\v{z}}ys}, {ter Linden}, {Terrell}, {Theil}, {Tiede}, {Troisi}, {Tsalmantza}, {Tur}, {Vaccari}, {Vachier}, {Valles}, {Van Hamme}, {Veltz}, {Virtanen}, {Wallut}, {Wichmann}, {Wilkinson}, {Ziaeepour}, and {Zschocke}}]{Gaia_2016}
{Gaia Collaboration}, {Prusti} T, {de Bruijne} JHJ, et~al (2016) {The Gaia mission}. \aap 595:A1. \doi{10.1051/0004-6361/201629272}, {\href{https://arxiv.org/abs/1609.04153}{{arXiv:1609.04153}}} {[astro-ph.IM]}

\bibitem[{{Gaia Collaboration} et~al(2018){Gaia Collaboration}, {Katz}, {Antoja}, {Romero-G{\'o}mez}, {Drimmel}, {Reyl{\'e}}, {Seabroke}, {Soubiran}, {Babusiaux}, {Di Matteo}, {Figueras}, {Poggio}, {Robin}, {Evans}, {Brown}, {Vallenari}, {Prusti}, {de Bruijne}, {Bailer-Jones}, {Biermann}, {Eyer}, {Jansen}, {Jordi}, {Klioner}, {Lammers}, {Lindegren}, {Luri}, {Mignard}, {Panem}, {Pourbaix}, {Randich}, {Sartoretti}, {Siddiqui}, {van Leeuwen}, {Walton}, {Arenou}, {Bastian}, {Cropper}, {Lattanzi}, {Bakker}, {Cacciari}, {Casta n}, {Chaoul}, {Cheek}, {De Angeli}, {Fabricius}, {Guerra}, {Holl}, {Masana}, {Messineo}, {Mowlavi}, {Nienartowicz}, {Panuzzo}, {Portell}, {Riello}, {Tanga}, {Th{\'e}venin}, {Gracia-Abril}, {Comoretto}, {Garcia-Reinaldos}, {Teyssier}, {Altmann}, {Andrae}, {Audard}, {Bellas-Velidis}, {Benson}, {Berthier}, {Blomme}, {Burgess}, {Busso}, {Carry}, {Cellino}, {Clementini}, {Clotet}, {Creevey}, {Davidson}, {De Ridder}, {Delchambre}, {Dell'Oro}, {Ducourant}, {Fern{\'a}ndez-Hern{\'a}ndez}, {Fouesneau},
  {Fr{\'e}mat}, {Galluccio}, {Garc{\'\i}a-Torres}, {Gonz{\'a}lez-N{\'u}{\~n}ez}, {Gonz{\'a}lez-Vidal}, {Gosset}, {Guy}, {Halbwachs}, {Hambly}, {Harrison}, {Hern{\'a}ndez}, {Hestroffer}, {Hodgkin}, {Hutton}, {Jasniewicz}, {Jean-Antoine-Piccolo}, {Jordan}, {Korn}, {Krone-Martins}, {Lanzafame}, {Lebzelter}, {L{\"o}ffler}, {Manteiga}, {Marrese}, {Mart{\'\i}n-Fleitas}, {Moitinho}, {Mora}, {Muinonen}, {Osinde}, {Pancino}, {Pauwels}, {Petit}, {Recio-Blanco}, {Richards}, {Rimoldini}, {Sarro}, {Siopis}, {Smith}, {Sozzetti}, {S{\"u}veges}, {Torra}, {van Reeven}, {Abbas}, {Abreu Aramburu}, {Accart}, {Aerts}, {Altavilla}, {{\'A}lvarez}, {Alvarez}, {Alves}, {Anderson}, {Andrei}, {Anglada Varela}, {Antiche}, {Arcay}, {Astraatmadja}, {Bach}, {Baker}, {Balaguer-N{\'u}{\~n}ez}, {Balm}, {Barache}, {Barata}, {Barbato}, {Barblan}, {Barklem}, {Barrado}, {Barros}, {Barstow}, {Bartholom{\'e} Mu{\~n}oz}, {Bassilana}, {Becciani}, {Bellazzini}, {Berihuete}, {Bertone}, {Bianchi}, {Bienaym{\'e}}, {Blanco-Cuaresma}, {Boch}, {Boeche},
  {Bombrun}, {Borrachero}, {Bossini}, {Bouquillon}, {Bourda}, {Bragaglia}, {Bramante}, {Breddels}, {Bressan}, {Brouillet}, {Br{\"u}semeister}, {Brugaletta}, {Bucciarelli}, {Burlacu}, {Busonero}, {Butkevich}, {Buzzi}, {Caffau}, {Cancelliere}, {Cannizzaro}, {Cantat-Gaudin}, {Carballo}, {Carlucci}, {Carrasco}, {Casamiquela}, {Castellani}, {Castro-Ginard}, {Charlot}, {Chemin}, {Chiavassa}, {Cocozza}, {Costigan}, {Cowell}, {Crifo}, {Crosta}, {Crowley}, {Cuypers}, {Dafonte}, {Damerdji}, {Dapergolas}, {David}, {David}, {de Laverny}, {De Luise}, {De March}, {de Souza}, {de Torres}, {Debosscher}, {del Pozo}, {Delbo}, {Delgado}, {Delgado}, {Diakite}, {Diener}, {Distefano}, {Dolding}, {Drazinos}, {Dur{\'a}n}, {Edvardsson}, {Enke}, {Eriksson}, {Esquej}, {Eynard Bontemps}, {Fabre}, {Fabrizio}, {Faigler}, {Falc a}, {Farr{\`a}s Casas}, {Federici}, {Fedorets}, {Fernique}, {Filippi}, {Findeisen}, {Fonti}, {Fraile}, {Fraser}, {Fr{\'e}zouls}, {Gai}, {Galleti}, {Garabato}, {Garc{\'\i}a-Sedano}, {Garofalo}, {Garralda}, {Gavel},
  {Gavras}, {Gerssen}, {Geyer}, {Giacobbe}, {Gilmore}, {Girona}, {Giuffrida}, {Glass}, {Gomes}, {Granvik}, {Gueguen}, {Guerrier}, {Guiraud}, {Guti{\'e}}, {Haigron}, {Hatzidimitriou}, {Hauser}, {Haywood}, {Heiter}, {Helmi}, {Heu}, {Hilger}, {Hobbs}, {Hofmann}, {Holland}, {Huckle}, {Hypki}, {Icardi}, {Jan{\ss}en}, {Jevardat de Fombelle}, {Jonker}, {Juh{\'a}sz}, {Julbe}, {Karampelas}, {Kewley}, {Klar}, {Kochoska}, {Kohley}, {Kolenberg}, {Kontizas}, {Kontizas}, {Koposov}, {Kordopatis}, {Kostrzewa-Rutkowska}, {Koubsky}, {Lambert}, {Lanza}, {Lasne}, {Lavigne}, {Le Fustec}, {Le Poncin-Lafitte}, {Lebreton}, {Leccia}, {Leclerc}, {Lecoeur-Taibi}, {Lenhardt}, {Leroux}, {Liao}, {Licata}, {Lindstr{\o}m}, {Lister}, {Livanou}, {Lobel}, {L{\'o}pez}, {Managau}, {Mann}, {Mantelet}, {Marchal}, {Marchant}, {Marconi}, {Marinoni}, {Marschalk{\'o}}, {Marshall}, {Martino}, {Marton}, {Mary}, {Massari}, {Matijevi{\v{c}}}, {Mazeh}, {McMillan}, {Messina}, {Michalik}, {Millar}, {Molina}, {Molinaro}, {Moln{\'a}r}, {Montegriffo}, {Mor},
  {Morbidelli}, {Morel}, {Morris}, {Mulone}, {Muraveva}, {Musella}, {Nelemans}, {Nicastro}, {Noval}, {O'Mullane}, {Ord{\'e}novic}, {Ord{\'o}{\~n}ez-Blanco}, {Osborne}, {Pagani}, {Pagano}, {Pailler}, {Palacin}, {Palaversa}, {Panahi}, {Pawlak}, {Piersimoni}, {Pineau}, {Plachy}, {Plum}, {Poujoulet}, {Pr{\v{s}}a}, {Pulone}, {Racero}, {Ragaini}, {Rambaux}, {Ramos-Lerate}, {Regibo}, {Riclet}, {Ripepi}, {Riva}, {Rivard}, {Rixon}, {Roegiers}, {Roelens}, {Rowell}, {Royer}, {Ruiz-Dern}, {Sadowski}, {Sagrist{\`a} Sell{\'e}s}, {Sahlmann}, {Salgado}, {Salguero}, {Sanna}, {Santana-Ros}, {Sarasso}, {Savietto}, {Schultheis}, {Sciacca}, {Segol}, {Segovia}, {S{\'e}gransan}, {Shih}, {Siltala}, {Silva}, {Smart}, {Smith}, {Solano}, {Solitro}, {Sordo}, {Soria Nieto}, {Souchay}, {Spagna}, {Spoto}, {Stampa}, {Steele}, {Steidelm{\"u}ller}, {Stephenson}, {Stoev}, {Suess}, {Surdej}, {Szabados}, {Szegedi-Elek}, {Tapiador}, {Taris}, {Tauran}, {Taylor}, {Teixeira}, {Terrett}, {Teyssandier}, {Thuillot}, {Titarenko}, {Torra Clotet},
  {Turon}, {Ulla}, {Utrilla}, {Uzzi}, {Vaillant}, {Valentini}, {Valette}, {van Elteren}, {Van Hemelryck}, {van Leeuwen}, {Vaschetto}, {Vecchiato}, {Veljanoski}, {Viala}, {Vicente}, {Vogt}, {von Essen}, {Voss}, {Votruba}, {Voutsinas}, {Walmsley}, {Weiler}, {Wertz}, {Wevers}, {Wyrzykowski}, {Yoldas}, {{\v{Z}}erjal}, {Ziaeepour}, {Zorec}, {Zschocke}, {Zucker}, {Zurbach}, and {Zwitter}}]{Gaia_2018}
{Gaia Collaboration}, {Katz} D, {Antoja} T, et~al (2018) {Gaia Data Release 2. Mapping the Milky Way disc kinematics}. \aap 616:A11. \doi{10.1051/0004-6361/201832865}, {\href{https://arxiv.org/abs/1804.09380}{{arXiv:1804.09380}}} {[astro-ph.GA]}

\bibitem[{{Gillon} and {Magain}(2006)}]{Gillon_2006}
{Gillon} M, {Magain} P (2006) {High precision determination of the atmospheric parameters and abundances of the COROT main targets}. \aap 448(1):341--350. \doi{10.1051/0004-6361:20053965}, {\href{https://arxiv.org/abs/astro-ph/0511099}{{arXiv:astro-ph/0511099}}} {[astro-ph]}

\bibitem[{{Girardi} et~al(2000){Girardi}, {Bressan}, {Bertelli}, and {Chiosi}}]{Girardi_2000}
{Girardi} L, {Bressan} A, {Bertelli} G, et~al (2000) {Evolutionary tracks and isochrones for low- and intermediate-mass stars: From 0.15 to 7 M$_{sun}$, and from Z=0.0004 to 0.03}. \aaps 141:371--383. \doi{10.1051/aas:2000126}, {\href{https://arxiv.org/abs/astro-ph/9910164}{{arXiv:astro-ph/9910164}}} {[astro-ph]}

\bibitem[{{Goswami}(2005)}]{Goswami_2005}
{Goswami} A (2005) {CH stars at high Galactic latitudes}. \mnras 359(2):531--544. \doi{10.1111/j.1365-2966.2005.08917.x}, {\href{https://arxiv.org/abs/astro-ph/0507202}{{arXiv:astro-ph/0507202}}} {[astro-ph]}

\bibitem[{{Goswami} and {Aoki}(2010)}]{Goswami_2010}
{Goswami} A, {Aoki} W (2010) {HD 209621: abundances of neutron-capture elements*}. \mnras 404(1):253--264. \doi{10.1111/j.1365-2966.2010.16265.x}, {\href{https://arxiv.org/abs/1002.4477}{{arXiv:1002.4477}}} {[astro-ph.SR]}

\bibitem[{{Goswami} and {Prantzos}(2000)}]{Goswami_2000}
{Goswami} A, {Prantzos} N (2000) {Abundance evolution of intermediate mass elements (C to Zn) in the Milky Way halo and disk}. \aap 359:191--212. \doi{10.48550/arXiv.astro-ph/0005179}, {\href{https://arxiv.org/abs/astro-ph/0005179}{{arXiv:astro-ph/0005179}}} {[astro-ph]}

\bibitem[{{Goswami} et~al(2006){Goswami}, {Aoki}, {Beers}, {Christlieb}, {Norris}, {Ryan}, and {Tsangarides}}]{Goswami_2006}
{Goswami} A, {Aoki} W, {Beers} TC, et~al (2006) {A high-resolution spectral analysis of three carbon-enhanced metal-poor stars}. \mnras 372(1):343--356. \doi{10.1111/j.1365-2966.2006.10877.x}, {\href{https://arxiv.org/abs/astro-ph/0608106}{{arXiv:astro-ph/0608106}}} {[astro-ph]}

\bibitem[{{Goswami} et~al(2007){Goswami}, {Bama}, {Shantikumar}, and {Devassy}}]{Goswami_2007}
{Goswami} A, {Bama} P, {Shantikumar} NS, et~al (2007) {Low-resolution spectroscopy of high Galactic latitude objects: A search for CH stars}. Bulletin of the Astronomical Society of India 35:339

\bibitem[{{Goswami} et~al(2010){Goswami}, {Karinkuzhi}, and {Shantikumar}}]{goswami_2010b}
{Goswami} A, {Karinkuzhi} D, {Shantikumar} NS (2010) {The CH fraction of carbon stars at high Galactic latitudes}. \mnras 402(2):1111--1125. \doi{10.1111/j.1365-2966.2009.15939.x}, {\href{https://arxiv.org/abs/0912.4347}{{arXiv:0912.4347}}} {[astro-ph.SR]}

\bibitem[{{Goswami} et~al(2016){Goswami}, {Aoki}, and {Karinkuzhi}}]{Goswami_2016}
{Goswami} A, {Aoki} W, {Karinkuzhi} D (2016) {Subaru/HDS study of CH stars: elemental abundances for stellar neutron-capture process studies}. \mnras 455(1):402--422. \doi{10.1093/mnras/stv2011}, {\href{https://arxiv.org/abs/1510.07814}{{arXiv:1510.07814}}} {[astro-ph.SR]}

\bibitem[{{Goswami} et~al(2021){Goswami}, {Rathour}, and {Goswami}}]{Partha_2021}
{Goswami} PP, {Rathour} RS, {Goswami} A (2021) {Spectroscopic study of CEMP-(s \& r/s) stars. Revisiting classification criteria and formation scenarios, highlighting i-process nucleosynthesis}. \aap 649:A49. \doi{10.1051/0004-6361/202038258}, {\href{https://arxiv.org/abs/2101.09518}{{arXiv:2101.09518}}} {[astro-ph.SR]}

\bibitem[{{Hansen} et~al(2016{\natexlab{a}}){Hansen}, {Nordstr{\"o}m}, {Hansen}, {Kennedy}, {Placco}, {Beers}, {Andersen}, {Cescutti}, and {Chiappini}}]{Hansen_2016c}
{Hansen} CJ, {Nordstr{\"o}m} B, {Hansen} TT, et~al (2016{\natexlab{a}}) {Abundances of carbon-enhanced metal-poor stars as constraints on their formation}. \aap 588:A37. \doi{10.1051/0004-6361/201526895}, {\href{https://arxiv.org/abs/1511.07812}{{arXiv:1511.07812}}} {[astro-ph.SR]}

\bibitem[{{Hansen} et~al(2011){Hansen}, {Andersen}, {Nordstr{\"o}m}, {Buchhave}, and {Beers}}]{hansen.2011ApJ...743L...1H}
{Hansen} T, {Andersen} J, {Nordstr{\"o}m} B, et~al (2011) {The Binary Frequency of r-Process-element-enhanced Metal-poor Stars and Its Implications: Chemical Tagging in the Primitive Halo of the Milky Way}. \apjl 743(1):L1. \doi{10.1088/2041-8205/743/1/L1}, {\href{https://arxiv.org/abs/1110.4536}{{arXiv:1110.4536}}} {[astro-ph.GA]}

\bibitem[{{Hansen} et~al(2016{\natexlab{b}}){Hansen}, {Andersen}, {Nordstr{\"o}m}, {Beers}, {Placco}, {Yoon}, and {Buchhave}}]{hansenetal.2016A&A...588A...3H}
{Hansen} TT, {Andersen} J, {Nordstr{\"o}m} B, et~al (2016{\natexlab{b}}) {The role of binaries in the enrichment of the early Galactic halo. III. Carbon-enhanced metal-poor stars - CEMP-s stars}. \aap 588:A3. \doi{10.1051/0004-6361/201527409}, {\href{https://arxiv.org/abs/1601.03385}{{arXiv:1601.03385}}} {[astro-ph.SR]}

\bibitem[{{Hekker} and {Mel{\'e}ndez}(2007)}]{Hekker_2007}
{Hekker} S, {Mel{\'e}ndez} J (2007) {Precise radial velocities of giant stars. III. Spectroscopic stellar parameters}. \aap 475(3):1003--1009. \doi{10.1051/0004-6361:20078233}, {\href{https://arxiv.org/abs/0709.1145}{{arXiv:0709.1145}}} {[astro-ph]}

\bibitem[{{Herwig}(2005)}]{herwig.2005ARA&A..43..435H}
{Herwig} F (2005) {Evolution of Asymptotic Giant Branch Stars}. \araa 43(1):435--479. \doi{10.1146/annurev.astro.43.072103.150600}

\bibitem[{{Jofr{\'e}} et~al(2015){Jofr{\'e}}, {Petrucci}, {Saffe}, {Saker}, {Artur de la Villarmois}, {Chavero}, {G{\'o}mez}, and {Mauas}}]{Jofre_2015}
{Jofr{\'e}} E, {Petrucci} R, {Saffe} C, et~al (2015) {Stellar parameters and chemical abundances of 223 evolved stars with and without planets}. \aap 574:A50. \doi{10.1051/0004-6361/201424474}, {\href{https://arxiv.org/abs/1410.6422}{{arXiv:1410.6422}}} {[astro-ph.EP]}

\bibitem[{{Jofr{\'e}} et~al(2014){Jofr{\'e}}, {Heiter}, {Soubiran}, {Blanco-Cuaresma}, {Worley}, {Pancino}, {Cantat-Gaudin}, {Magrini}, {Bergemann}, {Gonz{\'a}lez Hern{\'a}ndez}, {Hill}, {Lardo}, {de Laverny}, {Lind}, {Masseron}, {Montes}, {Mucciarelli}, {Nordlander}, {Recio Blanco}, {Sobeck}, {Sordo}, {Sousa}, {Tabernero}, {Vallenari}, and {Van Eck}}]{Jofre_2014}
{Jofr{\'e}} P, {Heiter} U, {Soubiran} C, et~al (2014) {Gaia FGK benchmark stars: Metallicity}. \aap 564:A133. \doi{10.1051/0004-6361/201322440}, {\href{https://arxiv.org/abs/1309.1099}{{arXiv:1309.1099}}} {[astro-ph.GA]}

\bibitem[{{Johnson} et~al(2007){Johnson}, {Herwig}, {Beers}, and {Christlieb}}]{Johnson_2007}
{Johnson} JA, {Herwig} F, {Beers} TC, et~al (2007) {A Search for Nitrogen-enhanced Metal-poor Stars}. \apj 658(2):1203--1216. \doi{10.1086/510114}, {\href{https://arxiv.org/abs/astro-ph/0608666}{{arXiv:astro-ph/0608666}}} {[astro-ph]}

\bibitem[{{Jorissen} et~al(2016{\natexlab{a}}){Jorissen}, {Hansen}, {Van Eck}, {Andersen}, {Nordstr{\"o}m}, {Siess}, {Torres}, {Masseron}, and {Van Winckel}}]{jorrisen.2016A&A...586A.159J}
{Jorissen} A, {Hansen} T, {Van Eck} S, et~al (2016{\natexlab{a}}) {HE 0017+0055: A probable pulsating CEMP-rs star and long-period binary}. \aap 586:A159. \doi{10.1051/0004-6361/201526993}, {\href{https://arxiv.org/abs/1510.06045}{{arXiv:1510.06045}}} {[astro-ph.SR]}

\bibitem[{{Jorissen} et~al(2016{\natexlab{b}}){Jorissen}, {Hansen}, {Van Eck}, {Andersen}, {Nordstr{\"o}m}, {Siess}, {Torres}, {Masseron}, and {Van Winckel}}]{Jorrisen_2016}
{Jorissen} A, {Hansen} T, {Van Eck} S, et~al (2016{\natexlab{b}}) {HE 0017+0055: A probable pulsating CEMP-rs star and long-period binary}. \aap 586:A159. \doi{10.1051/0004-6361/201526993}, {\href{https://arxiv.org/abs/1510.06045}{{arXiv:1510.06045}}} {[astro-ph.SR]}

\bibitem[{{Karakas} and {Lattanzio}(2014)}]{karakas.2014PASA...31...30K}
{Karakas} AI, {Lattanzio} JC (2014) {The Dawes Review 2: Nucleosynthesis and Stellar Yields of Low- and Intermediate-Mass Single Stars}. \pasa 31:e030. \doi{10.1017/pasa.2014.21}, {\href{https://arxiv.org/abs/1405.0062}{{arXiv:1405.0062}}} {[astro-ph.SR]}

\bibitem[{{Karinkuzhi} and {Goswami}(2014)}]{Karinkuzhi_2014}
{Karinkuzhi} D, {Goswami} A (2014) {Chemical analysis of CH stars - I. Atmospheric parameters and elemental abundances}. \mnras 440(2):1095--1113. \doi{10.1093/mnras/stu148}, {\href{https://arxiv.org/abs/1410.0111}{{arXiv:1410.0111}}} {[astro-ph.SR]}

\bibitem[{{Karinkuzhi} and {Goswami}(2015)}]{Karinkuzhi_2015}
{Karinkuzhi} D, {Goswami} A (2015) {Chemical analysis of CH stars - II. Atmospheric parameters and elemental abundances}. \mnras 446(3):2348--2362. \doi{10.1093/mnras/stu2079}, {\href{https://arxiv.org/abs/1412.3548}{{arXiv:1412.3548}}} {[astro-ph.SR]}

\bibitem[{{Kennedy} et~al(2011){Kennedy}, {Sivarani}, {Beers}, {Lee}, {Placco}, {Rossi}, {Christlieb}, {Herwig}, and {Plez}}]{Kennedy_2011}
{Kennedy} CR, {Sivarani} T, {Beers} TC, et~al (2011) {[O/Fe] Estimates for Carbon-enhanced Metal-poor Stars from Near-infrared Spectroscopy}. \aj 141(3):102. \doi{10.1088/0004-6256/141/3/102}, {\href{https://arxiv.org/abs/1101.2260}{{arXiv:1101.2260}}} {[astro-ph.SR]}

\bibitem[{{Koleva} and {Vazdekis}(2012)}]{Koleva_2012}
{Koleva} M, {Vazdekis} A (2012) {Stellar population models in the UV. I. Characterisation of the New Generation Stellar Library}. \aap 538:A143. \doi{10.1051/0004-6361/201118065}, {\href{https://arxiv.org/abs/1111.5449}{{arXiv:1111.5449}}} {[astro-ph.CO]}

\bibitem[{{Lambert} et~al(1986){Lambert}, {Gustafsson}, {Eriksson}, and {Hinkle}}]{Lambert_1986}
{Lambert} DL, {Gustafsson} B, {Eriksson} K, et~al (1986) {The Chemical Composition of Carbon Stars. I. Carbon, Nitrogen, and Oxygen in 30 Cool Carbon Stars in the Galactic Disk}. \apjs 62:373. \doi{10.1086/191145}

\bibitem[{{Liu} et~al(2014){Liu}, {Tan}, {Wang}, {Zhao}, {Sato}, {Takeda}, and {Li}}]{Liu_2014}
{Liu} YJ, {Tan} KF, {Wang} L, et~al (2014) {The Lithium Abundances of a Large Sample of Red Giants}. \apj 785(2):94. \doi{10.1088/0004-637X/785/2/94}, {\href{https://arxiv.org/abs/1404.1687}{{arXiv:1404.1687}}} {[astro-ph.SR]}

\bibitem[{{Lucatello} et~al(2005){Lucatello}, {Tsangarides}, {Beers}, {Carretta}, {Gratton}, and {Ryan}}]{Lucatello_2005}
{Lucatello} S, {Tsangarides} S, {Beers} TC, et~al (2005) {The Binary Frequency Among Carbon-enhanced, s-Process-rich, Metal-poor Stars}. \apj 625(2):825--832. \doi{10.1086/428104}, {\href{https://arxiv.org/abs/astro-ph/0412422}{{arXiv:astro-ph/0412422}}} {[astro-ph]}

\bibitem[{{Lugaro} et~al(2003){Lugaro}, {Herwig}, {Lattanzio}, {Gallino}, and {Straniero}}]{lugaro.2003ApJ...586.1305L}
{Lugaro} M, {Herwig} F, {Lattanzio} JC, et~al (2003) {s-Process Nucleosynthesis in Asymptotic Giant Branch Stars: A Test for Stellar Evolution}. \apj 586(2):1305--1319. \doi{10.1086/367887}, {\href{https://arxiv.org/abs/astro-ph/0212364}{{arXiv:astro-ph/0212364}}} {[astro-ph]}

\bibitem[{{Maldonado} et~al(2012){Maldonado}, {Eiroa}, {Villaver}, {Montesinos}, and {Mora}}]{Maldonado_2012}
{Maldonado} J, {Eiroa} C, {Villaver} E, et~al (2012) {Metallicity of solar-type stars with debris discs and planets{\ensuremath{\star}}}. \aap 541:A40. \doi{10.1051/0004-6361/201218800}, {\href{https://arxiv.org/abs/1202.5884}{{arXiv:1202.5884}}} {[astro-ph.EP]}

\bibitem[{{Massarotti} et~al(2008){Massarotti}, {Latham}, {Stefanik}, and {Fogel}}]{Massarotti_2008}
{Massarotti} A, {Latham} DW, {Stefanik} RP, et~al (2008) {Rotational and Radial Velocities for a Sample of 761 HIPPARCOS Giants and the Role of Binarity}. \aj 135(1):209--231. \doi{10.1088/0004-6256/135/1/209}

\bibitem[{{McClure} and {Woodsworth}(1990)}]{McClure&Woodsworth1990ApJ...352..709M}
{McClure} RD, {Woodsworth} AW (1990) {The Binary Nature of the Barium and CH Stars. III. Orbital Parameters}. \apj 352:709. \doi{10.1086/168573}

\bibitem[{{Mishenina} et~al(2012){Mishenina}, {Soubiran}, {Kovtyukh}, {Katsova}, and {Livshits}}]{Mishenina_2012}
{Mishenina} TV, {Soubiran} C, {Kovtyukh} VV, et~al (2012) {Activity and the Li abundances in the FGK dwarfs{\ensuremath{\star}}}. \aap 547:A106. \doi{10.1051/0004-6361/201118412}, {\href{https://arxiv.org/abs/1210.6843}{{arXiv:1210.6843}}} {[astro-ph.SR]}

\bibitem[{{Montes} et~al(2018){Montes}, {Gonz{\'a}lez-Peinado}, {Tabernero}, {Caballero}, {Marfil}, {Alonso-Floriano}, {Cort{\'e}s-Contreras}, {Gonz{\'a}lez Hern{\'a}ndez}, {Klutsch}, and {Moreno-J{\'o}dar}}]{Montes_2018}
{Montes} D, {Gonz{\'a}lez-Peinado} R, {Tabernero} HM, et~al (2018) {Calibrating the metallicity of M dwarfs in wide physical binaries with F-, G-, and K-primaries - I: High-resolution spectroscopy with HERMES: stellar parameters, abundances, and kinematics}. \mnras 479(1):1332--1382. \doi{10.1093/mnras/sty1295}, {\href{https://arxiv.org/abs/1805.05394}{{arXiv:1805.05394}}} {[astro-ph.SR]}

\bibitem[{{Pereira} and {Junqueira}(2003)}]{Perea.2003}
{Pereira} CB, {Junqueira} S (2003) {Spectroscopic analysis of two CH subgiant stars: HD 50264 and HD 87080}. \aap 402:1061--1071. \doi{10.1051/0004-6361:20030209}

\bibitem[{{Placco} et~al(2015){Placco}, {Beers}, {Ivans}, {Filler}, {Imig}, {Roederer}, {Abate}, {Hansen}, {Cowan}, {Frebel}, {Lawler}, {Schatz}, {Sneden}, {Sobeck}, {Aoki}, {Smith}, and {Bolte}}]{Placco_2015}
{Placco} VM, {Beers} TC, {Ivans} II, et~al (2015) {Hubble Space Telescope Near-Ultraviolet Spectroscopy of Bright CEMP-s Stars}. \apj 812(2):109. \doi{10.1088/0004-637X/812/2/109}, {\href{https://arxiv.org/abs/1508.05872}{{arXiv:1508.05872}}} {[astro-ph.SR]}

\bibitem[{{Preston} and {Sneden}(2001)}]{preston&sneden2001AJ....122.1545P}
{Preston} GW, {Sneden} C (2001) {The Incidence of Binaries among Very Metal-poor Carbon Stars}. \aj 122(3):1545--1560. \doi{10.1086/322082}

\bibitem[{{Prugniel} et~al(2011){Prugniel}, {Vauglin}, and {Koleva}}]{Prugniel_2011}
{Prugniel} P, {Vauglin} I, {Koleva} M (2011) {The atmospheric parameters and spectral interpolator for the MILES stars}. \aap 531:A165. \doi{10.1051/0004-6361/201116769}, {\href{https://arxiv.org/abs/1104.4952}{{arXiv:1104.4952}}} {[astro-ph.CO]}

\bibitem[{{Purandardas} and {Goswami}(2021{\natexlab{a}})}]{Purandardas_2021a}
{Purandardas} M, {Goswami} A (2021{\natexlab{a}}) {Chemical Analysis of Two Extremely Metal-poor Stars HE 2148-2039 and HE 2155-2043}. \apj 912(1):74. \doi{10.3847/1538-4357/abec45}, {\href{https://arxiv.org/abs/2103.07075}{{arXiv:2103.07075}}} {[astro-ph.SR]}

\bibitem[{{Purandardas} and {Goswami}(2021{\natexlab{b}})}]{Purandardas_2021b}
{Purandardas} M, {Goswami} A (2021{\natexlab{b}}) {Observational evidence points at AGB stars as possible progenitors of CEMP-s \& r/s stars}. arXiv e-prints arXiv:2108.06075. {\href{https://arxiv.org/abs/2108.06075}{{arXiv:2108.06075}}} {[astro-ph.SR]}

\bibitem[{{Purandardas} et~al(2019){Purandardas}, {Goswami}, {Goswami}, {Shejeelammal}, and {Masseron}}]{Purandardas_2019a}
{Purandardas} M, {Goswami} A, {Goswami} PP, et~al (2019) {Chemical analysis of CH stars - III. Atmospheric parameters and elemental abundances}. \mnras 486(3):3266--3289. \doi{10.1093/mnras/stz759}, {\href{https://arxiv.org/abs/1904.03904}{{arXiv:1904.03904}}} {[astro-ph.SR]}

\bibitem[{{Ram{\'\i}rez} et~al(2012){Ram{\'\i}rez}, {Fish}, {Lambert}, and {Allende Prieto}}]{Ramirez_2012}
{Ram{\'\i}rez} I, {Fish} JR, {Lambert} DL, et~al (2012) {Lithium Abundances in nearby FGK Dwarf and Subgiant Stars: Internal Destruction, Galactic Chemical Evolution, and Exoplanets}. \apj 756(1):46. \doi{10.1088/0004-637X/756/1/46}, {\href{https://arxiv.org/abs/1207.0499}{{arXiv:1207.0499}}} {[astro-ph.SR]}

\bibitem[{{Ram{\'\i}rez} et~al(2013){Ram{\'\i}rez}, {Allende Prieto}, and {Lambert}}]{Ramirez_2013}
{Ram{\'\i}rez} I, {Allende Prieto} C, {Lambert} DL (2013) {Oxygen Abundances in Nearby FGK Stars and the Galactic Chemical Evolution of the Local Disk and Halo}. \apj 764(1):78. \doi{10.1088/0004-637X/764/1/78}, {\href{https://arxiv.org/abs/1301.1582}{{arXiv:1301.1582}}} {[astro-ph.SR]}

\bibitem[{{Roriz} et~al(2023){Roriz}, {Pereira}, {Junqueira}, {Lugaro}, {Drake}, and {Sneden}}]{roriz.2023}
{Roriz} MP, {Pereira} CB, {Junqueira} S, et~al (2023) {High-resolution spectroscopic analysis of four new chemically peculiar stars}. \mnras 518(4):5414--5443. \doi{10.1093/mnras/stac3378}, {\href{https://arxiv.org/abs/2211.08627}{{arXiv:2211.08627}}} {[astro-ph.SR]}

\bibitem[{{Rosswog} et~al(2014){Rosswog}, {Korobkin}, {Arcones}, {Thielemann}, and {Piran}}]{rosswog.2014MNRAS.439..744R}
{Rosswog} S, {Korobkin} O, {Arcones} A, et~al (2014) {The long-term evolution of neutron star merger remnants - I. The impact of r-process nucleosynthesis}. \mnras 439(1):744--756. \doi{10.1093/mnras/stt2502}, {\href{https://arxiv.org/abs/1307.2939}{{arXiv:1307.2939}}} {[astro-ph.HE]}

\bibitem[{{Shappee} et~al(2017){Shappee}, {Simon}, {Drout}, {Piro}, {Morrell}, {Prieto}, {Kasen}, {Holoien}, {Kollmeier}, {Kelson}, {Coulter}, {Foley}, {Kilpatrick}, {Siebert}, {Madore}, {Murguia-Berthier}, {Pan}, {Prochaska}, {Ramirez-Ruiz}, {Rest}, {Adams}, {Alatalo}, {Ba{\~n}ados}, {Baughman}, {Bernstein}, {Bitsakis}, {Boutsia}, {Bravo}, {Di Mille}, {Higgs}, {Ji}, {Maravelias}, {Marshall}, {Placco}, {Prieto}, and {Wan}}]{shappee.2017Sci...358.1574S}
{Shappee} BJ, {Simon} JD, {Drout} MR, et~al (2017) {Early spectra of the gravitational wave source GW170817: Evolution of a neutron star merger}. Science 358(6370):1574--1578. \doi{10.1126/science.aaq0186}, {\href{https://arxiv.org/abs/1710.05432}{{arXiv:1710.05432}}} {[astro-ph.HE]}

\bibitem[{{Smith} et~al(1993){Smith}, {Coleman}, and {Lambert}}]{1993ApJ...417..287S}
{Smith} VV, {Coleman} H, {Lambert} DL (1993) {Abundances in CH Subgiants: Evidence of Mass Transfer onto Main-Sequence Companions}. \apj 417:287. \doi{10.1086/173311}

\bibitem[{{Sneden}(1973)}]{Sneden_1973}
{Sneden} CA (1973) {Carbon and Nitrogen Abundances in Metal-Poor Stars.} PhD thesis, THE UNIVERSITY OF TEXAS AT AUSTIN.

\bibitem[{{Soubiran} et~al(2016){Soubiran}, {Le Campion}, {Brouillet}, and {Chemin}}]{Soubiran_2016}
{Soubiran} C, {Le Campion} JF, {Brouillet} N, et~al (2016) {The PASTEL catalogue: 2016 version}. \aap 591:A118. \doi{10.1051/0004-6361/201628497}, {\href{https://arxiv.org/abs/1605.07384}{{arXiv:1605.07384}}} {[astro-ph.SR]}

\bibitem[{{Yang} et~al(2016){Yang}, {Liang}, {Spite}, {Chen}, {Zhao}, {Zhang}, {Liu}, {Liu}, {Liu}, {Deng}, {Spite}, {Hill}, and {Zhang}}]{Yang_2016}
{Yang} GC, {Liang} YC, {Spite} M, et~al (2016) {Chemical abundance analysis of 19 barium stars}. Research in Astronomy and Astrophysics 16(1):19. \doi{10.1088/1674-4527/16/1/019}, {\href{https://arxiv.org/abs/1602.08704}{{arXiv:1602.08704}}} {[astro-ph.SR]}

\bibitem[{{Yong} et~al(2013){Yong}, {Norris}, {Bessell}, {Christlieb}, {Asplund}, {Beers}, {Barklem}, {Frebel}, and {Ryan}}]{Yong_2013}
{Yong} D, {Norris} JE, {Bessell} MS, et~al (2013) {The Most Metal-poor Stars. II. Chemical Abundances of 190 Metal-poor Stars Including 10 New Stars with [Fe/H] <= -3.5}. \apj 762(1):26. \doi{10.1088/0004-637X/762/1/26}, {\href{https://arxiv.org/abs/1208.3003}{{arXiv:1208.3003}}} {[astro-ph.GA]}

\bibitem[{{Yoon} et~al(2016){Yoon}, {Beers}, {Placco}, {Rasmussen}, {Carollo}, {He}, {Hansen}, {Roederer}, and {Zeanah}}]{Yoon_2016}
{Yoon} J, {Beers} TC, {Placco} VM, et~al (2016) {Observational Constraints on First-star Nucleosynthesis. I. Evidence for Multiple Progenitors of CEMP-No Stars}. \apj 833(1):20. \doi{10.3847/0004-637X/833/1/20}, {\href{https://arxiv.org/abs/1607.06336}{{arXiv:1607.06336}}} {[astro-ph.SR]}

\bibitem[{{Zamora} et~al(2009){Zamora}, {Abia}, {Plez}, {Dom{\'\i}nguez}, and {Cristallo}}]{Zamora_2009}
{Zamora} O, {Abia} C, {Plez} B, et~al (2009) {The chemical composition of carbon stars. The R-type stars}. \aap 508(2):909--922. \doi{10.1051/0004-6361/200912843}, {\href{https://arxiv.org/abs/0909.4222}{{arXiv:0909.4222}}} {[astro-ph.SR]}

\end{thebibliography}

\begin{appendices}

\section{}\label{secA1}

{\footnotesize
\begin{table*}
\caption{Temperatures from photometry}\label{tabA1}
\resizebox{1.1\textwidth}{!}{\begin{tabular}{lcccccccccc}
\hline
star name	&	T$_{eff}$	& T$_{eff}$ &	T$_{eff}$	&	T$_{eff}$	&	T$_{eff}$	&	T$_{eff}$ & T$_{eff}$ &T$_{eff}$ &	T$_{eff}$	&	T$_{eff}$	\\
		      &	          &	(-0.5)	  & (-1.0)    &	(-1.5)	  &	(-2.0)    &	(-0.5)	  & (-1.0)    &	(-1.5)	 &	(-2.0)    &	(H$_{\alpha}$)	\\		
	        &	(J-K)	    &	(J-H)	    &	(J-H)     & (J-H)	    &	(J-H)	    &	(V-K)	    &	(v-k)     & (V-K)	   &	(V-K)	    &	\\
	\hline
HE 0002+0053	&	4070.13	&	4392.57	&	4409.57	&	4407.08	&	4385.19	&	4095.07	&	4081.52	&	4072.94	&	4069.28	&	4250	\\
HE 0017+0055	&	4261.44	&	4462.02	&	4479.96	&	4477.8	&	4455.61	&	4105.98	&	4092.46	&	4083.94	&	4080.36	&	4500	\\
HE 0037-0654    &   5458.60 & 5146.13  &  5174.46 & 5176.03 & 5150.80 & 5328.81 & 5326.75 & 5332.36 & 5345.71 & - \\
HE 0039-2635    &   4911.71 &  5064.04  &  5091.02 &  5092.09  &  5067.23  &  4820.78  & 4808.88  &  4803.26 & 5100 \\
HE 0429+0232	&	4375.27	&	4612.36	&	4632.42	&	4630.99	&	4608.13	&	4215.14	&	4201.89	&	4193.9	&	4191.12	&	4600	\\
HE 0443-1847	&	4559.4	&	4735.79	&	4757.66	&	4756.88	&	4733.47	&	4270.62	&	4257.49	&	4249.78	&	4247.39	&	4300	\\
HE 0503-2009	&	4678.27	&	4774.69	&	4797.14	&	4796.57	&	4772.99	&	4476.99	&	4464.29	&	4457.53	&	4456.63	&	4450	\\
HE 0507-1653	&	4935.03	&	5077.52	&	5104.72	&	5105.86	&	5080.95	&	4854.93	&	4843.61	&	4838.67	&	4840.07	&	5200	\\
HE 0039-2635	&	4911.71	&	5064.044&	5091.02	&	5092.09	&	5067.23	&	4820.78	&	4808.88	&	4803.26	&	4803.89	&	5100	\\
HE 0206-1916    & 4885.40   & 4967.59   & 4993.02   & 4993.53   & 4969.09   & 4706.30   & 4692.49   & 4684.68   & 4682.83   & 4650 \\
HE 0113+0110	&	4307.89	&	4279.09	&	4294.58	&	4291.59	&	4270.19	&	4356.06	&	4343.12	&	4335.79	&	4334.03	&	-	\\
HE 0155-2221	&	4266.03	&	4325.55	&	4341.65	&	4338.86	&	4317.26	&	3993.42	&	3979.57	&	-	&	-	&	4500	\\
HE 0228-0256	&	3926.25	&	4204.16	&	4218.68	&	4215.36	&	4194.3	&	4102.14	&	4088.61	&	4080.07	&	4076.47	&	4200	\\
HE 0237-0835	&	4487.1	&	4605.11	&	4625.07	&	4623.61	&	4600.78	&	4468.31	&	4455.6	&	4448.79	&	4447.83	&	-	\\
HE 0251-2118	&	4655.01	&	4778.62	&	4801.13	&	4800.58	&	4776.98	&	4579.4	&	4566.91	&	4560.6	&	4560.44	&	4400	\\
HE 0258-0218	&	4586.99	&	4509.73	&	4528.34	&	4526.41	&	4503.99	&	5058.64	&	5051.06	&	5050.39	&	5056.63	&	4800	\\
HE 0319-0215    &   -       &   4615.99  &  4636.10  &  4634.70  &  4611.82  &  4529.61  &  4517.02 &  4510.49  &  4509.97 & 4700\\
HE 0322-1504	&	4370.36	&	4597.89	&	4617.74	&	4616.24	&	4593.45	&	4228.24	&	4215.03	&	4207.1	&	4204.42	&	4600	\\
HE 0323-2702	&	4701.82	&	4705.19	&	4726.6	&	4701.82	&	4702.39	&	4917.14	&	4906.93	&	4903.25	&	4906.08	&	4500	\\
HE 0326-2603    &   -       &  5183.58  &  5212.52  &  5214.33  &  5188.94  &  5148.05  &  5142.22  &  5143.56  &  5152.06 & - \\
HE 0333-1819    &  4387.60  &   4572.81 &  4592.30  &   4590.68 & 4567.99  & 4293.84 & 4280.77 & 4273.16 & 4270.95 & - \\
HE 0341-0314	&	4298.49	&	4576.37	&	4595.92	&	4594.32	&	4571.62	&	4230.76	&	4217.55	&	4209.64	&	4206.97	&	-	\\
HE 0359-0141	&	4243.24	&	4402.35	&	4419.47	&	4417.03	&	4395.09	&	4112.59	&	4099.09	&	4090.6	&	4087.07	&	4450	\\
HE 0417-0513	&	4326.86	&	4350.81	&	4367.24	&	4364.56	&	4342.86	&	4026.51	&	4012.76	&	4003.83	&	3999.65	&	4100	\\
HE 0422-2518    &   4529.54 &   4810.32 &   4833.31 &   4832.93 &   4809.19 &     -     &    -      &     -     &     -     &    -    \\
HE 0443-1847    &   4559.40 & 4735.79   &   4757.66 &  4756.88  &  4733.47  &  4270.62  & 4257.49   & 4249.78   & 4247.40   & 4300 \\
HE 0503-2009    &   4678.27 &  4774.69  &   4797.14 &   4796.57 &   4772.99 &  4476.99  &  4464.29  &  4457.53  & 4456.63   &  4450\\
HE 0507-1430	&	4407.51	&	4472.15	&	4490.23	&	4488.12	&	4465.87	&	4312.81	&	4299.78	&	4292.25	&	4290.18	&	4600	\\
HE 0507-1653    &   4935.03 &   5077.52  &  5104.72 &   5105.87 &   5080.95 &  4854.93  &   4843.61 &  4838.67  &   4840.07 & 5200 \\
HE 0508-1604	&	4432.75	&	4605.11	&	4625.06	&	4623.61	&	4600.78	&	4569.11	&	4556.59	&	4550.25	&	4550.01	&	4650	\\
HE 0516-2515	&	4062.07	&	4300.63	&	4316.39	&	4313.5	&	4296.01	&	3917.3	&	3903.19	&	-	&	-	&	-	\\
HE 0518-1751	&	4589.77	&	4697.61	&	4718.91	&	4717.92	&	4694.69	&	4312.81	&	4299.78	&	4292.25	&	4290.18	&	4350	\\
HE 0518-2322	&	4885.4	&	4963.31	&	4988.67	&	4989.15	&	4964.73	&	4825.93	&	4814.11	&	4808.59	&	4809.34	&	5100	\\
HE 0519-2053	&	4623.51	&	4619.63	&	4639.79	&	4638.41	&	4615.52	&	4643.32	&	4628.52	&	4619.56	&	4616.39	&	4400	\\
HE 0919+0200	&	4487.01	&	4619.63	&	4639.79	&	4638.41	&	4615.52	&	4439.58	&	4426.81	&	4419.87	&	4418.7	&	4250	\\
HE 0926-0417	&	4731.7	&	4778.62	&	4801.13	&	4800.58	&	4776.98	&	4743.88	&	4730.68	&	4723.58	&	4722.53	&	4500	\\
HE 0930-0945	&	4672.42	&	4682.53	&	4703.6	&	4702.54	&	4679.37	&	4654.83	&	4640.2	&	4631.45	&	4628.52	&	-	\\
HE 0930-0018	&	4163.85	&	4353.99	&	4370.46	&	4367.8	&	4346.08	&	4001.55	&	3987.73	&	3978.67	&	3974.3	&	4400	\\
HE 1008-0946	&	4473.93	&	4523.58	&	4542.38	&	4540.51	&	4518.04	&	4033	&	4019.27	&	4010.38	&	4006.25	&	-	\\
HE 1023-1504  & 4062.07 & 4143.52 & 4157.27 & 4153.69 & 4132.90 & 4211.43 & 4198.17 & 4190.17 & 4187.36 & - \\ 
HE 1032-1655	&	4484.39	&	4506.29	&	4524.84	&	4522.89	&	4500.5	&	-	&	-	&	-	&	-	&	-	\\
HE 1045-1434    & 4564.8    & 4933.53  &  4958.42  & 4958.73 & 4934.44 & 4689.67 & 4675.59 & 4667.48 & 4665.28 & 4800 \\
HE 1058-1300	&	4448.07	&	4597.89	&	4617.74	&	4616.24	&	4593.45	&	4399.63	&	4386.78	&	4379.66	&	4378.2	&	-	\\
HE 1104-0957    &  -        &  4180.82  &  4195.04  &  4191.63  &  4170.67  &  3964.96  &  3951.02  &  -  & -  & - \\
HE 1112-2557    &  4540.34  &   4708.99 &  4730.46  &  4729.54  &  4706.25  &  4364.64  &  4351.72  &  4344.43 &  4342.73 & - \\
HE 1130-1956	&	4655.01	&	4774.69	&	4797.14	&	4796.57	&	4772.99	&	4436.51	&	4423.74	&	4416.79	&	4415.59	&	-	\\
HE 1150-2049	&	4589.77	&	4637.93	&	4658.36	&	4657.07	&	4634.09	&	4496.15	&	4483.49	&	4476.82	&	4476.06	&	-	\\
HE 1150-2546    &   4660.79 &   4708.99 &   4730.46 &  4729.54  &  4706.25  &   4407.06 &   4394.23 &   4387.14 & 4385.74  & - \\
HE 1152-0355    &  4050.00  & 4260.83 & 4276.08 & 4273.00 & 4251.69 & 4199.16 & 4185.87 & 4177.81 & 4174.91 & - \\
HE 1157-0518	&	4921.67	&	4871.05	&	4894.97	&	4894.92	&	4870.92	&	4770.93	&	4758.18	&	4751.59	&	4751.13	&	4700	\\
HE 1158-0708	&	4078.22	&	4004.33	&	4016.38	&	-	&	-	&	4547.94	&	4535.39	&	4528.94	&	4528.56	&	-	\\
HE 1205-0417	&	3983.95	&	4046.81	&	4059.37	&	4055.42	&	4035.04	&	3747.61	&	3732.73	&	-	&	-	&	-	\\
HE 1205-0521    & 4768.22  & 5109.28 & 5137.00 & 5138.35 & 5113.29 & 4483.35 & 4470.66 & 4463.92 & 4463.07 & 5000 \\
HE 1205-0849	&	4367.91	&	4451.95	&	4469.75	&	4467.54	&	4445.39	&	4505.85	&	4493.21	&	4486.58	&	4485.88	&	4600	\\
HE 1208-1247	&	4341.23	&	4369.97	&	4386.66	&	4384.07	&	4362.28	&	4410.05	&	4397.22	&	4390.14	&	4388.76	&	-	\\
HE 1212-0323	&	4415.04	&	4458.66	&	4476.55	&	4474.38	&	4452.19	&	4395.19	&	4382.34	&	4375.19	&	4373.71	&	-	\\
HE 1212-1414	&	4365.47	&	4431.96	&	4449.48	&	4447.18	&	4425.12	&	-	&	-	&	-	&	-	&	-	\\
\hline
\end{tabular}}

\end{table*}
}

{\footnotesize
\begin{table*}
\resizebox{1.1\textwidth}{!}{\begin{tabular}{lcccccccccc}
\hline
star name	&	T$_{eff}$	& T$_{eff}$ &	T$_{eff}$	&	T$_{eff}$	&	T$_{eff}$	&	T$_{eff}$ & T$_{eff}$ &T$_{eff}$ &	T$_{eff}$	&	T$_{eff}$	\\
		      &	          &	(-0.5)	  & (-1.0)    &	(-1.5)	  &	(-2.0)    &	(-0.5)	  & (-1.0)    &	(-1.5)	 &	(-2.0)    &	(H$_{\alpha}$)	\\		
	        &	(J-K)	    &	(J-H)	    &	(J-H)     & (J-H)	    &	(J-H)	    &	(V-K)	    &	(v-k)     & (V-K)	   &	(V-K)	    &	\\
	\hline
HE 1221-0651	&	4458.37	&	4451.95	&	4469.75	&	4467.54	&	4445.39	&	4566.55	&	4554.03	&	4547.67	&	4547.42	&	-	\\
HE 1236-0337	&	4289.15	&	4447.18	&	4449.48	&	4289.15	&	4425.12	&	4219.49	&	4206.25	&	4198.28	&	4195.54	&	-	\\
HE 1238-1714	&	4617.85	&	4743.52	&	4765.49	&	4764.76	&	4741.32	&	4505.04	&	4492.4	&	4485.76	&	4485.06	&	-	\\
HE 1241-0337    & 4189.85 & 4382.86 & 4399.71 & 4397.19 & 4375.34 & 4046.67 & 4032.98 & 4024.16 & 4020.13 & 4000\\
HE 1247-2554	&	4343.64	&	4502.85	&	4521.35	&	4519.39	&	4497.01	&	-	&	-	&	-	&	-	&	-	\\
HE 1251-2313	&	4830.66	&	4980.5	&	5006.14	&	5006.72	&	4982.23	&	4796.24	&	4783.92	&	4777.82	&	4777.91	&	-	\\
HE 1255-2324	&	4660.79	&	4925.09	&	4949.85	&	4950.12	&	4925.87	&	4455.8	&	4443.06	&	4436.19	&	4435.15	&	-	\\
HE 1305+0007  & 4687.06 & 4899.98 & 4924.35 & 4924.47 & 4900.34 & 4460.48 & 4440.91 & 4447.75 & 4439.89 & -\\
HE 1308-1012	&	5010.14	&	5104.72	&	5132.36	&	5133.68	&	5108.64	&	5263.57	&	5260.11	&	5264.15	&	5275.7	&	-	\\
HE 1318-1657	&	4551.21	&	4826.35	&	4849.58	&	4849.29	&	4825.48	&	4489.73	&	4477.06	&	4470.35	&	4469.55	&	-	\\
HE 1318-2451	&	4543.05	&	4660.11	&	4680.86	&	4679.68	&	4656.61	&	4381.99	&	4369.11	&	4361.91	&	4360.33	&	-	\\
HE 1319-1935    & 4678.00 & 4623.28 & 4643.49 & 4642.13 & 4619.22 & 4546.27 & 4533.71 & 4527.25 & 4526.85 & - \\
HE 1319-2340	&	4632.05	&	4778.62	&	4801.13	&	4800.58	&	4800.58	&	4326.57	&	4313.57	&	4306.11	&	4306.11	&	-	\\
HE 1336+0200	&	3739.45	&	3950.12	&	3961.53	&	-	&	-	&	3687.35	&	3672.08	&	-	&	-	&	-	\\
HE 1345-2616	&	4392.55	&	4572.81	&	4592.3	&	4590.68	&	4567.99	&	-	&	-	&	-	&	-	&	-	\\
HE 1354-2552	&	4397.52	&	4548.05	&	4567.18	&	4565.44	&	4542.87	&	4232.02	&	4218.81	&	4210.91	&	4208.25	&	-	\\
HE 1406-2016    &  4623.51  &  4739.65  &   4761.57 &   4760.82 &   4737.39 &  4346.13  &  4333.18  &   4325.81 &  4323.97 & - \\
HE 1429-0551    & 4609.4  & 5037.33 & 5063.87 & 5064.79 & 5040.04 & 4527.13 & 4514.54 & 4507.99 & 4507.46 & 4800 \\
HE 1430+0227	&	4500.19	&	4708.99	&	4730.46	&	4729.54	&	4706.25	&	4486.53	&	4473.86	&	4467.13	&	4466.3	&	-	\\
HE 1431-0245	&	4915.03	&	4663.83	&	4684.63	&	4683.47	&	4660.39	&	4736.91	&	4723.59	&	4716.37	&	4715.16	&	-	\\
HE 1442-0346	&	4719.69	&	4875.16	&	4899.14	&	4899.12	&	4875.09	&	4757.87	&	4744.9	&	4738.07	&	4737.31	&	4950	\\
HE 1523-1155    &  4768.20  & 4770.77 & 4793.16 & 4792.56 & 4769.00 & 4580.26 & 4567.77 & 4561.47 & 4561.32 & 5000 \\
HE 1528-0409    &  4872.30  &  4916.69  &  4941.32  &  4941.53  &  4917.33 &  4648.11  &  4633.38  &  4624.51 & 4621.44 & - \\
HE 2138-1616	&	4891.95	&	5068.53	&	5095.58	&	5096.68	&	5071.79	&	4493.54	&	4481.89	&	4475.19	&	4474.42	&	4600	\\
HE 2145-1715	&	4341.23	&	4254.78	&	4269.95	&	4266.85	&	4245.57	&	4233.91	&	4220.71	&	4212.82	&	4210.17	&	4100	\\
HE 2150-1800    &  5017.11  &   5095.62 &   5123.11 &   5124.37 &   5099.37 &  4976.52  &   4967.40 & 4964.96 &  4969.20 & 4900 \\
HE 2218+0127	&	5623.27	&	5785.56	&	5825.33	&	-	&	-	&	4505.04	&	4492.4	&	4485.76	&	4485.06	&	5400	\\
HE 2224-0330	&	4968.82	&	5068.53	&	5095.58	&	5096.68	&	5071.79	&	4805.42	&	4793.25	&	4787.33	&	4787.62	&	4750	\\
HE 2331-1329    &  -   &  3792.69 &  3802.31  &  -  &  -  &  3802.71  & 3788.13  &  -  & - & - \\
HE 2339-0837    & 4830.7  & 4774.69 &  4797.14  &  4796.57 &  4772.99 & 5092.67  &  5085.75 &  5085.84  &  5092.93 & 4600 \\
HE 2352-1906	&	4672.42	&	4834.41	&	4857.76	&	4857.51	&	4833.67	&	4516.44	&	4503.83	&	4497.24	&	4496.62	&	4450	\\

\hline
\end{tabular}}

The numbers in the parenthesis below T$_{eff}$ indicate the metallicity values at which the temperatures are calculated. Temperatures are given in Kelvin.
\end{table*}
}

\pagebreak

\section{}\label{secA2}

{\footnotesize
\begin{table*}
\caption{Metallicity from CaT lines for stars used for our analysis compared with literature high-resolution spectroscopic [Fe/H] estimates} \label{tableB2}
\resizebox{1.1\textwidth}{!}{\begin{tabular}{lcccccccccccc}
\hline
Star name	&	V	&	Parallax(mas)	&	Mv	&	W8498	&	W8542	&	W8662	&	[Fe/H]             &	[Fe/H] 	     &	[Fe/H] 	  &	[Fe/H] 	&	lit.value	&	reference	\\
          &   &               &      &      &       &       & (W8498+W8542+W8662)& (W8498+W8542) & (W8542+W8662) & (W8498+W8662) & & \\ 
          \hline
HD 4395	&	7.68	&	11.14$\pm$0.05	&	2.91	&	1.22	&	2.53	&	2.39	&	-0.16	&	-1.31	&	-0.74	&	-1.38	&	-0.27	&	1	\\
	      &		    &		    &		    &		    &		    &		    &		    &		    &		    &		    &	-0.38	&	2	\\
	      &		&		&		&		&		&		&		&		&		&		&	-0.16	&	3	\\
HD 5395	&	4.62	&	17.28$\pm$0.19	&	0.81	&	0.69	&	2.15	&	1.6	&	-1.48	&	-2.22	&	-1.79	&	-2.49	&	-0.4	&	4	\\
	&		&		&		&		&		&		&		&		&		&		&	-0.51	&	5	\\
	&		&		&		&		&		&		&		&		&		&		&	-0.45	&	6	\\
	&		&		&		&		&		&		&		&		&		&		&	-0.35	&	7	\\
	&		&		&		&		&		&		&		&		&		&		&	-0.24	&	3	\\
HD 16458	&	5.78	&	6.8$\pm$0.07	&	-0.06	&	0.85	&	2.46	&	2.21	&	-1.24	&	-2.19	&	-1.6	&	-2.31	&	-0.35	&	2	\\
	&		&		&		&		&		&		&		&		&		&		&	-0.65	&	3	\\
HD 22049	&	3.73	&	312.2$\pm$0.16	&	6.2	&	1.13	&	2.82	&	2.58	&	0.96	&	-0.44	&	0.35	&	-0.57	&	-0.15	&	8	\\
	&		&		&		&		&		&		&		&		&		&		&	-0.09	&	9	\\
HD 35410	&	5.07	&	16.7$\pm$0.16	&	1.18	&	1.29	&	2.83	&	2.59	&	-0.39	&	-1.54	&	-0.95	&	-1.65	&	-0.33	&	6	\\
	&		&		&		&		&		&		&		&		&		&		&	-0.28	&	2	\\
HD 43587	&	5.7	&	51.8$\pm$0.11	&	4.27	&	1.16	&	2.47	&	2.07	&	-0.005	&	-1.06	&	-0.59	&	-1.27	&	-0.12	&	1	\\
	&		&		&		&		&		&		&		&		&		&		&	-0.03	&	10	\\
	&		&		&		&		&		&		&		&		&		&		&	0.02	&	11	\\
HD 48565	&	7.18	&	20.15$\pm$0.09	&	3.7	&	0.81	&	1.96	&	1.49	&	-0.87	&	-1.64	&	-1.28	&	-1.9	&	-0.63	&	4	\\
	&		&		&		&		&		&		&		&		&		&		&	-0.59	&	3	\\
	&		&		&		&		&		&		&		&		&		&		&	-0.7	&	12	\\
HD 49738	&	5.67	&	2.85$\pm$0.09	&	-2.05	&	2.04	&	4.49	&	3.8	&	0.037	&	-1.38	&	-0.73	&	-1.65	&	-0.05	&	13	\\
HD 55496	&	8.4	&	1.99$\pm$0.04	&	-0.097	&	0.67	&	2.05	&	2.44	&	-1.4	&	-2.47	&	-1.69	&	-2.29	&	-1.49	&	14	\\
	&		&		&		&		&		&		&		&		&		&		&	-1.55	&	12	\\
HD 89668	&	9.41	&	31.17$\pm$0.06	&	6.87	&	0.83	&	2.31	&	2.02	&	0.39	&	-0.75	&	-0.07	&	-0.92	&	-0.13	&	14	\\
HD 90508	&	6.43	&	43.65$\pm$0.43	&	4.63	&	1.27	&	2.81	&	2.61	&	0.59	&	-0.74	&	-0.05	&	-0.85	&	-0.28	&	15	\\
	&		&		&		&		&		&		&		&		&		&		&	-0.37	&	16	\\
HD 104979	&	4.12	&	19.98$\pm$0.22	&	0.62	&	1.49	&	3.41	&	2.72	&	-0.16	&	-1.33	&	-0.79	&	-1.63	&	-0.26	&	14	\\
	&		&		&		&		&		&		&		&		&		&		&	-0.35	&	17	\\
	&		&		&		&		&		&		&		&		&		&		&	-0.51	&	5	\\
	&		&		&		&		&		&		&		&		&		&		&	-0.45	&	6	\\
HD 107574	&	8.54	&	7.47$\pm$0.06	&	2.9	&	0.66	&	1.72	&	1.73	&	-1.14	&	-2.02	&	-1.46	&	-2.01	&	-0.65	&	14	\\
HD 111395	&	6.29	&	58.48$\pm$0.05	&	5.12	&	1.06	&	2.52	&	1.98	&	0.15	&	-0.89	&	-0.4	&	-1.19	&	0.1	&	18	\\
	&		&		&		&		&		&		&		&		&		&		&	0.22	&	19	\\
HD 111721	&	7.97	&	5.08$\pm$0.06	&	1.5	&	1.24	&	3.11	&	2.31	&	-0.32	&	-1.36	&	-0.87	&	-1.73	&	-1.11	&	14	\\
	&		&		&		&		&		&		&		&		&		&		&	-1.39	&	20	\\
	&		&		&		&		&		&		&		&		&		&		&	-1.54	&	21	\\
HD 125079	&	8.63	&	3.38$\pm$0.06	&	1.27	&	1	&	2.82	&	2.42	&	-0.57	&	-1.65	&	-1.01	&	-1.84	&	-0.16	&	22	\\
	&		&		&		&		&		&		&		&		&		&		&	-0.18	&	3	\\
HD 126681	&	9.53	&	17.78$\pm$0.07	&	5.78	&	0.66	&	2.07	&	1.66	&	-0.3	&	-1.22	&	-0.66	&	-1.47	&	-0.9	&	14	\\
	&		&		&		&		&		&		&		&		&		&		&	-1.3	&	23	\\
HD 148897	&	5.24	&	48.7$\pm$0.13	&	-1.32	&	1.28	&	3.36	&	2.85	&	-0.8	&	-1.93	&	-1.3	&	-2.13	&	-1.02	&	3	\\
	&		&		&		&		&		&		&		&		&		&		&	-0.5	&	2	\\
HD 167768	&	6.0	&	9.29$\pm$0.09	&	0.84	&	1.2	&	3.05	&	2.46	&	-0.49	&	-1.56	&	-1.01	&	-1.83	&	-0.51	&	14	\\
	&		&		&		&		&		&		&		&		&		&		&	-0.77	&	13	\\
	&		&		&		&		&		&		&		&		&		&		&	-0.67	&	12	\\
HD 188650	&	5.78	&	2.65$\pm$0.05	&	-2.1	&	1.09	&	2.61	&	2.24	&	-1.63	&	-2.49	&	-2.04	&	-2.64	&	-0.87	&	24	\\
	&		&		&		&		&		&		&		&		&		&		&	-0.67	&	6	\\
	&		&		&		&		&		&		&		&		&		&		&	-0.4	&	12	\\
	&		&		&		&		&		&		&		&		&		&		&	-0.46	&	3	\\
HD 201626	&	8.16	&	4.51$\pm$0.06	&	1.43	&	0.64	&	1.79	&	1.55	&	-1.54	&	-2.29	&	-1.84	&	-2.43	&	-2.1	&	25	\\
	&		&		&		&		&		&		&		&		&		&		&	-1.5	&	26	\\
	&		&		&		&		&		&		&		&		&		&		&	-1.39	&	3	\\
HD 203638	&	5.36	&	12.91$\pm$0.15	&	0.91	&	0.96	&	2.97	&	2.58	&	-0.55	&	-1.69	&	-0.97	&	-1.87	&	0.27	&	27	\\
	&		&		&		&		&		&		&		&		&		&		&	0.17	&	4	\\
HD 204613	&	8.22	&	15.9$\pm$0.29	&	4.23	&	0.85	&	1.35	&	1.79	&	-0.88	&	-1.86	&	-1.33	&	-1.6	&	-0.24	&	14	\\
	&		&		&		&		&		&		&		&		&		&		&	-0.51	&	12	\\
	&		&		&		&		&		&		&		&		&		&		&	-0.32	&	28	\\
HD 212943	&	4.8	&	21$\pm$0.24	&	1.41	&	1.09	&	2.74	&	2.18	&	-0.63	&	-1.62	&	-1.12	&	-1.88	&	-0.45	&	1	\\
	&		&		&		&		&		&		&		&		&		&		&	-0.2	&	17	\\
HD 216219	&	7.44	&	9.88$\pm$0.05	&	2.41	&	0.96	&	2.45	&	2.2	&	-0.55	&	-1.59	&	-0.99	&	-1.72	&	-0.4	&	1	\\
	&		&		&		&		&		&		&		&		&		&		&	-0.36	&	4	\\
	&		&		&		&		&		&		&		&		&		&		&	-0.17	&	3	\\
 HE 0106-0136 	&	11.77	&	 2.98$\pm$0.06 	&	4.15	&	0.34	&	0.75	&	0.47	&	-2.31	&	-2.76	&	-2.6	&	-3.08	&	-1.85	&	29	\\
HE 0117-0201 	&	12.65	&	 0.26$\pm$0.06 	&	-0.25	&	0.33	&	0.47	&	0.39	&	-3.63	&	-4.21	&	-4.07	&	-4.38	&	-2.63	&	29	\\
HE 0110-0406 	&	12.48	&	 0.65$\pm$0.07 	&	1.56	&	0.86	&	1.04	&	1.01	&	-0.47	&	-1.25	&	-1.04	&	-1.99	&	-1.3	&	30	\\
HE 0315-1749 	&	10.18	&	 6.60$\pm$0.03 	&	4.28	&	0.24	&	0.48	&	0.36	&	-2.75	&	-3.35	&	-3.11	&	-3.64	&	-2.39	&	29	\\
HE 0852-0139 	&	10.8	&	 9.07$\pm$0.04 	&	5.58	&	1.1	&	1.98	&	1.38	&	0.27	&	-0.54	&	-0.66	&	-0.96	&	-1.03	&	29	\\
HE 0920-0506 	&	10.95	&	 2.37$\pm$0.06 	&	2.83	&	0.26	&	1.13	&	1	&	-2.63	&	-3.21	&	-2.82	&	-3.38	&	-1.01	&	29	\\
HE 0959-1714 	&	12.29	&	 0.47$\pm$0.04 	&	0.67	&	0.37	&	0.75	&	0.52	&	-3.05	&	-3.5	&	-3.32	&	-3.78	&	-2.37	&	29	\\
HE 1008-0938 	&	9.98	&	 7.65$\pm$0.08 	&	4.39	&	0.69	&	1.7	&	1.45	&	-1.33	&	-1.91	&	-1.61	&	-2.2	&	-1.01	&	29	\\
HE 1119-0218 	&	11.42	&	 2.39$\pm$0.04 	&	3.32	&	0.48	&	0.96	&	0.82	&	-2.21	&	-2.7	&	-2.51	&	-2.85	&	-1.67	&	29	\\
HE 1304-0651 	&	10.85	&	 2.06$\pm$0.05 	&	2.42	&	0.3	&	0.73	&	0.63	&	-2.61	&	-3.2	&	-2.84	&	-3.23	&	-2.19	&	29	\\
HE 1316-1236 	&	10.3	&	 2.88$\pm$0.04 	&	2.6	&	0.39	&	0.96	&	0.62	&	-2.44	&	-2.83	&	-2.67	&	-3.35	&	-1.41	&	29	\\
HE 1348-0135 	&	12.4	&	 0.58$\pm$0.05 	&	1.24	&	0.27	&	0.56	&	0.55	&	-3.15	&	-3.63	&	-3.45	&	-4.03	&	-2.61	&	29	\\
HE 1442-0848 	&	10.31	&	 4.41$\pm$0.10 	&	3.53	&	0.36	&	0.78	&	0.69	&	-2.71	&	-3.15	&	-3.14	&	-3.48	&	-0.95	&	29	\\
HE 1521-0138 	&	11.15	&	 1.64$\pm$0.04 	&	2.23	&	0.49	&	0.99	&	0.81	&	-2.39	&	-2.96	&	-2.64	&	-3	&	-1.59	&	29	\\
HE 2137-1240 	&	11.14	&	 0.18$\pm$0.10 	&	-2.56	&	0.39	&	0.73	&	0.58	&	-3.51	&	-3.94	&	-3.76	&	-4.09	&	-2.99	&	29	\\
HE 2144-1832 	&	10.97	&	 0.42$\pm$0.04 	&	-0.92	&	1.11	&	1.59	&	1.5	&	-0.96	&	-1.73	&	-1.66	&	-2.24	&	-1.7	&	31	\\
HE 2349-1709 	&	9.81	&	 14.66$\pm$0.07	&	5.64	&	0.84	&	1.96	&	1.58	&	-1.05	&	-1.71	&	-1.41	&	-1.88	&	-1.06	&	29	\\
HE 2320-0313 	&	10.66	&	 6.45$\pm$0.03 	&	4.71	&	0.75	&	1.81	&	1.48	&	-1.26	&	-1.89	&	-1.58	&	-2.12	&	-1.03	&	29	\\

\hline
\end{tabular}}

1. \cite{Boeche_2016}, 2. \cite{Soubiran_2016}, 3. \cite{Karinkuzhi_2014}, 4. \cite{Prugniel_2011}, 5. \cite{Massarotti_2008}, 6. \cite{Liu_2014},7. \cite{Jofre_2015}, 8. \cite{Delgado_2017}, 9. \cite{Jofre_2014}, 10. \cite{Montes_2018}, 11. \cite{Gillon_2006}, 12. \cite{Cenarro_2007}, 13. \cite{Hekker_2007}, 14. \cite{Karinkuzhi_2015}, 15. \cite{Brewer_2016}, 16. \cite{Ramirez_2013}, 17. \cite{Silva_2015}, 18. \cite{Mishenina_2012}, 19. \cite{Maldonado_2012}, 20. \cite{Beers_2014}, 21. \cite{Fulbright_2000}, 22. \cite{1993ApJ...417..287S}, 23. \cite{2016A&A...586A..49B}, 24. \cite{Adamczak_2014}, 25. \cite{Abate_2015}, 26. \cite{Placco_2015}, 27. \cite{Koleva_2012},  28. \cite{Ramirez_2012}, 29. \cite{Frebel_2006}, 30. \cite{Purandardas_2019a}, 31. \cite{Hansen_2016c}

\end{table*}}

\end{appendices}

\end{document}